\pdfmapfile{mypdftex.map}
\pdfmapfile{+pdftex.map}

\documentclass[twoside]{IEEEtran}

\usepackage{amsmath,amssymb}
\usepackage{amsthm} \theoremstyle{definition}
\usepackage[varg]{txfonts}
\usepackage[normalem]{ulem}
\usepackage[]{cite,url}
\usepackage{multirow,bigdelim}

\def\Eq#1{(\ref{#1})}
\def\Thm#1{Theorem~\ref{#1}}
\def\Lma#1{Lemma~\ref{#1}}
\def\Sect#1{Section~\ref{#1}}

\def\Table#1{Table~\ref{#1}}
\def\Prop#1{Proposition~\ref{#1}}
\def\Coro#1{Corollary~\ref{#1}}
\def\Def#1{Definition~\@\ref{#1}}
\def\ie{{i.\@e.\@,~}}
\def\eg{{e.\@g.\@,~}}

\newcommand{\argmin}{\mathop{\rm arg\,min}\limits}

\def\rank{\mathrm{rank\,}}
\def\F{\mathbb{F}}

\def\E{\mathrm{E}}
\def\c{\mathcal{C}}
\def\dc{\mathcal{C}^{\perp}}
\def\d{\mathcal{D}}
\def\dd{\mathcal{D}^{\perp}}
\def\row#1{\mathrm{row}\left(#1\right)}
\def\colq#1{\mathfrak{S}(#1)}
\def\spanvec#1{\langle #1 \rangle}
\def\dimq{\mathrm{dim}_{\F_q}}

\def\colinv{\Gamma(\F_{q^m}^n)}
\def\colinvi#1{\Gamma_{#1}(\F_{q^m}^n)}
\def\colsubmat{\Lambda(\F_{q^m}^n)}
\def\colsubmati#1{\Lambda_{#1}(\F_{q^m}^n)}
\def\colgcv{\Upsilon_\mathcal{F}}

\def\uequivocation{{\Theta}}
\def\ustrong{{\omega}}
\def\maxustrong{{\Omega}}

\def\bari#1{\,\overline{{ \{#1\} }}\,}
\def\vecxi{Z_{\bari{i}}}

\newtheorem{theorem}{Theorem}
\newtheorem{lemma}[theorem]{Lemma}
\newtheorem{proposition}[theorem]{Proposition}
\newtheorem{corollary}[theorem]{Corollary}
\newtheorem{remark}[theorem]{Remark}
\newtheorem{definition}[theorem]{Definition}

\allowdisplaybreaks
\begin{document}
\title{Relative Generalized Rank Weight of Linear Codes and Its Applications to Network Coding}
\author{Jun~Kurihara,~\IEEEmembership{Member,~IEEE},
Ryutaroh~Matsumoto,~\IEEEmembership{Member,~IEEE},\\
and Tomohiko~Uyematsu,~\IEEEmembership{Senior Member,~IEEE}%
\thanks{Manuscript received January 24, 2013; revised December 27, 2013; revised again December 2, 2014; accepted April 18, 2015.
This research was partially supported by
the Japan Society for the Promotion of Science Grant Nos.\ 23246071 and 26289116,
and the Villum Foundation through their VELUX Visiting Professor Programme.
The material in this paper was presented in part at the 2012 IEEE International Symposium on Information Theory, Cambridge, MA, USA, Jul. 2012 \cite{Kurihara2012}, and in part at the 50th Annual Allerton Conference on Communication, Control, and Computing, Monticello, IL, USA, Oct. 2012 \cite{Kurihara2012a}.}
\thanks{J. Kurihara is with KDDI R\&D Laboratories, Inc., 2--1--15 Ohara, Fujimino-shi, Saitama, 356--8502 Japan (e-mail: kurihara@ieee.org).}%
\thanks{T. Uyematsu and R. Matsumoto are with Department of Communications and Integrated Systems, Tokyo Institute of Technology 2--12--1 Ookayama, Meguro-ku, Tokyo, 152--8550 Japan (e-mail: uyematsu@ieee.org; ryutaroh@rmatsumoto.org).}%
\thanks{R. Matsumoto is also with Department of Mathematical Sciences,
Aalborg University, Fredrik Bajers Vej 7G, 9220 Aalborg \O, Denmark.}
\thanks{This paper is registered to the ORCID of Ryutaroh Matsumoto: http://orcid.org/0000-0002-5085-8879.}
\thanks{\copyright\ 2015 IEEE. Personal use of this material is permitted. Permission from IEEE must be obtained for all other uses, in any current or future media, including reprinting/republishing this material for advertising or promotional purposes, creating new collective works, for resale or redistribution to servers or lists, or reuse of any copyrighted component of this work in other works.}
\thanks{Digital Object Identifier {http://dx.doi.org/10.1109/TIT.2015.2429713}.}
}

\maketitle
\begin{abstract}
By extending the notion of \textit{minimum rank distance},
this paper introduces two new relative code parameters of a linear code $\c_1$
of length $n$ over a field extension $\F_{q^m}$
 and its subcode $\c_2 \subsetneqq \c_1$.
One is called
the \textit{relative dimension/intersection profile} (RDIP), and
the other is called
the \textit{relative generalized rank weight} (RGRW).
We clarify
their basic properties and the relation between the RGRW and the minimum rank distance.
As applications of the RDIP and the RGRW,
the security performance and the error correction
capability of secure network coding,
guaranteed independently of the underlying network code,
are analyzed and clarified.
We propose a construction of secure network coding scheme,
and analyze its security performance and error correction capability as an example of applications of the RDIP and the RGRW.
Silva and Kschischang showed the existence of a secure network coding
in which no part of the secret message is revealed to
the adversary even if any $\dim \c_1 -1$ links are wiretapped,
which is guaranteed over any underlying network code.
However, the explicit construction of such a scheme remained an open problem.
Our new construction is just one instance of secure network coding that solves this open problem.
\end{abstract}

\section{Introduction}\label{sect:intro}
\textit{Secure network coding} was first introduced by Cai and Yeung \cite{Cai2011},
and further investigated by Feldman et al.\@ \cite{Feldman2004}.
In the scenario of secure network coding,
a source node transmits $n$ packets from $n$ outgoing links to sink
nodes through a network that
implements network coding \cite{Ahlswede2000a,Koetter2003,Li2003},
and each sink node receives $N$ packets from $N$ incoming links.
In the network, there is
an adversary who eavesdrops $\mu$ links.
The problem of secure network coding is how to encode a secret message into $n$ transmitted
packets at the source node, in such a way that the adversary obtains
as little information as possible
about the message in terms of information theoretic security.

As shown in \cite{Chan2011a,ElRouayheb2012}, secure network coding can be seen as
a generalization of \textit{secret sharing schemes} \cite{Shamir1979,Blakley1979}
or the \textit{wiretap channel II} \cite{Ozarow1984}
to network coding.
The problem of secret sharing schemes
is how to encode a secret message into $n$ information symbols called \textit{shares}
in such a way that the message can be recovered only from certain subsets of shares.
In order to solve both problems of secure network coding and secret sharing schemes,
the \textit{nested coset coding scheme} \cite{Zamir2002}
is commonly used to encode a secret message to shares/transmitted packets,
\eg it has been used in \cite{Duursma2010,Shamir1979,Ozarow1984,ElRouayheb2012,Silva2011,Ngai2011}.
The nested coset coding scheme is defined by
a linear code $\c_1\subseteq \F_{q^m}^n$
and its subcode $\c_2\subsetneqq\c_1$ with $\dim \c_2=\dim \c_1 -l$ ($l\geq 1$) over $\F_{q^m}$,
where $\F_{q^m}$ denotes an $m$-degree ($m > 0$) field extension of a field $\F_q$ of order $q$.
From a secret message of $l$ elements in $\F_{q^m}$,
it generates each transmitted packet/each share defined as an element of $\F_{q^m}$.

Duursma and Park \cite{Duursma2010} defined the \textit{coset distance}
as a relative code parameter of $\c_1$ and $\c_2$.
The coset distance is the minimum value of the Hamming weight of codewords in $\c_1\backslash\c_2$.
They investigated the mathematical properties of the coset distance, and proved
that in the case of secret sharing schemes using the nested coset coding scheme,
the security guarantee of the scheme is exactly expressed in terms of the coset distance
when the message consists of one information symbol, \ie $l=1$.
Motivated by their result using the coset distance,
we \cite{Kurihara2012b} generalized their analysis to
 a secret message
consisting of multiple ($l \geq 1$) information symbols.
In \cite{Kurihara2012b},
it was clarified that
the minimum uncertainty of the message given $\mu (< n)$ shares is 
exactly expressed in terms of
a relative code parameter of $\c_1$ and $\c_2$,
called the \textit{relative dimension/length profile} (RDLP) \cite{Luo2005}.
The paper \cite{Kurihara2012b}
also introduced a definition of the security in secret sharing schemes
for the information leakage of every possible subset of elements composing the message
by generalizing the 
security definition of \textit{strongly secure ramp threshold secret sharing schemes} \cite{yamamoto}.
It was revealed in \cite{Kurihara2012b} that this security is also
exactly expressed in terms of
a relative code parameter of dual codes of $\c_1$ and $\c_2$,
called the \textit{relative generalized Hamming weight} (RGHW) \cite{Luo2005},
where the coset distance coincides with the first RGHW.
We note that in \cite{Luo2005},
the mathematical properties of the RDLP and the RGHW are extensively
investigated in a systematic manner.

\subsection{Our Aims and Motivations}
The main aim of this paper is to extend the work in \cite{Kurihara2012b}
to the security analysis of secure network coding based on the nested coset coding schemes,
and
to demonstrate its security performance guaranteed over \textit{any} underlying network coded network.
Namely, the security performance is guaranteed even over the \textit{random network coding} \cite{Ho2006}.
On the other hand,
the adversary in the scenario of network coding might be able to not only eavesdrop
but also inject erroneous packets anywhere in the network,
and the network may suffer from a rank deficiency
of the transfer matrix at a sink node.
Hence the second aim of this paper is to reveal the error correction capability
of secure network coding based on the nested coset coding schemes with $\c_1$ and $\c_2$
which is guaranteed over \textit{any} underlying network coded network as well as the security performance.

Simultaneously, we also aim to investigate the security performance
and error correction capability in a general manner
with no restriction on parameters.
In particular, we aim to study them
for a smaller extension degree $m$ of $\F_{q^m}$, \ie the packet length.
To see the reason why smaller $m$ deserve investigation in its own right,
consider the case where the secure network coding is implemented
as an application layer overlay network \cite{zhu04} on the Internet.
Recall that the Internet protocol allows an intermediate router to split single
packet into multiple fragments and route those fragments over different paths
\cite{stevens94}.
When the packet length $m$ is larger than the path MTU \cite[Section 2.9]{stevens94} (the maximum size that can avoid
fragmentation on every link to a sink),
symbols in a packet can be routed on different paths.
Shioji et al.\@ \cite{Shioji2010} demonstrated that existing security proofs cannot
ensure the promised security when symbols in a packet are routed on
different paths, because such a case is equivalent to the situation
that an adversary changes the set of eavesdropped links
according to
the position of a symbol in a packet.
Then, the maximum possible $m$ is the path MTU,
and hence the packet length $m$ may have to be small, \eg $m < n$.

Furthermore, consider the software implementation of the encoder and decoder of the nested coset coding scheme.
Recall that
the encoding and decoding operations
are executed
not over the base field $\F_q$ but over the field extension $\F_{q^m}$
in the secure network coding scenario.
In general, the operations over the smaller field work faster on software,
and hence the size of $\F_{q^m}$ should be as small as possible for the fast encoding and decoding operations.
On the other hand, the intermediate network nodes execute $\F_q$-linear operations on packets
as the underlying network coding operations.
Considering the case where the random network coding \cite{Ho2006} is employed as the underlying network coding,
the size of $\F_q$ should be appropriately large
to guarantee the feasibility of the underlying network coded network with high probability near $1$.
Moreover, we may be unable to change the size of $\F_q$
if the network coded network is already in-use,
and the packet length $m$ may be naturally only the parameter that can be changed by the source node.
Especially in such cases,
the $m$ may have to be small like $m < n$ in order to attain the system requirements
for high-speed data processing at software encoder and decoder.

From these observations,
although the majority of existing researches of secure network coding, \eg \cite{Silva2011,Oggier2012}
have assumed $m \geq n$,
it is necessary to consider the secure network coding and its security performance
and error correction capability
for an arbitrary $m$
in a general manner.

\subsection{Our Contributions}
To these ends,
this paper first investigates mathematical properties of
new relative code parameters of a linear code $\c_1 \subseteq \F_{q^m}^n$
and its subcode $\c_2 \subsetneqq \c_1$ in a similar manner to \cite{Luo2005} on the RDLP and the RGHW.
In \Sect{sect:defrdiprgrw} of this paper,
we introduce two new relative code parameters
called the \textit{relative dimension/intersection profile} (RDIP)
and the \textit{relative generalized rank weight} (RGRW),
and give some basic properties of the RDIP and the RGRW.
Similar to the aim of this paper,
Ngai et al.\@ \cite{Ngai2011} introduced a code parameter
called the \textit{network generalized Hamming weight} (Network-GHW),
and later Zhang et al.\@ \cite{Zhang2009} extended Network-GHW to
the \textit{relative generalized network Hamming weight} (R-Network-GHW).
The value of the (R-)Network-GHW depends on the underlying network topology
and the network code,
and hence the security performance expressed in terms of the (R-)Network-GHW is not 
guaranteed independently of the underlying network code.
We will clarify the relation between the R-Network-GHW and the proposed parameters in
\Sect{sect:defparameters}.
We note that the \textit{generalized rank weight} \cite{Oggier2012} was
introduced by Oggier and Sboui concurrently and independently of
the conference version \cite{Kurihara2012a} of this paper,
and that the generalized rank weight is a special case of the RGRW.
In \Sect{sect:defrdiprgrw}, we also
point out that the RGRW can be viewed as a generalization of the \textit{minimum rank distance}
\cite{Gabidulin1985} of a linear code.

In order to measure the security performance of secure network coding,
we first define a criterion called
the \textit{universal equivocation} $\uequivocation_{\mu,P_{S,X}}$,
in \Sect{sect:universalsecurity},
which is the minimum uncertainty of the message under observation with $\mu$ links
for the joint distribution $P_{S,X}$ of the secret message $S$ and the transmitted packets $X$.
Although $P_{S,X}$ have been assumed to be uniform
 for the definition of $\uequivocation_{\mu,P_{S,X}}$
in the conference version of this paper \cite{Kurihara2012a},
we make no assumption regarding $P_{S,X}$ in this paper.
In \cite{ElRouayheb2012,Ngai2011,Cai2011,Zhang2009},
the minimum uncertainty of the message was analyzed,
but their analyses depend on the underlying network code.
In contrast, $\uequivocation_{\mu,P_{S,X}}$
is guaranteed independently of the underlying network code.
Hence, it is called universal in the sense of \cite{Silva2011}.
Next, we introduce the second criterion.
Consider the case where $\uequivocation_{\mu,P_{S,X}}$ is less than
the Shannon entropy \cite[Ch. 2, p. 13]{Cover2006} of the secret message.
Then, some part of the secret message could
be uniquely determined by the adversary.
It is clearly desirable that no part of the
secret message is deterministically revealed
and that every part is kept hidden,
even if some information of the secret
message leaks to the adversary.
From this observation,
we define the \textit{universal $\ustrong$-strong security}
to be the condition where
the mutual information between any $r$ $\F_{q^m}$-symbols of the secret message
and observed packets from arbitrary $\ustrong-r+1$ tapped links $(1\leq r \leq l)$
is always zero.
Note that $\ustrong$ is defined independently of the underlying network code and universal.
The universal strong security defined in \cite{Kurihara2012,Silva2009}
is a special case for $\ustrong = n-1$.

The rest of \Sect{sect:universalsecurity} of this paper gives the main contribution
of the paper: \textit{we demonstrate that the universal security performance of secure network coding
 based on the nested coset coding scheme with $\c_1$ and $\c_2$
 is exactly expressed in terms of our new code parameters, the RDIP and the RGRW.}
This section first presents
 the upper and lower bound of the mutual information
leaked from a set of tapped links with an arbitrary distribution $P_{S,X}$.
By using this analysis,
we demonstrate that
the upper and lower bounds of $\uequivocation_{\mu,P_{S,X}}$ are expressed in terms
of the RDIP of $\c_1$ and $\c_2$ for arbitrary $P_{S,X}$,
and the maximum possible value of $\ustrong$,
defined as the \textit{universal maximum strength} $\maxustrong$,
is expressed in terms of the RGRW of dual codes of $\c_1$ and $\c_2$.
Moreover, in terms of $\maxustrong$,
we express the upper bound of the maximum mutual information between
a part of the secret message and observed packets for arbitrary $P_{S,X}$,
which is independent of the underlying network code.
In a later section, we give an example of this analysis for specific parameters of $\c_1$ and $\c_2$.

For the error correction problem of secure network coding,
in \Sect{sect:errorcorrection},
we define the universal error correction capability
against at most $t$ injected error packets
and at most $\rho$ rank deficiency of the transfer matrix of a sink node.
This is called universal because it is guaranteed independently of the underlying network code,
as well as $\uequivocation_{\mu,P_{S,X}}$ and $\maxustrong$.
Then, \Sect{sect:errorcorrection} gives the other main contribution of the paper:
\textit{We clarified that in secure network coding based on the nested coset coding scheme
with $\c_1$ and $\c_2$,
the universal error correction capability against $t$ errors and $\rho$ rank deficiency
is exactly expressed in terms of the first RGRW of $\c_1$ and $\c_2$.}
Although the conference version \cite{Kurihara2012a} of this paper considered
only the case where the transfer matrix is completely known to each sink node,
the analysis in this paper includes not only the case of known transfer matrices
but also the case where the transfer matrix is unknown to every sink node.

As an example of applications of the above analyses by the RDIP and the RGRW,
\Sect{sect:example} of this paper also proposes a universal strongly secure network coding
constructed from the nested coset coding schemes with $\c_1$ and $\c_2$ for fixed parameters,
and provides its analysis.
An explicit construction of the nested coset coding scheme
that always achieves $\maxustrong = \dim \c_1 - 1$ had remained an open problem \cite{Silva2009}.
Inspired by Nishiara et al.\@'s strongly secure threshold ramp secret sharing scheme
\cite{Nishiara2009} using a Reed-Solomon code and its systematic generator matrix,
\Sect{sect:example} of this paper proposes an explicit construction
with $\maxustrong = \dim \c_1 - 1$ using
an maximum rank distance (MRD) code \cite{Gabidulin1985}
and its systematic generator matrix,
and solves the open problem.
The earlier version of the proposed scheme
was presented in the conference paper \cite{Kurihara2012}.
We note that in \cite{Kurihara2012},
the error correction in the scheme was not considered at all.
With the addition of error correction,
the scheme proposed in this paper is an extension of the earlier version.
Also note that the proposed scheme completely solves
the open problem posed at the end of Section V-B
of the survey paper of Cai and Chan on secure network coding \cite{Chan2011a}.

The analysis of universal security performance of the proposed scheme
is provided as an example of applications of the RDIP and the RGRW
by means of the approach in \Sect{sect:universalsecurity},
which is a different from the analysis
in the conference version \cite{Kurihara2012}.
We also provide an analysis of the universal error correction capability of our scheme
as an application of the RGRW
by the approach of \Sect{sect:errorcorrection}.
Note that by the analyses in \Sect{sect:universalsecurity} and \Sect{sect:errorcorrection},
the universal equivocation and error correction capability of the scheme of Silva and Kschischang \cite{Silva2011}
can be also easily explained in terms of the RDIP and the RGRW for
MRD codes $\c_1$ and $\c_2$.
We shall briefly explain the difference between our sample scheme and \cite{Silva2011}
in the next subsection.

\begin{table*}
\begin{center}
\caption{
Comparison between the scheme in \cite{Silva2011} and our scheme in \Sect{sect:example}
with the following criteria:
The maximum possible number of tapped links $\mu$
with no information leakage of the secret message;
The guarantee of the universal maximum strength $\maxustrong = \dim \c_1 -1$;
The condition to correctly decode the secret message
against $t$ injected error packets
and $\rho$ rank deficiency of the transfer matrix at the sink node.}
\label{tbl:comparison}
\begin{tabular}{|c||c|c||c||c|}
\multicolumn{1}{c}{}
&
\multicolumn{2}{c}{\textbf{Universal Security Performance}}
&
\multicolumn{1}{c}{\textbf{Universal Error Correction Capability}}
&
\multicolumn{1}{c}{}
\\
\hline
& Maximum Possible $\mu$ with
& Guarantee of
& Condition to Correctly Decode the Message
& Necessary Condition
\\
& No Information Leakage
& $\maxustrong = \dim \c_1 - 1$
& against $t$ Errors and $\rho$ Rank Deficiency
& on the Size of $m$
\\
\hline
\hline
\multirow{2}{*}{\cite{Silva2011}}
& \multirow{2}{*}{$\dim \c_2$}
& Not Always
& \multirow{2}{*}{$n-\dim \c_1 \geq 2t+\rho$}
& \multirow{2}{*}{$m \geq n$}
\\
&&($\maxustrong \leq \dim \c_1 - 1$)
&
&\\
\hline
\Sect{sect:example} 
& $\dim \c_2$
& Always
& $n-\dim \c_1 \geq 2t+\rho$
& $m \geq l+n$\\
\hline
\end{tabular}
\end{center}
\end{table*}

\subsection{Difference Between Our Scheme and \cite{Silva2011}}
In \cite{Silva2011}, Silva and Kschischang \cite{Silva2011} proposed a secure network coding scheme
based on the nested coset coding scheme
with MRD codes
$\c_1$ and $\c_2$ \cite{Gabidulin1985}.
Although our scheme is based on the nested coset coding scheme using an MRD code as well as their schemes,
there exist several differences between these two schemes.
\Table{tbl:comparison} summarizes
the comparison between the scheme in \cite{Silva2011} and the proposed scheme
for the universal security performance and the universal error correction capability.

Both in the proposed scheme and in the scheme of Silva and Kschischang \cite{Silva2011},
it is guaranteed that
the universal equivocation $\uequivocation_{\mu,P_{S,X}}$ for $\mu \leq \dim \c_2$
equals the Shannon entropy $H(S)$ \cite[Ch.~2, p.~13]{Cover2006} of the secret message $S$
when the distribution of the transmitted packets $X$ is conditionally uniform given $S$.
This implies that no information about the message leaks out
even if any $\dim\c_2$ links are observed by an adversary.
Both schemes also guarantee
that the secret message is correctly decodable against any $t$ error packets injected somewhere
in the network and $\rho$ rank deficiency of the transfer matrix of the sink node
whenever $n-\dim\c_1+1 > 2t+\rho$ holds.

Our only assumption is that
the network must transport packets of size $m\geq l+n$ symbols.
Although this necessary condition on the packet length is greater than the that
of the scheme of Silva and Kschischang given as $m \geq n$,
our scheme has an advantage on the universal maximum strength over their scheme.
Unlike Silva et al.\@'s scheme \cite{Silva2011},
our scheme always guarantees the universal maximum strength
$\maxustrong = \dim \c_1 - 1=n-1$ for $\c_1 = \F_{q^m}^n$.
This implies that in our scheme
no information about any $r$ $\F_{q^m}$-symbols of the secret message
is obtained by the adversary with $\mu = \dim \c_1 - r$ tapped links
$(1 \leq r \leq l)$.
In \cite{Silva2009}, Silva and Kschischang proved that there exist cases where their scheme \cite{Silva2011}
has $\maxustrong = \dim \c_1 - 1=n-1$,
and showed that the sufficient condition on the existence of such a case is given as
$m \geq (l+n)^2/8+ \log_q 16l$ for the packet length.
However, $\maxustrong = \dim\c_1-1=n-1$ is not always guaranteed in their scheme even if the sufficient condition is satisfied,
and an explicit construction of the scheme that always has $\maxustrong=\dim\c_1-1=n-1$
had been an open problem, as stated in the previous subsection.

\subsection{Organization}
Here again, we briefly show the structure of this paper.
The remainder of this paper is organized as follows.
\Sect{sect:defrdiprgrw} defines the RDIP and RGRW of linear codes,
and introduces their basic properties.
We also show their relations to the existing code parameters in this section.
\Sect{sect:universalsecurity} defines the universal security performance
 over the wiretap network model,
 and reveals that the universal security performance of secure network coding
 is exactly expressed in terms of the RDIP and the RGRW.
In \Sect{sect:errorcorrection},
we also reveal that the universal error correction capability of secure network coding
is exactly expressed in terms of the RGRW.
As an example, an explicit construction of 
strongly secure network coding is proposed in \Sect{sect:example},
and its security performance and error correction capability are analyzed by the RDIP and the RGRW.
Finally, \Sect{sect:conc} presents our conclusions.

\section{New Parameters of Linear Codes and Their Properties}\label{sect:defrdiprgrw}

\subsection{Notations and Preliminaries}
Let $\F_q$ be a finite field containing $q$ elements
and $\F_{q^m}$ be an $m$-degree field extension of $\mathbb{F}_q$
 ($m \geq 1$).
Let $\F_q^n$ denote an $n$-dimensional row vector space over $\F_q$.
Similarly, $\F_{q^m}^n$ denotes an $n$-dimensional row vector space
over $\F_{q^m}$.
Unless otherwise stated,
we consider subspaces, ranks, dimensions, etc, over the field extension $\F_{q^m}$
instead of the base field $\F_q$.

An $[n,k]$ linear code $\c$ over $\F_{q^m}^n$
is a $k$-dimensional subspace of $\F_{q^m}^n$.
Let $\dc$ denote the \textit{dual code} of a code $\c$ \cite[Ch.~1, p.~26]{MacWilliams1977}.
A subspace of a code is called a \textit{subcode}.
For $\c \subseteq \F_{q^m}^n$,
we denote by $\c|\F_{q}$ a \textit{subfield subcode} $\c \cap \F_q^n$
\cite[Ch.~7, p.~207]{MacWilliams1977}.
Observe that $\dim \c$ means the dimension of $\c$ as a vector space
over $\F_{q^m}$ whereas $\dim \c|\F_q$ is the dimension of $\c|\F_q$
over $\F_q$.

For a vector $v=[v_1,\dots,v_n]\in\F_{q^m}^n$
and a subspace $V \subseteq \F_{q^m}^n$,
we denote $v^{\,q} = [v_1^q,\dots,v_n^q]$
and $V^q = \{v^{\,q} : v \in V\}$.
For a subspace $V\subseteq\F_{q^m}^n$,
we define by $V^* \triangleq \sum_{i=0}^{m-1} V^{q^i}$ the sum of
 subspaces $V,V^q,\dots,V^{q^{m-1}}$.
Define a family of subspaces $V\subseteq\F_{q^m}^n$ satisfying $V = V^q$
by
\begin{align*}
\colinv \triangleq \left\{\text{$\F_{q^m}$-linear subspace } V \subseteq \F_{q^m}^n : V = V^q \right\}.
\end{align*}
Also define
\begin{align*}
\colinvi{i} \triangleq \{V\in\colinv : \dim V = i\}.
\end{align*}
For $\colinv$, we have the following lemmas given in \cite{Stichtenoth1990}.
\begin{lemma}[{\cite[Lemma 1]{Stichtenoth1990}}]\label{lma:stichtenothlemma1}
Let $V \subseteq \F_{q^m}^n$ be a subspace.
Then, the followings are equivalent;
1) $V \in \colinv$,
2) There is a basis of $V$ consisting of vectors in $\F_q^n$.
In particular, $V \in \colinv$ if and only if $\dim V|\F_q = \dim V$.
\end{lemma}
\begin{lemma}[{\cite{Stichtenoth1990}}]\label{lma:stichtenothlemma2}
For a subspace $V \subseteq \F_{q^m}^n$,
$V^\ast$ is the smallest subspace in $\colinv$, containing $V$.
\end{lemma}
\begin{lemma}[{\cite{Stichtenoth1990}}]\label{lma:stichtenothlemma3}
For a subspace $V \subseteq \F_{q^m}^n$,
$\dim V^\ast \leq m \cdot \dim V$.
\end{lemma}

\subsection{Definitions of New Parameters}\label{sect:defparameters}
We first define the \textit{relative dimension/intersection profile} (RDIP)
of linear codes as follows.
\begin{definition}[Relative Dimension/Intersection Profile]\label{def:rdip}
Let $\c_1 \subseteq \F_{q^m}^n$ be a linear code and $\c_2\subsetneqq\c_1$ be its
 subcode.
Then, the $i$-th relative dimension/intersection profile ($i$-th RDIP) of $\c_1$ and $\c_2$ is
the greatest difference between dimensions of
intersections, defined as
\begin{align}
K_{R,i} (\c_1,\c_2) \triangleq
\max_{V \in \colinvi{i}}
\left\{
\dim(\c_1 \cap V) - \dim(\c_2 \cap V)
\right\}, \label{eq:defrdip}
\end{align}
for $0 \leq i \leq n$.
\end{definition}
Next, we define the \textit{relative generalized rank weight} (RGRW) of linear codes
as follows.
\begin{definition}[Relative Generalized Rank Weight]\label{def:rgrw}
Let $\c_1 \subseteq \F_{q^m}^n$ be a linear code and $\c_2\subsetneqq\c_1$ be its
 subcode.
Then, the $i$-th relative generalized rank weight ($i$-th RGRW) of $\c_1$ and
 $\c_2$ is defined by
\begin{align}
&M_{R,i}(\c_1,\c_2)\nonumber\\
& \triangleq
\min
\left\{ \dim V : 
V \in \colinv,
\dim(\c_1 \cap V) - \dim(\c_2 \cap V) \geq i
\right\},\label{eq:defrgrw}
\end{align}
for $0 \leq i \leq \dim (\c_1/\c_2)$.
\end{definition}

In \cite{Oggier2012},
Oggier and Sboui proposed the \textit{generalized rank weight}
under the restriction of the degree $m \geq n$.
The generalized rank weight
can be viewed as a special case of the RGRW with $\c_2 = \{0\}$ when $m \geq n$.
In the first version of this paper,
we pointed out this fact, but we did not give its proof.
Later, Ducoat proved this in \cite{Ducoat2013}.

Here we briefly explain the relation between these new parameters
and the existing \textit{relative} parameters defined by a code and its subcode.
For an index set $\mathcal{I}\subseteq\{1,\dots,n\}$, define a subspace
 $\mathcal{E}_\mathcal{I} \triangleq \left\{x=[x_1,\dots,x_n] \in \F_{q^m}^n : \text{$x_i = 0$ for $i \notin \mathcal{I}$}\right\} \subseteq \F_{q^m}^n$.
We have $\dim \mathcal{E}_\mathcal{I} = |\mathcal{I}|$.
Let $\colsubmat$ and $\colsubmati{i}$ for $0 \leq i \leq n$
be collections of $\F_{q^m}$-linear subspaces of $\F_{q^m}^n$, defined by
\begin{align*}
\colsubmat &\triangleq
\left\{
\mathcal{E}_\mathcal{I} \subseteq \F_{q^m}^n : \mathcal{I} \subseteq \{1,\dots,n\}
\right\},
\\
\colsubmati{i} &\triangleq
\left\{
 \mathcal{E}_\mathcal{I} \in \colsubmat
: \dim \mathcal{E}_{\mathcal{I}}=i
\right\}.
\end{align*}
The $i$-th \textit{relative dimension/length profile} (RDLP)
defined by Luo et al.\@ \cite{Luo2005}
is obtained by replacing $\colinvi{i}$ in \Eq{eq:defrdip} with $\colsubmati{i}$.
Also, the \textit{relative generalized Hamming weight} (RGHW) \cite{Luo2005}
is given by replacing $\colinv$ in \Eq{eq:defrgrw} with $\colsubmat$.
Additionally, the \textit{generalized Hamming weight} (GHW) \cite{Wei1991}
is obtained by replacing $\colinv$ in \Eq{eq:defrgrw} with $\colsubmat$ and setting $\c_2=\{0\}$.

\begin{remark}
For an arbitrary index set $\mathcal{I} \subseteq \{1,\dots,n\}$,
a basis of $\mathcal{E}_\mathcal{I}$
is $\{ e^{(i)}=[e^{(i)}_1,\dots,e^{(i)}_n ] : i \in \mathcal{I} \}$ from the definition of $\mathcal{E}_\mathcal{I}$,
where $e^{(i)}_j = 1$ if $i =j$, and $e^{(i)}_j = 0$ if $i \neq j$.
This implies that a basis of $\mathcal{E}_\mathcal{I}$ consists of vectors in $\F_q^n$,
and hence we have $\mathcal{E}_\mathcal{I} \in \colinv$ from \Lma{lma:stichtenothlemma1}.
We thus have $\colsubmat \subseteq \colinv$ and $\colsubmati{i} \subseteq \colinvi{i}$.
This implies that the RDIP of linear codes is always greater than or equal to the RDLP of the codes,
and that the RGRW of linear codes is always smaller than or equal to the RGHW of the codes.
\end{remark}

We also show the relation between
the RGRW and the \textit{relative network generalized Hamming weight} (R-Network-GHW) \cite{Zhang2009}.
Let $\mathcal{F}$ be a set of some one-dimensional subspaces of $\F_{q}^n$.
Each subspace in $\mathcal{F}$ was defined as a space spanned
 by a global coding vector \cite[Ch.~2, p.~18]{Fragouli2007}
 of each link in the network coded network.
(For the definition of global coding vectors, see \Sect{sect:basicnetwork} or \cite{Fragouli2007}).
Let $2^{\mathcal{F}}$ be the power set of $\mathcal{F}$.
For $2^{\mathcal{F}}$,
define a set of direct sums of subspaces by
\begin{align*}
\colgcv \triangleq
\left\{
W \subseteq \F_{q}^n :
W= \sum_{V \in \mathcal{J}} V,
\mathcal{J} \in 2^{\mathcal{F}}
\right\}.
\end{align*}
We restrict the degree $m$ of field extension $\F_{q^m}$ to $m=1$,
\ie $\c_1$ and $\c_2$ are $\F_q$-linear subspaces of $\F_q^n$.
Then, the R-Network-GHW of $\c_1$ and $\c_2$ for the network is obtained
by replacing $\colinv$ in \Eq{eq:defrgrw} with $\colgcv$.
In addition, the \textit{network generalized Hamming weight} (Network-GHW) \cite{Ngai2011}
is obtained by replacing $\colinv$ in \Eq{eq:defrgrw} with $\colgcv$ and set $\c_2=\{0\}$,
as the relation between the RGHW and the GHW.

\begin{remark}
Note that in the definitions of R-Network-GHW and Network-GHW,
the field over which the global coding vectors are defined
must coincide with the field over which linear codes $\c_1$ and $\c_2$
are defined.
Hence, we restricted the degree $m$ to $1$ of the field extension $\F_{q^m}$
over which $\c_1$ and $\c_2$ are defined.
In the case of $m=1$, we have $\colgcv \subseteq \colinv$.
Thus, the RGRW of $\c_1$ and $\c_2$ for $m=1$ is always smaller than or equal to the R-Network-GHW.
\end{remark}

\subsection{Basic Properties of the RDIP and the RGRW}
This subsection introduces some basic properties of the RDIP and the RGRW.
They will be used for expressions of
the universal security performance (\Sect{sect:universalsecurity})
and the universal error correction capability (\Sect{sect:errorcorrection}) of secure network coding.

\begin{theorem}[Monotonicity of the RDIP]\label{thm:monotonerdip}
Let $\c_1 \subseteq \F_{q^m}^n$ be a linear code and $\c_2 \subsetneqq \c_1$ be
 its subcode.
Then, the $i$-th RDIP $K_{R,i}(\c_1,\c_2)$ is nondecreasing with $i$
 from $K_{R,0}(\c_1,\c_2)=0$ to $K_{R,n}(\c_1,\c_2)=\dim(\c_1/\c_2)$,
and
$0 \leq K_{R,i+1}(\c_1,\c_2)-K_{R,i}(\c_1,\c_2)\leq 1$ holds.
\end{theorem}
\begin{IEEEproof}
$K_{R,0}(\c_1,\c_2)=0$ and $K_{R,n}(\c_1,\c_2)=\dim(\c_1/\c_2)$
 are obvious from \Def{def:rdip}.
By \Lma{lma:stichtenothlemma1},
for any subspace $V_1 \in \colinvi{i+1}$,
some $V_2$'s satisfying $V_2 \in \colinvi{i}$ and
$V_2 \subsetneqq V_1$ always exist.
This yields $K_{R,i}(\c_1,\c_2)\leq K_{R,i+1}(\c_1,\c_2)$.

Next we show that the increment at each step is at most $1$.
Consider arbitrary subspaces $V, V' \in \colinv$
 such that $\dim V'=\dim V + 1$ and $V \subsetneqq V'$.
Let
$f = \dim(\c_1 \cap V) - \dim(\c_2 \cap V)$ and $g = \dim(\c_1 \cap V') - \dim(\c_2 \cap V')$.
Since
\begin{align*}
 \dim (\c_1 \cap V) +1 \geq \dim (\c_1 \cap V') \geq \dim (\c_1 \cap V),
\end{align*}
holds and $\c_2 \subsetneqq \c_1$,
we have $f+1 \geq g \geq f$ and hence
$K_{R,i}(\c_1,\c_2)+1 \geq K_{R,i+1}(\c_1,\c_2) \geq K_{R,i}(\c_1,\c_2)$.
\end{IEEEproof}\vspace{1.5ex}
We note that if we replace $\colinvi{i}$ with $\colsubmati{i}$ in \Thm{thm:monotonerdip},
it coincides with \cite[Proposition 1]{Luo2005} for the monotonicity of the RDLP.

\begin{lemma}\label{lma:rgrw}
Let $\c_1 \subseteq \F_{q^m}^n$ be a linear code and $\c_2 \subsetneqq \c_1$
 be its subcode.
Then, the $i$-th RGRW $M_{R,i}(\c_1,\c_2)$ is strictly increasing with $i$.
Moreover, $M_{R,0}(\c_1,\c_2)=0$ and
\begin{align*}
&M_{R,i}(\c_1,\c_2)\\
&=
\min
\left\{ j : 
K_{R,j}(\c_1,\c_2)
 = i
\right\}\\
 &=
\min
\left\{ \dim V : V \in \colinv,
\dim(\c_1 \cap V) - \dim(\c_2 \cap V)
 = i
\right\},
\end{align*}
for $0\leq i \leq \dim (\c_1/\c_2)$.
\end{lemma}
\begin{IEEEproof}
First we have
\begin{align*}
&\min\left\{ j : 
K_{R,j}(\c_1,\c_2)
 \geq i
\right\}\\
&=
\min
\big\{
j : \exists V \in \colinvi{j},\\
&\qquad\qquad
 \text{such that }
 \dim(\c_1 \cap V)
-\dim(\c_2 \cap V) \geq i
\big\}\\
&=
\min\left\{
\dim V : V \in \colinv, \dim(\c_1 \cap V)
-\dim(\c_2 \cap V) \geq i
\right\}\\
&= M_{R,i}(\c_1,\c_2).
\end{align*}
We also have
\begin{align*}
\left\{ j : 
K_{R,j}(\c_1,\c_2)
 = i
\right\}
\cap
\left\{ j : 
K_{R,j}(\c_1,\c_2)
 \geq i+1
\right\}
=\emptyset, 
\end{align*}
by the basic set theory.
Recall that from \Thm{thm:monotonerdip},
$K_{R,j}(\c_1,\c_2)$ is nondecreasing function of $j$
and 
$\left\{ j :  K_{R,j}(\c_1,\c_2) = i \right\}\neq \emptyset$
for all $i \in \left\{0,\dots,\dim(\c_1/\c_2)\right\}$.
We thus have
\begin{align*}
M_{R,i}(\c_1,\c_2) &=
\min\left\{
j : K_{R,j}(\c_1,\c_2)
 \geq i
\right\}\\
&=
\min\left\{
j : K_{R,j}(\c_1,\c_2)= i
\right\}.
\end{align*}
Therefore the RGRW is strictly increasing with $i$ and thus
\begin{align*}
&M_{R,i}(\c_1,\c_2)\\
&=
\min
\left\{
\dim V : V \in \colinv, \dim(\c_1 \cap V)
-\dim(\c_2 \cap V) = i
\right\},
\end{align*}
is established.
\end{IEEEproof}\vspace{1.5ex}
In \cite[Lemma 1]{Oggier2012},
it was shown that
in the case of $\c_2 = \{0\}$,
the second RGRW $M_{R,2}(\c_1,\{0\})$ is greater than the first RGRW $M_{R,1}(\c_1,\{0\})$.

We note that if we replace $\colinv$ and $K_{R,j}(\c_1,\c_2)$ in \Lma{lma:rgrw}
 with $\colsubmat$ and the $j$-th RDLP,
the lemma coincides with \cite[Theorem 3]{Luo2005} for the properties of RGHW.
Also, if we replace $\colinv$ in \Lma{lma:rgrw} with $\colgcv$,
the property of strictly increasing the RGRW shown in the lemma
also becomes the property of the R-Network-GHW \cite[Theorem 3.2]{Zhang2009}.

Now we present the following upper bound of the RGRW.
\begin{proposition}
\label{prop:generalizedsingleton}
Let $\c_1 \subseteq \F_{q^m}^n$ be a linear code and $\c_2 \subsetneqq \c_1$
 be its subcode.
Then, the RGRW of $\c_1$ and $\c_2$ is upper bounded by
\begin{align}
M_{R,i}(\c_1,\c_2)
\leq \min\left\{ n- \dim \c_1, (m-1) \dim \c_1/\c_2 \right\} + i, \label{eq:singleton}
\end{align}
for $1 \leq i \leq \dim(\c_1/\c_2)$.
\end{proposition}
\begin{IEEEproof}
We can assume
that $\c_2$ is a systematic code without loss of generality.
That is, we can choose a basis of $\c_2$ in such a way that
the set of subvectors consisting of the first $\dim \c_2$ coordinates of the chosen basis
coincides with the canonical basis of $\F_{q^m}^{\dim \c_2}$.
Let $\mathcal{S}\subsetneqq\F_{q^m}^n$ be a linear code such that $\c_1$ is the
 direct sum of $\c_2$ and $\mathcal{S}$.
Then, after suitable permutation of coordinates,
a basis of $\mathcal{S}$ can be chosen such that its first $\dim \c_2$ coordinates are zero.
Hence, a code $\mathcal{S}$ can be regarded as a code of length $n-\dim\c_2$,
and we have
$M_{R,\dim \mathcal{S}}(\mathcal{S},\{0\}) \leq n-\dim \c_2$
from the definition of the RGRW.
On the other hand, since
$M_{R,\dim \mathcal{S}} (\mathcal{S},\{0\}) = \dim \mathcal{S}^*$ from the definition of the RGRW and \Lma{lma:stichtenothlemma2},
and $\dim \mathcal{S}^* \leq m \cdot \dim \mathcal{S}$
 from \Lma{lma:stichtenothlemma3},
we have $M_{R,\dim \mathcal{S}}(\mathcal{S},\{0\}) \leq m \cdot \dim \mathcal{S}= m \cdot \dim \c_1/\c_2$.
We thus have
\begin{align*}
M_{R,\dim \mathcal{S}}(\mathcal{S},\{0\})
 \leq \min\left\{ n  - \dim \c_2, m \cdot \dim \c_1/\c_2 \right\}.
\end{align*}

We shall use the mathematical induction on $t$.
We see that 
\begin{align}
M_{R,t}(\mathcal{S},\{0\}) \leq \min \left\{ n- \dim \c_1, (m-1) \dim \c_1/\c_2 \right\} + t,
\label{eq:midproofsingleton}
\end{align}
is true for $t=\dim \mathcal{S} = \dim \c_1 - \dim \c_2$.
Assume that for some $t \geq 1$,
\Eq{eq:midproofsingleton} is true.
Then, since the $M_i(\mathcal{S},\{0\})$
 is strictly increasing with $i$ from \Lma{lma:rgrw},
we have
\begin{align*}
M_{R,t - 1}(\mathcal{S},\{0\})
&\leq M_{R,t}(\mathcal{S},\{0\})-1\\
&\leq \min \left\{ n- \dim \c_1, (m-1) \dim \c_1/\c_2 \right\} +t-1,
\end{align*}
holds.
Thus, it is proved by mathematical induction that
\Eq{eq:midproofsingleton} holds for $1 \leq t \leq \dim (\c_1/\c_2)$.

Lastly, we prove \Eq{eq:singleton} by the above discussion about the RGRW
 of $\mathcal{S}$ and $\{0\}$.
For an arbitrarily fixed subspace $V \subseteq \F_{q^m}^n$,
we have
$\dim (\c_1 \cap V) 
\geq \dim (\mathcal{S} \cap V) + \dim (\c_2 \cap V)$,
because $\c_1$ is a direct sum of $\mathcal{S}$ and $\c_2$.
Hence,
$\dim (\c_1 \cap V) -  \dim (\c_2 \cap V)\geq \dim (\mathcal{S} \cap V)$
holds, and 
we have $M_{R,i}(\c_1,\c_2) \leq M_{R,i}(\mathcal{S},\{0\})$
 for $1 \leq i \leq \dim (\c_1/\c_2)$ from the definition of the RGRW.
Therefore, from the foregoing proof,
we have 
\begin{align}
M_{R,i}(\c_1,\c_2)
&\leq M_{R,i}(\mathcal{S},\{0\})
\nonumber \\
&\leq \min \left\{ n- \dim \c_1, (m-1) \dim \c_1/\c_2 \right\}+i, \label{eq:xxxxxxxxxx}
\end{align}
for $1 \leq i \leq \dim (\c_1/\c_2)$, and the proposition is proved.
\end{IEEEproof}\vspace{1.5ex}
If $n-\dim\c_2 \leq m \cdot \dim \c_1/\c_2$ holds
and $\colinv$ is replaced with $\colsubmat$,
this lemma coincides with the generalized Singleton bound for the
RGHW \cite[Theorem 4]{Luo2005}.
Also, if $n-\dim\c_2 \leq m \cdot \dim \c_1/\c_2$ holds
and $\colinv$ is replaced with $\colgcv$, \ie the RGRW is replaced to the R-Network-GHW,
it becomes \cite[Theorem 3.4]{Zhang2009}.

\subsection{Relation between the Rank Distance and the RGRW}
Next, we show the
relation between the rank distance \cite{Gabidulin1985} and the RGRW.
We will use the relation to express
the universal security performance (\Sect{sect:universalsecurity})
and the universal error correction capability (\Sect{sect:errorcorrection}) of secure network coding.

For a vector $x=[x_1,\dots,x_n] \in \F_{q^m}^n$,
we denote  by $\colq{x} \subseteq \F_{q^m}$
an $\F_q$-linear subspace of $\F_{q^m}$ spanned by $x_1,\dots,x_n$.
The rank distance \cite{Gabidulin1985} between two vectors
$x,y\in\F_{q^m}^n$ is given by
$d_R(x,y) \triangleq \dimq \colq{y-x}$,
where $\dimq$ denotes the dimension over the base field $\F_q$.
In other words,
it is the maximum number of coordinates in $(y-x)$ that are linearly independent over $\F_q$.
The minimum rank distance \cite{Gabidulin1985} of a code
 $\c$ is given as
\begin{align*}
d_R(\c)
&\triangleq
\min\{d_R(x,y): x,y \in \c, x \neq y\}\\
&= \min\{d_R(x,0): x \in \c, x \neq 0\}.
\end{align*}

\begin{lemma}\label{lma:xxxx}
Let $b \in \F_{q^m}^n$ be an $n$-dimensional nonzero vector over $\F_{q^m}$,
and let $\spanvec{b} \subseteq \F_{q^m}^n$
be an $\F_{q^m}$-linear one-dimensional subspace of $\F_{q^m}^n$ spanned by $b$.
Then, we have $\dim \spanvec{b}^* = d_R(b,0)$.
\end{lemma}
\begin{IEEEproof}
Let $\{\gamma_1,\dots,\gamma_m\}$ be an $\F_q$-basis of $\F_{q^m}$.
Let $d = d_R(b,0) = \dimq \colq{b}$.
From the definition of the rank distance,
there exists a nonsingular matrix $P\in\F_q^{n \times n}$ satisfying
\begin{align*}
b = \underbrace{[\gamma_1,\dots,\gamma_{d},0,\dots,0]}_{\triangleq a \in \F_{q^m}^n} P.
\end{align*}
For $\alpha_1,\alpha_2 \in\F_q$, $\beta_1,\beta_2\in\F_{q^m}$,
we have $\alpha_1\beta_1^{q^i} + \alpha_2\beta_2^{q^i}=(\alpha_1\beta_1 + \alpha_2\beta_2)^{q^i}$
 ($0 \leq i \leq m-1$).
Thus, since $P$ is a matrix over $\F_q$, we have
$b^{q^i} = \left( a P \right)^{q^i} = a^{q^i} P$.
Let $\spanvec{b, b^q,\dots,b^{q^{m-1}}} \subseteq \F_{q^m}^n$
be an $\F_{q^m}$-linear subspace of $\F_{q^m}^n$ spanned by $m$ vectors $b, b^q,\dots,b^{q^{m-1}}$,
then
we have $\spanvec{b}^* = \spanvec{b, b^q, \dots, b^{q^{m-1}}}$.
Hence, since $P$ is nonsingular, we have
\begin{align*}
\dim \spanvec{b}^*
&= \dim \spanvec{b, b^q, \dots, b^{q^{m-1}}}
\\
&= \dim \spanvec{aP, a^q P, \dots, a^{q^{m-1}}P}
\\
&= \dim \spanvec{a, a^q, \dots, a^{q^{m-1}}}
\\
&= \rank
\underbrace{
\begin{bmatrix}
a\\
a^q\\
\vdots\\
a^{q^{m-1}}
\end{bmatrix}
}_{\triangleq T \in \F_{q^m}^{m \times n}}.
\end{align*}
Since the right $n-d$ columns of $T$ are zero columns,
we have $\rank T \leq d$.
On the other hand,
the upper-left $d \times d$ submatrix
$T'$ of $T$ is the generator matrix of Gabidulin code of length $d$ and dimension $d$ \cite{Gabidulin1985},
and hence we must have $\rank T' = d$.
Thus, we have $\rank T \geq d$.
Therefore, we have $\dim \spanvec{b}^* = \rank T = d$.
\end{IEEEproof}\vspace{1.5ex}

\begin{lemma}\label{lma:rankdistance}
For a code $\c_1 \subseteq \F_{q^m}^n$ and its subcode $\c_2 \subsetneqq \c_1$,
the first RGRW can be represented as
$M_{R,1}(\c_1,\c_2) 
= \min \left\{
d_R(x,0) : x \in \c_1\backslash\c_2
\right\}$.
\end{lemma}
\begin{IEEEproof}
From \Lma{lma:stichtenothlemma2},
$M_{R,1}(\c_1,\c_2)$ can be represented as
\begin{align*}
&M_{R,1}(\c_1,\c_2)\\*
&=
\min \left\{
\dim W : W \!\in\! \colinv,
\dim (\c_1 \!\cap\! W) - \dim (\c_2 \!\cap\! W)\geq 1
\right\}\\*
&=
\min \left\{
\dim W : W \in \colinv,
\exists v \in (\c_1 \cap W)\backslash \c_2
\right\}
\\*
&=
\min \left\{
\dim \spanvec{v}^* : v \in \c_1\backslash\c_2
\right\}.
\end{align*}
Therefore, since $\dim \spanvec{v}^* = d_R(v,0)$ for a vector $v \in \F_{q^m}^n$ from \Lma{lma:xxxx},
we have $M_{R,1}(\c_1,\c_2)= \min \left\{ d_R(v,0) : v \in \c_1\backslash\c_2 \right\}$.
\end{IEEEproof}\vspace{1.5ex}
\Lma{lma:rankdistance} immediately yields
that $M_{R,1}(\cdot,\{0\})$ coincides with $d_R(\cdot)$.

\begin{corollary}\label{coro:rankdistance}
For a linear code $\c$, $d_{R}(\c) = M_{R,1}(\c,\{0\})$ holds.

\hfill\IEEEQED
\end{corollary}

Here we introduce the Singleton-type bound of rank distance \cite{Gabidulin1985,Loidreau2008}.
\begin{proposition}[Singleton-Type Bound of Rank Distance {\cite{Gabidulin1985,Loidreau2008}}]\label{prop:singleton}
Let $\c \subseteq \F_{q^m}^n$ be a linear code.
Then, the minimum rank distance of $\c$ is upper bounded by
\begin{align}
d_R(\c)
\leq \min\left\{ 1, \frac{m}{n} \right\}(n-\dim\c) + 1. \label{eq:ineqsingleton}
\end{align}
\end{proposition}
Note that the right-hand side of \Eq{eq:ineqsingleton}
is $n - \dim \c + 1$ if $m \geq n$ and $\frac{m}{n}(n-\dim\c)+1$ if $m < n$.
A code satisfying the equality of \Eq{eq:ineqsingleton} is called
 a \textit{maximum rank distance} (MRD) code \cite{Gabidulin1985}.
The Gabidulin code \cite{Gabidulin1985} is known as an MRD code.

In the following, we shall present some extra properties
 of the RGRW $M_{R,i}(\cdot,\cdot)$ and the minimum rank distance $d_R(\cdot)$
by using the relation between $M_{R,i}(\cdot,\cdot)$ and $d_R(\cdot)$ shown above
and the properties of the RGRW described in the previous subsection.
In the case where $m \geq n$,
\Coro{coro:mrdrgrw} gives a generalization of the Singleton-type bound
 of rank distance \cite{Gabidulin1985,Loidreau2008} of $\c \subseteq \F_{q^m}^n$,
and \Coro{coro:mrdrgrw2} 
shows that the RGRW of $\c_1\subseteq\F_{q^m}^n$ and $\c_2\subsetneqq\c_1$
 depends only on $\c_1$ when $\c_1$ is MRD.
\Prop{prop:firstrgrwsingleton} presents an upper bound of the first RGRW by combining the Singleton-type bound
 of rank distance \cite{Gabidulin1985,Loidreau2008} of $\c \subseteq \F_{q^m}^n$ for $m < n$
 and the upper bound of the RGRW given in \Prop{prop:generalizedsingleton}.
In the case where $m < n$,
\Coro{coro:firstrgrwsingleton2} gives a tighter upper bound of the minimum rank distance of $\c \subseteq \F_{q^m}^n$
for $m < n$ and $\dim \c = 1$ than that shown in \Prop{prop:singleton}.

First,
\Lma{lma:rgrw} and \Prop{prop:generalizedsingleton} yield
the following
corollary from \Coro{coro:rankdistance} and \Prop{prop:singleton}.
This corollary shows a generalization
 of the Singleton-type bound of rank distance \cite{Gabidulin1985,Loidreau2008}
 of $\c\subseteq\F_{q^m}^n$ in the case where $m \geq n$.
\begin{corollary}\label{coro:mrdrgrw}
For a linear code $\c \subseteq \F_{q^m}^n$ with $m \geq n$,
$M_{R,i}(\c,\{0\}) \leq (n-\dim \c)+i$
 for $1 \leq i \leq \dim \c$.
The equality holds for all $i$ if and only if $\c$ is an MRD code.
\end{corollary}
\begin{IEEEproof}
From \Prop{prop:generalizedsingleton}, $M_{R,i}(\c,\{0\}) \leq (n-\dim \c)+i$ is immediate.
The RGRW $M_{R,i}(\c,\{0\})$ is strictly increasing with $i$ from \Lma{lma:rgrw},
and $M_{R,\dim\c} (\c,\{0\}) \leq n$ holds.
Therefore, from \Coro{coro:rankdistance} and \Prop{prop:singleton},
$M_{R,i}(\c,\{0\}) = n-\dim\c + i$ for $1\leq i \leq \dim\c$ must hold
if and only if $\c$ is MRD with $m \geq n$.
\end{IEEEproof}\vspace{1.5ex}

Next, we give the following corollary of \Prop{prop:generalizedsingleton}
for the RGRW of $\c_1 \subseteq \F_{q^m}^n$ and $\c_2\subsetneqq\c_1$.
This corollary reveals that when $\c_1$ is an MRD code with $m \geq n$,
the $i$-th RGRW $M_{R,i}(\c_1,\c_2)$ always coincides with the maximum possible value of $M_{R,i}(\c_1,\{0\})$,
 shown in \Coro{coro:mrdrgrw},
 regardless of its subcode $\c_2$.
\begin{corollary}\label{coro:mrdrgrw2}
Let $m \geq n$.
Let $\c_1 \subseteq \F_{q^m}^n$ be an MRD code and $\c_2 \subsetneqq \c_1$
 be its arbitrary subcode.
Then, the RGRW of $\c_1$ and $\c_2$ is
$M_{R,i}(\c_1,\c_2) = n -\dim \c_1 + i$
for $1 \leq i \leq \dim(\c_1/\c_2)$.
\end{corollary}
\begin{IEEEproof}
By the definition of the RGRW in \Def{def:rgrw},
we first have $M_{R,i}(\c_1,\c_2) \geq M_{R,i}(\c_1,\{0\})$.
Hence, since $\c_1$ is MRD with $m \geq n$,
we have $M_{R,i}(\c_1,\c_2) \geq M_{R,i}(\c_1,\{0\}) = n-\dim \c_1 + i$ from \Coro{coro:mrdrgrw}.
On the other hand,
we have $M_{R,i}(\c_1,\c_2) \leq n-\dim\c_1+i$
from \Prop{prop:generalizedsingleton}.
Therefore,
we have $M_{R,i}(\c_1,\c_2) = n-\dim\c_1+i$.
\end{IEEEproof}\vspace{1.5ex}

By combining \Prop{prop:singleton} and \Prop{prop:generalizedsingleton},
we also have the following proposition only for the first RGRW.
This proposition presents an upper bound of the first RGRW,
 obtained by the Singleton-type bound of the rank distance of $\c \subseteq \F_{q^m}^n$ for $m < n$
 in \Prop{prop:singleton}.
\begin{proposition}\label{prop:firstrgrwsingleton}
The first RGRW of a linear code $\c_1 \subseteq \F_{q^m}^n$ and its subcode $\c_2 \subsetneqq \c_2$ is upper bounded by
\begin{align*}
&M_{R,1}(\c_1,\c_2) \\
&\leq
\min \left\{ n - \dim \c_1, (m-1)\dim \c_1/\c_2, \frac{m(n-\dim\c_1)}{n-\dim\c_2} \right\} +1.
\end{align*}
\end{proposition}
\begin{IEEEproof}
As in the proof of \Prop{prop:generalizedsingleton},
let $\mathcal{S}\subsetneqq\F_{q^m}^n$ be a linear code such that $\c_1 = \c_2 + \mathcal{S}$.
Also, we suppose that the first $\dim \c_2$ coordinates of $\mathcal{S}$ are zero without loss of generality.
Since $\mathcal{S}$ can be viewed as a code of length $n-\dim \c_2$,
we have the following inequality from \Prop{prop:singleton}.
\begin{align*}
d_R(\mathcal{S})
&=
M_{R,1}(\mathcal{S},\{0\})\\
&\leq
\frac{m}{n-\dim \c_2} \left\{ (n-\dim\c_2) - \dim \mathcal{S} \right\}+1\\
&=
\frac{m(n - \dim \c_1)}{n-\dim \c_2}+1.
\end{align*}
Thus, from \Eq{eq:xxxxxxxxxx},
\begin{align*}
M_{R,1}(\c_1,\c_2)
\leq
M_{R,1}(\mathcal{S},\{0\})
\leq
\frac{m(n - \dim \c_1)}{n-\dim \c_2}+1.
\end{align*}
Therefore, from \Prop{prop:generalizedsingleton}, the proposition is proved.
\end{IEEEproof}\vspace{1.5ex}

The following corollary is immediately obtained from \Prop{prop:firstrgrwsingleton}.
\begin{corollary}\label{coro:firstrgrwsingleton2}
Assume $m \geq 2$.
For a linear code $\c \subseteq \F_{q^m}^n$, we have the following inequalities.
\begin{align*}
d_R(\c) &=
M_{R,1}(\c,\{0\})\\
&\leq
\begin{cases}
n - \dim \c+1 & (n \leq m)\\
(m-1)\dim \c+1 & (n > m, \dim \c = 1) \\
\frac{m}{n}(n-\dim\c)+1 & (n > m, \dim \c \geq 2).
\end{cases}
\end{align*}
\hfill\IEEEQED
\end{corollary}
This corollary presents a tighter upper bound of $d_R(\c)$ for $\c \subseteq \F_{q^m}^n$
than that shown in \Prop{prop:singleton},
when $m < n$ and $\dim \c = 1$.

Lastly, 
by using the relation between the RGRW and the rank distance
\cite{Gabidulin1985} presented above,
we introduce an extra property of the RDIP $K_{R,i}(\c_1,\c_2)$ when $\c_1$ is MRD.
We define $\left[ x \right]^+ = \max\{0,x\}$.
\begin{proposition}\label{prop:mrdrdip}
Let $\c_1\subseteq\F_{q^m}^n$ be a linear code
and $\c_2\subsetneqq\c_1$ be a its subcode.
Assume $m \geq n$ and $\c_1$ be an MRD code.
Then, the RDIP of $\c_1$ and $\c_2$ is given by
$K_{R,\mu}(\c_1,\c_2) = \left[\mu- n+\dim \c_1\right]^+$
for $0 \leq \mu \leq n-\dim \c_2$.
\end{proposition}
\begin{IEEEproof}
From \Coro{coro:mrdrgrw2},
we have $M_{R,i}(\c_1,\c_2)=n-\dim\c_1+i$ for $0 \leq i \leq \dim(\c_1/\c_2)$.
Thus, from \Prop{lma:rgrw} for $i=1$, we have
\begin{align*}
\min \left\{\mu : K_{R,\mu}(\c_1,\c_2) = 1 \right\}
 &= n-\dim \c_1 + 1,
\end{align*}
and hence $K_{R,\mu}(\c_1, \c_2)=0$ for $0\leq \mu \leq n-\dim \c_1$
from \Thm{thm:monotonerdip}.
On the other hand,
from \Prop{lma:rgrw} for $i=\dim(\c_1/\c_2)$, we have
\begin{align*}
&\min \left\{\mu : K_{R,\mu}(\c_1,\c_2)= \dim(\c_1/\c_2) \right\}\\
&\quad = n-\dim \c_1+\underbrace{\dim(\c_1/\c_2)}_{=\dim \c_1 - \dim \c_2}\\
&\quad = n-\dim \c_2 ,
\end{align*}
and hence $K_{R,n-\dim \c_2}(\c_1, \c_2)=\dim(\c_1/\c_2)$.
Thus, since
\begin{align*}
K_{R,n-\dim \c_2}(\c_1, \c_2) - K_{R,n-\dim \c_1}(\c_1,\c_2)
&=\dim (\c_1/\c_2)\\
&= \dim \c_1 - \dim \c_2,
\end{align*}
holds,
$K_{R,\mu}(\c_1,\c_2)=\mu-n+\dim \c_1$ for $n-\dim \c_1 \leq \mu \leq n-\dim \c_2$
must hold from \Thm{thm:monotonerdip}.
Therefore, the proposition is established.
\end{IEEEproof}\vspace{1.5ex}

\section{Universal Security Performance of Secure Network Coding}\label{sect:universalsecurity}
This section derives the security performance
of secure network coding based on the \textit{nested coset coding scheme} \cite{Zamir2002},
which is guaranteed independently of the underlying network code construction.

This section first presents the network model with errors,
and introduces the wiretap network model and the nested coset coding scheme
in secure network coding.
Next, we define the universal equivocation, the universal $\ustrong$-strong security
and the universal maximum strength
as the universal security performance of secure network coding on the wiretap network
model.
We then give the main contribution of this paper: we exactly
express the universal security performance of secure network coding based on
the nested coset coding scheme in terms of the RDIP and the RGRW.

\subsection{Network Model with Errors}\label{sect:basicnetwork}
We first introduce the basic network model in which no errors occur in the network.
As in \cite{Silva2011,Ngai2011,Zhang2009,Cai2011,ElRouayheb2012},
we consider a multicast communication network represented by a directed acyclic
multigraph with unit capacity links, a single source node, and multiple
sink nodes.
We assume that \textit{linear network coding} \cite{Li2003,Koetter2003} is
employed over the network.
Elements of a column vector space $\F_q^{m \times 1}$ are called \textit{packets}.
Assume that each link in the network can carry a single $\F_q$-symbol per
one time slot, and that each link transports a single packet over $m$ time
slots without delays, erasures, or errors.

The source node produces $n$ packets
$X_1$, \ldots, $X_n\in \F_q^{m \times 1}$
and transmits $X_1$, \ldots, $X_n$ on $n$ outgoing links
over $m$ consecutive time slots.
Define the $m \times n$ matrix $X=[X_1,\dots,X_n]$.
The data flow on any link can be represented as an $\F_q$-linear
combination of packets $X_1,\dots,X_n \in \F_q^{m \times 1}$.
Namely, the information transmitted on a link $e$ can be denoted as
 $b_e X^{\rm T} \in \F_q^{1 \times m}$,
 where $b_e \in \mathbb{F}_q^n$ is called
 a \textit{global coding vector} \cite[Ch.~2, p.~18]{Fragouli2007} of $e$.
Suppose that a sink node has $N$ incoming links.
Then, the information received at a sink node can be represented as an
 $N \times m$ matrix $AX^{\rm T} \in\F_q^{N \times m}$,
 where $A\in\F_q^{N \times n}$ is
the transfer matrix of the network
 constructed by gathering the global coding vectors of $N$ incoming links.
The network code is called \textit{feasible}
 if each transfer matrix to each sink node
 has rank $n$ over $\F_q$, otherwise it is called rank deficient.
The \textit{rank deficiency} of the network coded network \cite{Silva2011,Silva2008,Silva2009a} is
defined by
\begin{align*}
\rho \triangleq n - \min \left\{ \rank A : \text{$A$ at each sink node}\right\},
\end{align*}
\ie the maximum column-rank deficiency of the transfer matrix $A$ among all sink nodes.
As in \cite{Silva2009a,Silva2008,Silva2011}, $\rho$ is also referred to as \textit{$\rho$ erasures}.

The above setup of the network coded network is referred
to as an \textit{$(n \times m)_q$ linear network} \cite{Silva2011}.
We may also call it a \textit{$\rho$-erasure $(n \times m)_q$ linear network}
when we need to indicate the rank deficiency $\rho$ of the network.

Now we extend the basic model of the $(n \times m)_q$ linear network defined above
to incorporate packet errors, as \cite{Silva2011,Silva2009a}.
We define the network model with errors as follows.
\begin{definition}[$t$-Error $(n \times m)_q$ Linear Network]\label{def:terror}
Suppose that the network is an $(n \times m)_q$ linear network.
Also suppose that at most $t$ error packets, represented by $Z\in\F_q^{m \times t}$, are injected
from $t$ links chosen arbitrarily in the network.
That is, the information transported over a link $e$ with the global coding vector $b_e$ is represented by
$b_e X^{\rm T} + f_e Z^{\rm T} \in \F_q^{1 \times m}$,
where $f_e \in\F_q^{1 \times t}$ corresponds to the overall linear transformation
applied to the injected error packets $Z$ on the route to the link $e$.
Then, the network is called a \textit{$t$-error $(n \times m)_q$ linear network}.
\end{definition}
This $t$-error $(n \times m)_q$ linear network may also be called
 a \textit{$t$-error-$\rho$-erasure $(n \times m)_q$ linear network}
for the rank deficiency $\rho$.
Note that in the $t$-error $(n \times m)_q$ linear network,
the information received at a sink node is expressed as
\begin{align}
Y^{\rm T} = A X^{\rm T} + D Z^{\rm T} \in \F_q^{N \times m}, \label{eq:receivedpackets}
\end{align}
where $D \in\F_q^{N \times t}$ is constructed by gathering $f_e$'s
of incoming links $e$'s to the sink node,
and hence $D$ corresponds to the overall linear transformation
applied to $Z$ on the route to the sink node.

The system of linear network coding is called \textit{coherent}
 if the transfer matrix $A$ is known to each
sink node, otherwise it is called \textit{noncoherent}.

\subsection{Wiretap Network Model and Nested Coset Coding Scheme}\label{sect:nestedcoding}
Following \cite{Silva2011,Zhang2009},
assume that in the $t$-error $(n \times m)_q$ linear network defined in \Def{def:terror},
there is an adversary who observes packets transmitted on any $\mu$ links.
We also assume that the adversary knows the coding scheme applied
 at the source node and all the global coding vectors in the network.

Let $\mathcal{W}$ be the set of $\mu$ links observed by the adversary,
and let $B_\mathcal{W} \in \F_q^{\mu \times n}$ be the transfer matrix whose
rows are the global coding vectors $b_e$'s associated with the links $e$'s in $\mathcal{W}$.
The information obtained by the adversary can be expressed by
\begin{align}
W^{\rm T} = B_\mathcal{W} X^{\rm T} + F_\mathcal{W}Z^{\rm T} \in \F_q^{\mu \times m},
\label{eq:erroneouswiretappedinf}
\end{align}
where $F_{\mathcal{W}}\in\F_q^{\mu \times t}$ is constructed by gathering $f_e$'s
of links $e$'s in $\mathcal{W}$,
and $F_\mathcal{W}Z^{\rm T} \in \F_q^{\mu \times m}$ corresponds to the errors.
In the following, we consider the reliable transmission of a secret message
through this wiretap network model.

The procedure of the secure
message transmission over the wiretap network model
is called \textit{secure network coding} \cite{Cai2011,ElRouayheb2012,Silva2011,Zhang2009,Ngai2011}.
In the scenario of secure network coding,
first regard an $m$-dimensional column vector space
$\F_q^{m \times 1}$ as $\F_{q^m}$,
and fix $l$ for $1 \leq l \leq n$.
Let $S = [S_1,\dots,S_l] \in \F_{q^m}^l$ be the secret message of $l$ packets.
Under the adversary's observation of $\mu$ links,
the source node wants to transmit $S$
as small information leakage to the adversary as possible.
To protect $S$ from the adversary,
the source node encodes $S$ to the transmitted vector $X=[X_1,\dots,X_n]\in\F_{q^m}^n$
 of $n$ packets according to some kind of coding scheme.
Then, the source node finally transmits $X$ as an $m \times n$ matrix
 over $\F_q$ to sink nodes through the network.
In this paper, we assume that the source node knows nothing
about the errors that occur in the network,
 as in the model of \cite{Silva2011,Zhang2009}.

In the secure network coding described in \cite{ElRouayheb2012,Silva2011,Zhang2009,Ngai2011},
$S$ is encoded by the \textit{nested coset coding scheme}
 \cite{Zamir2002,Subramanian2009,Chen2007,Duursma2010} at the source node.
In secure network coding based on the nested coset coding scheme,
$S$ is encoded to $X$ at the source node as follows.
\begin{definition}[Nested Coset Coding Scheme]\label{def:nestedcoding}
Let $\c_1 \subseteq \F_{q^m}^n$ be a linear code over $\F_{q^m}$ ($m \geq 1$),
and $\c_2 \subsetneqq \c_1$ be its subcode with dimension $\dim \c_2 = \dim \c_1 - l$ over $\F_{q^m}$.
Let $\psi:\F_{q^m}^l\rightarrow\c_1/\c_2$ be an arbitrary linear bijection.
For a secret message $S\in\F_{q^m}^l$,
we randomly choose $X$ from a coset $\psi(S) \in \c_1/\c_2$.
We make no assumption on the joint distribution $P_{S,X}$ unless otherwise stated.
\end{definition}
In \cite{Duursma2010,Chen2007,Kurihara2012b}, the nested coset coding scheme is
 called a \textit{secret sharing scheme based on linear codes}.
\Def{def:nestedcoding} includes the Ozarow-Wyner coset coding scheme \cite{Ozarow1984}
as a special case with $\c_1=\F_{q^m}^n$.

Corresponding to $X$ transmitted from the source node,
 the sink node receives a vector of $N$ packets
$Y \in \F_{q^m}^N$.
The decoding of $S$ from $Y$ will be discussed in \Sect{sect:errorcorrection}.

\subsection{Definition of the Universal Security Performance}\label{sect:defsecurity}
In order to measure the security performance of secure network coding in the
above model,
this subsection presents two criteria.
The security performance measured by our criteria is guaranteed
independently of the underlying network code,
hence we call them \textit{universal}.

Let $H(X)$ be the Shannon entropy for a random variable $X$,
$H(X|Y)$ be the conditional entropy of $X$ given $Y$,
 and $I(X;Y)$ be the mutual information between $X$ and $Y$
 \cite[Ch.~2, pp.~12--19]{Cover2006}.
The entropy and the mutual information are
always computed using $\log_{q^m}$.

\subsubsection{Universal Equivocation}
First, we define \textit{universal equivocation} as follows.
\begin{definition}[Universal Equivocation]\label{def:universalperformance}
Assume that the secret message $S$ is chosen according to an arbitrary distribution $P_S$ over $\F_{q^m}^l$,
and suppose that $S$ is encoded to the transmitted packets $X\in\F_{q^m}^n$ by a certain coding scheme.
We make no assumption on the joint distribution $P_{S,X}$.
Then, the \textit{universal equivocation} $\uequivocation_{\mu,P_{S,X}}$ of the coding scheme
is the minimum uncertainty of $S$
 given $BX^{\rm T}$ for all $B\in\F_q^{\mu \times n}$, defined as
\begin{align*}
\uequivocation_{\mu,P_{S,X}} &\triangleq \min_{B \in \F_q^{\mu \times n}}H(S|BX^{\rm T})
\\
&=
H(S) - \max_{B \in \F_q^{\mu \times n}}I(S;BX^{\rm T}).
\end{align*}
We will also call
${\displaystyle \max_{B \in \F_q^{\mu \times n}}I(S;BX^{\rm T})}$ in the above equation
the \textit{maximum amount of information leakage} to the adversary.
\end{definition}
Note that the conditional entropy
of $S$ given $BX^{\rm T}$ is considered in \Def{def:universalperformance}.
But, we need to consider the adversary
in the $t$-error $(n \times m)_q$ linear network as the model presented in \Sect{sect:nestedcoding},
\ie we need to consider the conditional entropy
of $S$ given $W$ that contains the errors,
as given in \Eq{eq:erroneouswiretappedinf}.
In order to justify this difference between \Def{def:universalperformance} and the wiretap network model,
 we derive the following proposition.
\begin{proposition}\label{prop:ngai}
Fix a matrix $B\in\F_q^{\mu \times n}$ arbitrarily.
Let $S \in \F_{q^m}^l$ be chosen according to an arbitrary distribution,
and let $X\in\F_{q^m}^n$ be chosen according to an arbitrary distribution
 such that $S$ is uniquely determined from $X$ by some surjection.
Suppose that $E \in \F_{q^m}^\mu$ is chosen according to an arbitrary distribution.
Then, for $W^{\rm T}=BX^{\rm T} + E^{\rm T}$,
$H(S|W) \geq H(S| BX^{\rm T})$
always holds.
\end{proposition}
\begin{IEEEproof}
Observe that $S \leftrightarrow BX^{\rm T} \leftrightarrow W$
forms a Markov chain.
By the data processing inequality \cite[Ch.~2, pp.~32--33]{Cover2006},
we have $I(S;BX^{\rm T}) \geq I(S;W)$, which implies
$H(S|W) \geq H(S|BX)$.
\end{IEEEproof}\vspace{1.5ex}
The statement equivalent to \Prop{prop:ngai} was given in \cite[Theorem 4.1]{Zhang2009}.
This proposition shows that for $\mu$ tapped links,
the uncertainty of at least $\uequivocation_{\mu,P_{S,X}}$
defined by \Def{def:universalperformance}
is always guaranteed even if errors occur in the network.
In other words, from \Prop{prop:ngai},
we can see that \Def{def:universalperformance} considers
 the most advantageous case for the adversary
 in the wiretap network model given in \Sect{sect:nestedcoding},
 as with the model considered in \cite{Silva2011,Zhang2009}.


As the security measure for secure network coding,
the maximum uncertainty of $S$ given $B_{\mathcal{W}} X^{\rm T}$
for all possible $\mathcal{W}$'s of tapped links
was considered in \cite{Ngai2011,Zhang2009,Cai2011,ElRouayheb2012}, where $m=1$.
However, the security measure 
in \cite{Ngai2011,Zhang2009,Cai2011,ElRouayheb2012}
is dependent on the underlying network coded network,
\ie it is not \textit{universal}.
On the other hand, as defined in \Def{def:universalperformance},
$\uequivocation_{\mu,P_{S,X}}$
does not depend on the set of possible $\mathcal{W}$'s of tapped links in
the network.
Thus, $\uequivocation_{\mu,P_{S,X}}$
is guaranteed on any underlying network code, and hence it is \textit{universal}.

Silva and Kschischang proposed a scheme based on
the nested coset coding scheme with MRD codes $\c_1,\c_2$ \cite{Silva2011}
with which
no information of $S$ is obtained from any $\dim \c_1-l=\dim \c_2$ links
for any distribution of $S$ when
the conditional distribution of $X$ given $S$ is uniform over $\psi(S)$, provided $m \geq n$.
That is, their scheme guarantees
the universal equivocation $\uequivocation_{\dim \c_1-l,P_{S,X}} = H(S)$
for any distribution of $S$.

\subsubsection{Universal $\ustrong$-Strong Security and Universal Maximum Strength}
\Def{def:universalperformance} defines the universal equivocation $\uequivocation_{\mu,P_{S,X}}$
as the security measure for all the components of a 
secret message $S=[S_1,\dots,S_l]$.
Consider the case where $\uequivocation_{\mu,P_{S,X}} < H(S)$,
\ie some information of the
secret message leaks to the adversary.
Then, some components of $S_1,\dots,S_l$
could be uniquely determined by the adversary.
It is clearly desirable that no component of $S_1,\dots,S_l$ is deterministically revealed
and that every symbol $S_i$ is kept hidden,
even if some information of $S$ leaks to the adversary.
Hence, we can say that the number of tapped links
such that every symbol $S_i$ is kept hidden
represents the resiliency or strength of the coding scheme against eavesdropping.
Now we focus on such security and give the following definition
as the resiliency of the coding scheme against eavesdropping.

\begin{definition}[Universal $\ustrong$-Strong Security and Universal Maximum Strength]\label{def:universalalpha}
Let $S_\mathcal{Z}=(S_i:i\in\mathcal{Z})$ be a tuple whose indices belong to a subset
 $\mathcal{Z}\subseteq\{1,\dots,l\}$.
We say that the coding scheme attains
\textit{universal $\ustrong$-strong security} if we have
\begin{align}
I(S_\mathcal{Z};BX^{\rm T})&=0,\quad
 \forall \mathcal{Z},
 \forall B \in \F_q^{(\ustrong-|\mathcal{Z}|+1) \times n},\label{eq:universalalpha}
\end{align}
for uniformly distributed $S$ and conditionally uniformly distributed $X$ given $S$.
The maximum possible value of $\ustrong$ in the scheme is called
\textit{universal maximum strength} $\maxustrong$ of the coding scheme,
defined by
\begin{align}
\maxustrong
\triangleq
\max \Big\{
\ustrong :\ &
 I(S_\mathcal{Z};BX^{\rm T})=0,\nonumber\\
& \forall \mathcal{Z} \subseteq \{1,\dots,l\},
 \forall B \in \F_q^{(\ustrong - |\mathcal{Z}| + 1) \times n}
\Big\}.\label{eq:defmaxustrong}
\end{align}
\end{definition}
The universal strong security defined in \cite{Silva2009}
is a special case of \Def{def:universalalpha} for $\maxustrong = n-1$
and $\c_1 = \F_{q^m}^n$.
Unlike the definition of $\uequivocation_{\mu,P_{S,X}}$ in \Def{def:universalperformance},
we have considered the case where the secret message $S$
is uniformly distributed and the transmitted packets $X$
are conditionally uniformly distributed given $S$.
This is because
the value of $I(S_{\mathcal{Z}};BX^{\rm T})$
is dependent on the conditional distribution of $X$ given $S_{\mathcal{Z}}$,
while $\uequivocation_{\mu,P_{S,X}}$
does not have such  dependence.
Without assuming a joint probability distribution
on $S$ and $X$, we cannot have a meaningful sufficient
condition for $I(S_\mathcal{Z};BX^{\rm T})=0$ in \Eq{eq:defmaxustrong}.

In \Def{def:universalalpha},
the mutual information between a part of $S$ and $BX^{\rm T}$ is considered,
and the errors $Z$ contained in the eavesdropped information are not considered.
This is based on the same reason that errors are not considered
in \Def{def:universalperformance}.
That is, from \Prop{prop:ngai},
the universal $\maxustrong$-strong security
is always guaranteed even if errors occur in the network.

As in \cite{Harada2008,Matsumoto2011,Silva2009},
a scheme with universal $\ustrong$-strong security does not leak
any $|\mathcal{Z}|$ components of $S$ even if at most $\ustrong-|\mathcal{Z}|+1$ links are
observed by the adversary,
provided that $P_{S,X}$ satisfies the assumption in \Def{def:universalalpha}.
Moreover, this guarantee holds over any underlying
network code as $\uequivocation_{\mu,P_{S,X}}$.
Hence, $\ustrong$ and $\maxustrong$ are also \textit{universal}.
In \Coro{coro:OmegaI},
we will present an upper bound of $I(S_\mathcal{Z};BX^{\rm T})$
with arbitrary $P_{S,X}$ in terms of $\maxustrong$.

\subsection{Expression of the Universal Security Performance in Terms of the RDIP and RGRW}\label{sect:expression}
In this subsection,
we express $\uequivocation_{\mu,P_{S,X}}$ and $\maxustrong$ given in
\Sect{sect:defsecurity} in terms of the RDIP and RGRW.

We first give a lemma for the mutual information
 between the message and information observed by the adversary.
From now on, for a matrix $M \in \F_{q^m}^{\mu \times n}$, we represent a row space of $M$ over $\F_{q^m}$
by $\row{M} \triangleq \{ u M : u \in \F_{q^m}^{\mu} \}\subseteq\F_{q^m}^n$.
For a set $\mathcal{A}$, denote by $U_{\mathcal{A}}$
the random variable uniform on
$\mathcal{A}$.
For a random variable $V$ and a set $\mathcal{A}(V)$
depending on $V$, denote by $U_{\mathcal{A}(V)}$
the random variable conditionally uniform on $\mathcal{A}(V)$
given $V$.
For three random variables $A$, $B$, $C$,
denote by $D(A \| B)$
the relative entropy \cite[Ch.~2, p.~18]{Cover2006}
between probability distributions $P_A$ and $P_B$.
Also we denote by $D(A \| B | C)$ 
the conditional relative entropy \cite[Ch.~2, p.~22]{Cover2006}
between two conditional probability distributions $P_{A|C}$ and $P_{B|C}$ given $C$.

\begin{lemma}\label{lma:lmamutual}
Let $\c_1 \subseteq \F_{q^m}^n$ be a linear code and $\c_2 \subsetneqq \c_1$ be its subcode
with $\dim \c_2 = \dim \c_1 - l$.
Assume that a random variable $S\in \F_{q^m}^l$
is chosen according to an arbitrary distribution over $\F_{q^m}^l$.
For a bijective function $\psi:\F_{q^m}^l \rightarrow \c_1/\c_2$ and given $S$,
let $X\in\F_{q^m}^n$ be
a random variable arbitrarily distributed over a coset $\psi(S) \in \c_1/\c_2$.
Fix a matrix $B \in \F_{q^m}^{\mu \times n}$ over $\F_{q^m}$ arbitrarily,
and let $W^{\rm T}=BX^{\rm T} \in \F_{q^m}^{\mu \times 1}$.
Then,
we have the following three statements.
\begin{enumerate}
\item For any distributions of $S$ and $X$,
we have
\begin{align}
I(S;W)
&\leq
\dim(\dc_2 \cap \row{B}) - \dim(\dc_1 \cap \row{B}) \nonumber\\*
&\quad+D(X\|U_{\psi(S)}|S), \label{eq:lmamutual}
\end{align}
and
\begin{align}
I(S;W) &\geq \dim(\dc_2 \cap \row{B}) - \dim(\dc_1 \cap \row{B})\nonumber\\*
&\quad-D(S\| U_{\F_{q^m}^l}). \label{eq:lmamutualxx}
\end{align}
\item If both $S$ and $X$ are uniformly distributed over $\F_{q^m}^l$ and $\c_1$ respectively,
the equalities in \Eq{eq:lmamutual} and \Eq{eq:lmamutualxx} hold,
\ie $I(S;W)=\dim(\dc_2 \cap \row{B}) - \dim(\dc_1 \cap \row{B})$.
\item If $I(S;W)=0$ holds for a distribution of $S$
that assigns a positive probability to every element in $\F_{q^m}^l$,
then we have $\dim(\dc_2 \cap \row{B}) - \dim(\dc_1 \cap \row{B})=0$.
\end{enumerate}
\end{lemma}
\begin{IEEEproof}
See Appendix~\ref{sect:lmamutual_proof}.
\end{IEEEproof}\vspace{1.5ex}

Note that the matrix $B$ in \Lma{lma:lmamutual} is defined over the field extension $\F_{q^m}$,
while the transfer matrix $B$ to the wiretapper
is restricted to the subfield $\F_{q}$
in the wiretap network model defined in \Sect{sect:nestedcoding}.

Cai and Yeung \cite{Cai2011,Cai2007,cai2009isit}
considered
the security condition of secure network coding
such that the adversary obtains no information about the secret message
when $S$ and $X$ are chosen according to
distributions that assign positive probabilities to all the secret messages and the transmitted packets
\cite[Lemma 3.1]{Chan2011a}.
Their security condition corresponds to
the statement 3) in \Lma{lma:lmamutual}
and the fact that $I(S;W)=0$
when $\dim(\dc_2 \cap \row{B}) = \dim(\dc_1 \cap \row{B})$ holds
and $X$ is conditionally uniform over $\psi(S)$ given $S$
from \Eq{eq:lmamutual}.
Note that in the statement 3) in \Lma{lma:lmamutual},
we made no assumption on the distribution of $X$, unlike \cite[Lemma 3.1]{Chan2011a}.
For an arbitrary joint distribution of $S$ and $X$,
Zhang and Yeung \cite{Zhang2009a} generalized the security condition
of \cite{Cai2011,Cai2007,cai2009isit}
that corresponds to the statement 3) in \Lma{lma:lmamutual}.
We should note that in the statement 3) in
\Lma{lma:lmamutual},
we assumed the distribution of $S$ that assigns a positive probability to every message,
in order to express the condition in terms of the dimensions of subspaces.
The proof of \Lma{lma:lmamutual} (See Appendix~\ref{sect:lmamutual_proof}) can be adapted to
the case of arbitrarily distributed $S$,
and we can easily show that if $I(S;W)=0$ for $P_S$,
we have $I(S';W')=0$ for any $P_{S'}$ satisfying
$\{s:P_{S'}(s)>0\} \subseteq \{s:P_{S}(s)>0\}$.
Further, unlike \cite{Cai2011,Cai2007,cai2009isit,Zhang2009a},
\Lma{lma:lmamutual} additionally derived the upper and lower bounds of
the mutual information leaked to the adversary by using the relative entropy
for arbitrarily distributed $S$ and $X$.
In the following,
we shall derive the security performance expressed in terms of the RDIP and the RGRW by using \Lma{lma:lmamutual},
which is guaranteed independently of the underlying network coded network.

Here we recall that if an $\F_{q^m}$-linear space $V \subseteq \F_{q^m}^n$
admits a basis in $\F_{q}^n$, then $V \in \colinv$
 by \Lma{lma:stichtenothlemma1}.
Since the transfer matrix $B$ is defined over the base field $\F_q$ in \Sect{sect:nestedcoding},
this implies
\begin{equation}
\row{B} \in \colinv. \label{eq:vb}
\end{equation}

We give the following theorem for the maximum amount of information leakage to the adversary,
defined in \Def{def:universalperformance}.
\begin{theorem}\label{thm:equivocation}
Consider the nested coset coding scheme
with $\c_1$, $\c_2$ and $\psi$ in \Def{def:nestedcoding}.
Then, the maximum amount of information leakage to the adversary,
defined in \Def{def:universalperformance},
is in the range of
\begin{align}
&K_{R,\mu}(\dc_2,\dc_1) -D(S \| U_{\F_{q^m}^l})\nonumber\\
&\qquad \leq
\max_{B \in \F_q^{\mu \times n}} I(S;BX^{\rm T})\nonumber\\
&\qquad \leq K_{R,\mu}(\dc_2,\dc_1) +D(X\|U_{\psi(S)}|S).
\label{eq:maxuequivocation}
\end{align}
If both the secret message $S$ and the transmitted packets $X$
 are uniformly distributed over $\F_{q^m}^l$ and $\c_1$ respectively,
the maximum amount of information leakage exactly equals
${\displaystyle \max_{B \in \F_q^{\mu \times n}} I(S;BX^{\rm T})} = K_{R,\mu}(\dc_2,\dc_1)$.
If the maximum amount of information leakage is exactly zero
for a distribution of $S$ that assigns a positive probability to every element in $\F_{q^m}^l$,
we have $K_{R,\mu}(\dc_2,\dc_1) = 0$, which corresponds to \cite[Lemma 3.1]{Chan2011a}.
\end{theorem}
\begin{IEEEproof}
By \Lma{lma:stichtenothlemma1} and \Eq{eq:vb}, we have
\begin{align}
\left\{ \row{B} : B\in \F_{q}^{\mu\times n}\right\} = \bigcup_{i\leq \mu} \colinvi{i}.
\label{eq:m3}
\end{align}
From \Lma{lma:lmamutual}, we have
\begin{align}
\max_{B \in \F_{q}^{\mu\times n}}
 I(S;BX^{\rm T})
 &\!\leq\! \max_{B\in \F_{q}^{\mu\times n}}
\left\{
\dim(\dc_2 \!\cap\! \row{B}) \!-\! \dim(\dc_1 \!\cap\! \row{B})
\right\}\nonumber\\*
&\quad+D(X\|U_{\psi(S)}|S),
\label{eq:samui1}
\end{align}
and
\begin{align}
\max_{B \in \F_{q}^{\mu\times n}}
 I(S;BX^{\rm T})
 &\!\geq\!
 \max_{B\in \F_{q}^{\mu\times n}}
\left\{
\dim(\dc_2 \!\cap\! \row{B}) \!-\! \dim(\dc_1 \!\cap\! \row{B})
\right\}\nonumber\\*
&\quad -D(S \| U_{\F_{q^m}^l}).
\label{eq:samui2}
\end{align}
For the first terms on the right-hand side of \Eq{eq:samui1} and \Eq{eq:samui2},
we have
\begin{align}
&\max_{B\in \F_{q}^{\mu\times n}}
\left\{
\dim(\dc_2 \cap \row{B}) - \dim(\dc_1 \cap \row{B})
\right\}
\nonumber\\
&\quad=
 \max_{V \in \bigcup_{i\leq \mu} \colinvi{i}}
\left\{
\dim(\dc_2 \cap V) - \dim(\dc_1 \cap V)
\right\}\text{\hspace{1ex} (by \Eq{eq:m3})}
\nonumber\\
&\quad=\max
\left\{
K_{R,i}(\dc_2,\dc_1)
: i \leq \mu
\right\}\text{\hspace{3ex} (by \Def{def:rdip})}
\nonumber\\
&\quad=
K_{R,\mu}(\dc_2,\dc_1).
\text{\hspace{15ex} (by \Thm{thm:monotonerdip})}
\label{eq:samui3}
\end{align}
Therefore, we have \Eq{eq:maxuequivocation}.

If $S$ and $X$ are uniform over $\F_{q^m}^l$ and $\c_1$ respectively,
we have $I(S;BX^{\rm T})=\dim(\dc_2 \cap \row{B}) - \dim(\dc_1 \cap \row{B})$ from \Lma{lma:lmamutual}.
Therefore, for the uniformly distributed $S$ and $X$,
we have $\max_{B \in \F_q^{\mu \times n}} I(S;BX^{\rm T}) = K_{R,\mu}(\dc_2,\dc_1)$
by \Eq{eq:samui3}.

Finally, we prove the last statement.
Assume that $\max_{B \in \F_q^{\mu \times n}} I(S;BX^{\rm T})=0$ holds
for a distribution of $S$ that assigns a positive probability to
every element in $\F_{q^m}^l$.
Then, we have
$\dim(\dc_2 \cap \row{B}) - \dim(\dc_1 \cap \row{B})=0$ simultaneously for all $B\in\F_q^{\mu \times n}$
from \Lma{lma:lmamutual}.
Therefore, we have $K_{R,\mu}(\dc_2,\dc_1)=0$.
\end{IEEEproof}\vspace{1.5ex}
This theorem includes the condition such that
the maximum amount of information leakage is exactly zero.
This corresponds to the security condition of secure network coding
given in \cite{cai2009isit,Cai2007,Cai2011,Chan2011a},
for all possible sets of tapped links,
as \Lma{lma:lmamutual} for one set of possible tapped links.
We note that while our condition is independent of the underlying network code, \ie universal,
their security condition is dependent on the underlying network code.
Further, as \Lma{lma:lmamutual}, we should note that 
the distribution of $X$ is arbitrary
in the last statement in \Thm{thm:equivocation}.

\Thm{thm:equivocation}
immediately yields the following proposition for the universal equivocation.

\begin{proposition}\label{prop:equivocation2}
Consider the nested coset coding scheme with $\c_1$, $\c_2$ and $\psi$
in \Def{def:nestedcoding}.
Then, the universal equivocation $\uequivocation_{\mu,P_{S,X}}$,
defined in \Def{def:universalperformance},
is in the range of
 \begin{align*}
&H(S) - D(X\|U_{\psi(S)}|S) - K_{R,\mu}(\dc_2,\dc_1)\\
&\qquad \leq
\uequivocation_{\mu,P_{S,X}}\\
&\qquad \leq
l - K_{R,\mu}(\dc_2,\dc_1).
\end{align*}
When both the secret message $S$ and the transmitted packets $X$
 are uniformly distributed over $\F_{q^m}^l$ and $\c_1$ respectively,
we have $\uequivocation_{\mu,P_{S,X}} = l - K_{R,\mu}(\dc_2,\dc_1)$.
\hfill\IEEEQED
\end{proposition}

Also from \Thm{thm:equivocation}, we obtain the following corollary.
\begin{corollary}\label{prop:perfectsecrecy}
Consider the transmission of $X\in\F_{q^m}^n$ over the wiretap network,
which is generated from the secret message $S$
by the nested coset coding scheme with $\c_1$, $\c_2$ and $\psi$,
defined in \Def{def:nestedcoding}.
Then, we have the following four statements for an arbitrarily fixed $j\in\{1,\dots,l\}$.
\begin{enumerate}
\item If the adversary observes $\mu < M_{R,j}(\dc_2,\dc_1)$ links,
the maximum amount of information leakage in \Def{def:universalperformance} is
at most $j-1+D(X\|U_{\psi(S)}|S)$ between $S$ and observed packets;
\item If the adversary observes $\mu \geq M_{R,j}(\dc_2,\dc_1)$ links,
the maximum amount of information leakage is at least $j-D(S\|U_{\F_{q^m}^l})$
between $S$ and observed packets.
\item If the adversary observes $\mu=M_{R,j}(\dc_2,\dc_1)$ links,
there exist $P_{S,X}$ and $B$ by which the adversary
obtains the mutual information $j$ between $S$
 and the observed packets $BX^{\rm T}$;
and
\item If the maximum amount of information leakage is exactly zero
for a distribution of $S$ that assigns a positive
probability to every element in $\F_{q^m}^l$,
the number of tapped links is $\mu < M_{R,1}(\dc_2,\dc_1)$,
which again corresponds to \cite[Lemma 3.1]{Chan2011a}.
\end{enumerate}
\end{corollary}
\begin{IEEEproof}
Since we have
\begin{align*}
 \min\left\{ \mu: K_{R,\mu}(\dc_2,\dc_1) = j \right\} = M_{R,j}(\dc_2,\dc_1),
\end{align*} from \Lma{lma:rgrw},
we obtain $K_{R,\mu}(\dc_2,\dc_1)=j$
and $K_{R,\mu-1}(\dc_2,\dc_1)=j-1$ for $\mu=M_{R,j}(\dc_2,\dc_1)$.
This implies that from \Thm{thm:equivocation},
the maximum amount of information leakage is at most $j-1+D(X\|U_{\psi(S)}|S)$
from less than $M_{R,j}(\dc_2,\dc_1)$ links.
Also, the maximum amount of information leakage is
at least $j-D(S\|U_{\F_{q^m}^l})$ from $M_{R,j}(\dc_2,\dc_1)$ or more links.
Also note that when $S$ and $X$ are uniformly distributed,
the maximum information leakage for $\mu$ is exactly equal to $K_{R,\mu}(\dc_2,\dc_1)$ by \Thm{thm:equivocation}, and recall again that $\min\left\{ \mu: K_{R,\mu}(\dc_2,\dc_1) = j \right\} = M_{R,j}(\dc_2,\dc_1)$.
Thus, statements 1)--3) are proved.

When $\max_{B \in \F_q^{\mu \times n}} I(S;BX^{\rm T})=0$ holds
for a distribution of $S$ that assigns a positive probability to
every element in $\F_{q^m}^l$,
we have $K_{R,\mu}(\dc_2,\dc_1)=0$ from \Thm{thm:equivocation},
and $M_{R,1}(\dc_2,\dc_1) = \min \left\{ \mu: K_{R,\mu}(\dc_2,\dc_1) = 1\right\}$ holds from \Lma{lma:rgrw}.
Therefore, statement 4) is proved.
\end{IEEEproof}\vspace{1.5ex}
Statement 1) in \Coro{prop:perfectsecrecy} shows that
if the transmitted packets $X$ are chosen uniformly at random from a coset $\psi(S)$ given $S$,
the adversary obtains no information of the message $S$ from any $M_{R,1}(\dc_2,\dc_1)-1$ links,
 independently of the distribution of $S$.
In contrast, if $X$ is not uniform and $D(X\|U_{\psi(S)}|S) > 0$,
the information of $S$ leaks out from less than $M_{R,1}(\dc_2,\dc_1)$ links.

In \cite[Proposition 2]{Oggier2012},
Oggier and Sboui introduced a special case of \Coro{prop:perfectsecrecy}
for $j=1$, $\c_1 = \F_{q^m}^n$ and uniformly distributed $X\in\c_1$,
in terms of the minimum rank distance.
Namely, they showed that the adversary obtains
 no information of $S$ from any $d_R(\dc_2)-1=M_{R,1}(\dc_2,\{0\})-1$ links.

Ngai et al.\@ \cite{Ngai2011} and Zhang et al.\@\cite{Zhang2009}
analyzed the lower bound of the uncertainty of the secret message
by the (R-)Network-GHW
in the case where the transmitted packets $X$ are uniformly distributed over $\psi(S)$.
\Prop{prop:equivocation2} and \Coro{prop:perfectsecrecy}
correspond to their results using (R-)Network-GHW.
We should note that unlike \cite{Ngai2011,Zhang2009},
we considered the case where both $S$ and $X$ are arbitrarily distributed,
and derived both upper and lower bounds.
Further, while our analyses using the RDIP and the RGRW are universal,
the analyses using (R-)Network-GHW in \cite{Ngai2011,Zhang2009} are dependent
on the underlying network code.

Lastly, we express $\maxustrong$ in
\Def{def:universalalpha} in terms of the RGRW.
In order to derive $\maxustrong$,
we first introduce the following proposition
that reveals the mutual information between
a part of the message $S$ and observed packets of the adversary
in the case where $S$ and the transmitted packets $X$ are arbitrarily distributed.

\begin{proposition}\label{prop:partmaxmutual}
Consider the nested coset coding scheme
and fix $\c_1$, $\c_2$ and $\psi$ in \Def{def:nestedcoding}.
For a subset $\mathcal{Z} \subseteq \{1,\dots,l\}$,
 let $S_\mathcal{Z} \triangleq (S_i: i \in \mathcal{Z})$,
 and $\c_{3,\mathcal{Z}}$ be a subcode of $\c_1$ defined by
\begin{align}
\c_{3,\mathcal{Z}} \triangleq
 \bigcup_{\substack{S_i=0: i \in \mathcal{Z},\\ S_j \in \F_{q^m}: j \notin \mathcal{Z}}}
 \psi([S_1,\dots,S_l]). \label{eq:defunion}
\end{align}
Also, define a bijective function
$\psi_\mathcal{Z} : \F_{q^m}^{|\mathcal{Z}|} \rightarrow \c_1/\c_{3,\mathcal{Z}}$
by
\begin{align}
\psi_\mathcal{Z}(S_{\mathcal{Z}}) \triangleq
 \bigcup_{\substack{S_j \in \F_{q^m}: j \notin \mathcal{Z}}}
 \psi([S_1,\dots,S_l]).\label{eq:bijectionz}
\end{align}
Then,
the maximum amount of the mutual information between $S_\mathcal{Z}$ and $BX^{\rm T}$
is in the range of
\begin{align}
&K_{R,\mu}(\dc_{3,\mathcal{Z}},\dc_1) -D(S_{\mathcal{Z}} \| U_{\F_{q^m}^{|\mathcal{Z}|}})\nonumber\\*
&\qquad \leq
\max_{B \in \F_q^{\mu \times n}} I(S_\mathcal{Z};BX^{\rm T})\nonumber\\*
&\qquad \leq
K_{R,\mu}(\dc_{3,\mathcal{Z}},\dc_1) +D(X\|U_{\psi_{\mathcal{Z}}(S_{\mathcal{Z}})}|S_\mathcal{Z}).
\label{eq:partmaxmutual}
\end{align}
If both $S$ and $X$ are uniformly distributed over $\F_{q^m}^l$ and $\c_1$ respectively,
it exactly equals
${\displaystyle \max_{B \in \F_q^{\mu \times n}} I(S_{\mathcal{Z}};BX^{\rm T})} = K_{R,\mu}(\dc_{3,\mathcal{Z}},\dc_1)$.
\end{proposition}
\begin{IEEEproof}
For a subset $\mathcal{Z}\subseteq\{1,\dots,l\}$,
$\c_{3,\mathcal{Z}}$ is an $\F_{q^m}$-linear subspace
satisfying $\c_2 \subseteq \c_{3,\mathcal{Z}} \subseteq \c_1$
and $\dim \c_{3,\mathcal{Z}} = \dim \c_1 - |\mathcal{Z}|$.
Observe that $S_\mathcal{Z}$ is chosen from $\F_{q^m}^{|\mathcal{Z}|}$ according to the distribution $P_{S_\mathcal{Z}}$,
and that $X$ can be regarded as a random variable chosen from a coset $\psi_{\mathcal{Z}}(S_\mathcal{Z})\in\c_1/\c_{3,\mathcal{Z}}$ according to the conditional distribution $P_{X|S_\mathcal{Z}}$ given $S_\mathcal{Z}$.
Also recall that $\psi_\mathcal{Z}$ is bijective.
Hence, the information leakage of $S_\mathcal{Z}$
in the nested coset coding with $\c_1$ and $\c_2$ according to $P_{S,X}$
is equal to the one in the nested coset coding scheme with $\c_1$ and $\c_{3,\mathcal{Z}}$
according to $P_{S_{\mathcal{Z}},X}$,
where $P_{S_\mathcal{Z},X}$ is the joint distribution of $S_\mathcal{Z}$ and $X$.
Thus, by \Thm{thm:equivocation},
\Eq{eq:partmaxmutual} holds.
Assume that $S$ and $X$ are uniformly distributed over $\F_{q^m}^l$ and $\c_1$, respectively.
Then, $S_\mathcal{Z}$ is uniform over $\F_{q^m}^{|\mathcal{Z}|}$,
and from the definition of $\psi_{\mathcal{Z}}$ in \Eq{eq:bijectionz},
$X$ is also uniform over $\psi_{\mathcal{Z}}(S_{\mathcal{Z}})$ given $S$.
Therefore, we have
$\max_{B \in \F_q^{\mu \times n}} I(S_{\mathcal{Z}};BX^{\rm T})
= K_{R,\mu}(\dc_{3,\mathcal{Z}},\dc_1)$.
\end{IEEEproof}\vspace{1.5ex}

From \Prop{prop:partmaxmutual}
and the definition of the universal maximum strength $\maxustrong$ in \Def{def:universalalpha},
we give the following upper bound of $\max_{B\in\F_q^{\mu \times n}} I(S_\mathcal{Z}; BX^{\rm T})$
for arbitrarily distributed $S$ and $X$,
which is expressed in terms of $\maxustrong$.
\begin{corollary}\label{coro:OmegaI}
Consider the nested coset coding scheme defined in \Def{def:nestedcoding}
with the universal maximum strength $\maxustrong$.
Then, for fixed $\mu$ and $\mathcal{Z} \subseteq \{1, \dots,l \}$,
we have
\begin{align*}
\max_{B\in\F_q^{\mu \times n}} I(S_{\mathcal{Z}};BX^{\rm T})
\leq
\left[ \mu-\maxustrong + |\mathcal{Z}| -1 \right]^+
+ D(X\|U_{\psi_{\mathcal{Z}}(S_\mathcal{Z})}|S_\mathcal{Z}),
\end{align*}
where $\psi_{\mathcal{Z}}(S_\mathcal{Z})$ is defined by \Eq{eq:bijectionz}.
\end{corollary}
\begin{IEEEproof}
When $S$ and $X$ are uniformly distributed,
we have
$\max_{B\in\F_q^{\mu \times n}} I(S_{\mathcal{Z}};BX^{\rm T}) = K_{R,\mu}(\dc_{3,\mathcal{Z}},\dc_1)$
from \Prop{prop:partmaxmutual}.
Recall that
the RDIP $K_{R,\mu}(\dc_{3,\mathcal{Z}},\dc_1)$ is monotonically increasing with $\mu$
 from \Thm{thm:monotonerdip},
and that $\max_{B\in\F_q^{\mu \times n}} I(S_{\mathcal{Z}};BX^{\rm T}) = 0$
if $\mu \leq \maxustrong - |\mathcal{Z}| + 1$ from \Def{def:universalalpha}.
We thus have
\begin{align*}
K_{R,\mu}(\dc_{3,\mathcal{Z}},\dc_1) \leq \left[ \mu-\maxustrong + |\mathcal{Z}| -1 \right]^+.
\end{align*}
Therefore, from \Prop{prop:partmaxmutual},
we have
\begin{align*}
\max_{B\in\F_q^{\mu \times n}} I(S_{\mathcal{Z}};BX^{\rm T})
&\leq
K_{R,\mu}(\dc_{3,\mathcal{Z}},\dc_1)
+ D(X\|U_{\psi_{\mathcal{Z}}(S_\mathcal{Z})}|S_\mathcal{Z})\\
&\leq
\left[ \mu-\maxustrong + |\mathcal{Z}| -1 \right]^+
 + D(X\|U_{\psi_{\mathcal{Z}}(S_\mathcal{Z})}|S_\mathcal{Z}),
\end{align*}
for arbitrarily distributed $S$ and $X$.
\end{IEEEproof}\vspace{1.5ex}

This corollary shows that if the universal maximum strength $\maxustrong$ is known,
the maximum amount of information leakage of $S_\mathcal{Z}$ to the adversary
can be estimated by calculating $D(X\|U_{\psi_{\mathcal{Z}}(S_\mathcal{Z})}|S_\mathcal{Z})$
depending on distributions of $S$ and $X$.
This also implies that
when $D(X\|U_{\psi_{\mathcal{Z}}(S_\mathcal{Z})}|S_\mathcal{Z}) > 0$
for some $\mathcal{Z}$,
a part of the secret message might be revealed to the wiretapper
from $\mu < \maxustrong-|\mathcal{Z}|$ tapped links.
Here, we note that $D(X\|U_{\psi_{\mathcal{Z}}(S_\mathcal{Z})}|S_\mathcal{Z})$
is always zero for any $\psi$ and any $\mathcal{Z}$
when $S$ and $X$ are uniformly distributed over $\F_{q^m}^l$ and $\c_1$, respectively.
From these observations,
we can say that since $S$ and $X$ are assumed to be uniform in \Def{def:universalalpha},
the universal maximum strength $\maxustrong$
represents the resiliency of the scheme against eavesdropping in the ideal environment
in which every part of the secret message is hidden.

By \Prop{prop:partmaxmutual}, we have the following theorem which
exactly expresses $\maxustrong$ in terms of the RGRW.

\begin{theorem}\label{thm:universalalpha}
Fix $\c_1,\c_2$ and $\psi$ in \Def{def:nestedcoding},
and consider the corresponding nested coset coding scheme
with uniformly distributed $S$ and $X$.
For a subset $\mathcal{Z} \subseteq \{1,\dots,l\}$,
 let $S_\mathcal{Z} \triangleq (S_i: i \in \mathcal{Z})$
 and $\c_{3,\mathcal{Z}}$ be a subcode of $\c_1$, defined by \Eq{eq:defunion}.
Then, the universal maximum strength $\maxustrong$ of the scheme,
defined in \Def{def:universalalpha}, is given by
\begin{align*}
\maxustrong
 = \min \left\{
M_{R,1}(\dc_{3,\mathcal{Z}},\dc_1)+|\mathcal{Z}| : \mathcal{Z} \subseteq \{1,\dots,l\}
\right\}-2.
\end{align*}
\end{theorem}
\begin{IEEEproof}
The universal maximum strength $\maxustrong$, \ie the maximum value of $\ustrong$, is given as
\begin{align*}
&\maxustrong\\*
&=
\max \left\{
\ustrong :
 I(S_\mathcal{Z};BX^{\rm T}) = 0,
 \forall \mathcal{Z} \subseteq \{1,\dots,l\},
 \forall B \in \F_q^{(\ustrong - |\mathcal{Z}| + 1)\times n}
\right\}
\\*
&=
\min\left\{ \mu : \mathcal{Z}\!\subseteq\!\{1,\dots,l \},
\exists B \!\in\! \F_{q}^{(\mu-|\mathcal{Z}|+1) \times n}, I(S_\mathcal{Z}; BX^{\rm T}) \!=\! 1
\right\}\!-\!1\\*
&\hspace{38ex} \text{(by \Def{def:universalalpha})}
\\*
&=
\!\min\!\left\{ \mu \!+\! |\mathcal{Z}| \!-\! 1 \!:\! \mathcal{Z}\!\subseteq\!\{1,\dots,l \},
\exists B \!\in\! \F_{q}^{\mu \times n}, I(S_\mathcal{Z}; BX^{\rm T})\!=\!1
\right\}\!-\! 1
\\*
&=
\min_{\mathcal{Z}\subseteq\{1,\dots,l \}}
\left\{ \min\left\{\mu : \exists B \!\in\! \F_{q}^{\mu \times n}, I(S_\mathcal{Z}; BX^{\rm T})\!=\!1 \right\}
 \!+\! |\mathcal{Z}| \!-\!1 \right\} \!-\! 1
\\*
&=
\min_{\mathcal{Z}\subseteq\{1,\dots,l \}}
\left\{ \min\left\{\mu : \max_{B\in\F_{q}^{\mu \times n}} I(S_\mathcal{Z};BX^{\rm T}) = 1 \right\}
 + |\mathcal{Z}|-1 \right\}-1
\\*
&=
\min_{\mathcal{Z}\subseteq\{1,\dots,l \}}
\left\{ \min\left\{\mu : K_{R,\mu}(\dc_{3,\mathcal{Z}},\dc_1)=1 \right\}
 + |\mathcal{Z}|-1 \right\}-1
\\* &\hspace{37ex}\text{(by \Prop{prop:partmaxmutual})}
\\*
&=
\min_{\mathcal{Z}\subseteq\{1,\dots,l \}}
\left\{
M_{R,1}(\dc_{3,\mathcal{Z}},\dc_1) + |\mathcal{Z}|
\right\}-2.
\hspace{8ex} \text{(by \Lma{lma:rgrw})}
\end{align*}
\end{IEEEproof}\vspace{1.5ex}

In order to derive the exact value of $\maxustrong$,
we must calculate the RGRW's of $\c_1$ and $\c_{3,\mathcal{Z}}$'s for all possible $\mathcal{Z}$'s
 as shown in \Thm{thm:universalalpha}.
Thus, the calculation of $\maxustrong$ involves the search for the minimum value of the RGRW
over the exponentially large set for $l$.
Here, we give the upper and lower bounds of $\maxustrong$.
The bounds can be obtained by calculating only $l$ values of RGRW's,
hence they are useful for estimating the value of $\maxustrong$ in nested coset coding schemes.
An upper bound of $\maxustrong$ is simply obtained by \Thm{thm:universalalpha} as follows.
\begin{proposition}\label{prop:universalalphaupper}
Fix $\c_1,\c_2$ and $\psi$ in \Def{def:nestedcoding},
 and consider the corresponding nested coset coding scheme
with uniformly distributed $S$ and $X$.
For $i \subseteq \{1,\dots,l\}$,
 let $\c_{3,\{i\}}$ be a subcode of $\c_1$, defined in \Eq{eq:defunion} for $\mathcal{Z}=\{i\}$.
Then, the universal maximum strength $\maxustrong$ of the scheme is upper bounded by
\begin{align*}
\maxustrong
\leq \min
\left\{
M_{R,1}(\dc_{3,\{i\}},\dc_{1}) : 1 \leq i \leq l
\right\} -1.
\end{align*}
\hfill\IEEEQED
\end{proposition}

We also give a lower bound of $\maxustrong$.
For a subset $\mathcal{J} \subseteq \{1,\dots,N\}$ and a vector $c=[c_1,\dots,c_N]\in\F_{q^m}^N$,
let $P_\mathcal{J}(c)$ be a vector of length $|\mathcal{J}|$ over $\F_{q^m}$,
obtained by removing the $t$-th components $c_t$ for $t \notin \mathcal{J}$.
For example for $\mathcal{J}=\{1,3\}$ and $c=[1,1,0,1]$ ($N=4$),
we have $P_\mathcal{J}(c)=[1,0]$.
The \textit{punctured code} $P_\mathcal{J}(\c)$ of a code $\c \subseteq \F_{q^n}^N$ is given by
$P_\mathcal{J}(\c) \triangleq \left\{P_\mathcal{J}(c) : c\in \c \right\}$.
The \textit{shortened code} $\c_\mathcal{J}$ of a code $\c \subseteq \F_{q^m}^N$
is defined by
$\c_\mathcal{J} \triangleq \left\{
P_\mathcal{J}(c)
: c=[c_1,\dots,c_N] \in \c, c_i = 0 \text{ for } i \notin \mathcal{J}
\right\}$.
For example for $\c=\{[0,0,0],[1,1,0],[1,0,1],[0,1,1]\}$ ($N=3$) and
$\mathcal{J}=\{2,3\}$,
we have $\c_\mathcal{J}=\{[0,0],[1,1]\}$.
\begin{proposition}\label{prop:universalalphalower}
Fix $\c_1,\c_2$ and $\psi$ in \Def{def:nestedcoding},
and consider the corresponding nested coset coding scheme
with uniformly distributed $S$ and $X$.
Define a lengthened code of $\c_1$ by
\begin{align*}
\c'_1 \triangleq \left\{[S,X]: S \in \F_{q^m}^l\text{
 and } X\in\psi(S)\right\} \subseteq \F_{q^m}^{l+n}.
\end{align*}
Let $\bari{i}\triangleq\{1,\dots,l+n\}\backslash\{i\}$.
For each index $1 \leq i \leq l$,
we define a punctured code $\d_{1,i}$ of $\c'_1$ as
 $\d_{1,i} \triangleq P_{\bari{i}}(\c'_1) \subseteq\F_{q^m}^{l+n-1}$,
and a shortened code $\d_{2,i}$ of $\c'_1$ as
 $\d_{2,i} \triangleq (\c'_1)_{\bari{i}} \subseteq\F_{q^m}^{l+n-1}$.
Then, the universal maximum strength $\maxustrong$ of the scheme
 is lower bounded by
\begin{align}
\maxustrong
\geq \min
\left\{
M_{R,1}(\dd_{2,i},\dd_{1,i}) : 1 \leq i \leq l
\right\} -1.
\label{eq:universalalphamin}
\end{align}
\end{proposition}
\begin{IEEEproof}
See Appendix~\ref{sect:universalalphalower_proof}.
\end{IEEEproof}\vspace{1.5ex}

\begin{remark}
In \cite{Kurihara2012b}, the security analysis of secret sharing schemes based on linear codes
was given in terms of the relative dimension/length profile and the relative generalized Hamming
weight \cite{Luo2005}.
By replacing the RDIP and the RGRW
 in all the theorems given in this section
 with the RDLP and the RGHW
and restricting shares to be uniformly distributed over $\c_1$,
we can obtain the theorems presented in \cite{Kurihara2012b}.
In particular,
\Thm{thm:equivocation} and \Prop{prop:equivocation2}
 become \cite[Theorem 4]{Kurihara2012b},
and
\Coro{prop:perfectsecrecy}
becomes
 \cite[Theorem 9, Corollary 11]{Kurihara2012b}.
Also, \Thm{thm:universalalpha} and \Thm{prop:universalalphalower}
become \cite[Theorem 12]{Kurihara2012b},
where we note that in the case of secret sharing schemes,
the exact value of $\maxustrong$ in \Thm{thm:universalalpha}
 coincides with its lower bound given in \Thm{prop:universalalphalower}.
\end{remark}

\begin{remark}
In \cite{Zhang2009,Ngai2011},
the security analysis of secure network coding in the case of packet length $m=1$
was given in terms of the (relative) network generalized Hamming weight (R-)Network-GHW.
By replacing the RGRW with the (R-)Network-GHW
and restricting transmitted packets to be uniformly distributed over $\c_1$,
\Coro{prop:perfectsecrecy} becomes \cite[Theorem 7]{Ngai2011},\cite[Lemma 4.3]{Zhang2009}.
Note that 
since the (R-)Network-GHW is determined according to the global coding vectors of all links
as we explained in \Sect{sect:defparameters},
their security analysis by the (R-)Network-GHW
is dependent on the underlying network code construction,
unlike our analysis by the RDIP and RGRW.
\end{remark}

\section{Universal Error Correction Capability of Secure Network Coding}\label{sect:errorcorrection}
This section reveals the error correction capability of the nested coset coding scheme
which is guaranteed independently of the underlying network code construction.
Here, recall that as described in the end of \Sect{sect:basicnetwork},
the system of network coding is called \textit{coherent} if the transfer matrix is known to each sink node and otherwise it is called \textit{noncoherent}.
In this section, we shall consider the error correction
not only over the coherent system of network coding
but also over the noncoherent system.
Here we note that the decoding of the secret message is executed
independently by each sink node in the network.
Hence, from now on,
only one sink node may be assumed without loss of generality
and for the sake of simplicity.


We first give a definition of error correction capability in secure network coding.
Now we consider a coding scheme that is a generalization
 of the nested coset coding scheme, described as follows.
\begin{definition}\label{def:generalcoding}
Let $\mathcal{S}$ be a set of possible secret messages.
Let $\mathcal{P}_\mathcal{S}$ be a collection of sets of $n$-dimensional vectors over $\F_{q^m}$
such that $|\mathcal{P}_\mathcal{S}|=|\mathcal{S}|$
and each element in $\mathcal{P}_\mathcal{S}$ is a non-empty set.
Assume that there exists a certain bijective function between $\mathcal{S}$
 and $\mathcal{P}_\mathcal{S}$.
The coding scheme first maps
a secret message $S \in \mathcal{S}$ to a unique set
 $\mathcal{X}_S \in \mathcal{P}_\mathcal{S}\ (\mathcal{X}_S \subseteq \F_{q^m}^n, |\mathcal{X}_S|>0)$
 of $n$-dimensional vectors by the bijective function.
Then, an element $X\in\mathcal{X}_S$ is chosen from $\mathcal{X}_S$
and served as $n$ packets transmitted through the network.
\end{definition}
Here we note that in the nested coset coding scheme with $\c_1$ and $\c_2$,
$\mathcal{S} = \F_{q^m}^l$,
$\mathcal{X}_S = \psi(S) \in \c_1/\c_2$ and
$\mathcal{P}_\mathcal{S} = \left\{\mathcal{X}_S : S \in \mathcal{S} \right\} = \c_1/\c_2$,
 as defined in \Def{def:nestedcoding}.
The reason we consider \Def{def:generalcoding} is that we need to analyze the error correction capability
in generalized fashion in the case of the noncoherent network coding system,
due to the modification to the nested coset coding scheme
as described later in \Sect{sect:noncoherent}.
For this generalized coding scheme, we define the following error correction capability in
the model of network coding described in \Sect{sect:basicnetwork}.
\begin{definition}[Universally $t$-Error-$\rho$-Erasure-Correcting]\label{def:errorcorrecting}
Consider the $t$-error-$\rho$-erasure $(n \times m)_q$ linear network in \Def{def:terror}.
Consider a coding scheme defined in \Def{def:generalcoding}.
Then, the coding scheme is called \textit{universally $t$-error-$\rho$-erasure-correcting}, if
every $S \in \mathcal{S}$ can be uniquely determined from
$Y^{\rm T}=AX^{\rm T} + DZ^{\rm T} \in \F_{q^m}^N$ for
$\forall A \in \F_q^{N \times n}: \rank A \geq n - \rho,
 \forall X \in \mathcal{X}_S,
 \forall D \in \F_q^{N \times t}, \forall Z \in \F_{q^m}^t$.
\end{definition}
As defined in \Def{def:errorcorrecting},
the capability of universally $t$-error-$\rho$-erasure-correcting
is guaranteed on any underlying network code, and hence it is called \textit{universal}.
Silva et al.\@'s secure network coding scheme \cite[Section VI]{Silva2011} uses MRD codes $\c_1$ and $\c_2$,
and it is universally $t$-error-$\rho$-erasure-correcting when the minimum rank distance
\cite{Gabidulin1985} of $\c_1$ is greater than $2t+\rho$.

In the following subsections,
we explain the coding scheme executed at the source node
 in the both cases of a coherent system and a noncoherent system,
and present the main theorems about universal error-correction capability for both cases.
The derivations of these main theorems are given in Appendix \ref{sect:appendix_errorcorrection},
and they are a natural generalization of the work in \cite{Silva2009a}
from the ordinary encoding scheme of a linear code and the rank distance
to the nested coset coding scheme and the RGRW.

\subsection{Case of Coherent System}\label{sect:coherent}
First we explain the fundamental case of a coherent network coding system,
 \ie the transfer matrix $A$ is known to the sink node.
In this case, the source node simply encodes a secret message $S \in \mathcal{S}=\F_{q^m}^l$
 to the transmitted $n$ packets $X \in \mathcal{X}_S = \psi(S)$
by the nested coset coding scheme with $\c_1,\c_2$, as explained in \Sect{sect:nestedcoding}.
And then, $\mathcal{P}_\mathcal{S} = \c_1/\c_2$.
Finally, $X\in\F_{q^m}^n$ is regarded as an $m \times n$ matrix over $\F_q$,
 and transmitted through the network.

In this setting over the coherent network coding system,
the universal error correction capability of the nested coset coding scheme
is exactly expressed in terms of the first RGRW $M_{R,1}(\c_1,\c_2)$ as follows.
\begin{theorem}\label{thm:errorcorrectioncap}
Consider the $t$-error $(n \times m)_q$ linear network in \Def{def:terror}.
Then, the nested coset coding scheme with $\c_1,\c_2$ in \Def{def:nestedcoding} is universally
 (\ie simultaneously for all $A \in \F_q^{N \times n}$ with rank deficiency at most $\rho$)
 $t$-error-$\rho$-erasure-correcting
if and only if $M_{R,1}(\c_1,\c_2) > 2t + \rho$.
\end{theorem}
\begin{IEEEproof}
See Appendix \ref{sect:appendix_errorcorrection},
where the detailed proof itself is given in Appendix \ref{sect:coherentcapability}.
\end{IEEEproof}\vspace{1.5ex}

\subsection{Case of Noncoherent System}\label{sect:noncoherent}
As described in \Sect{sect:basicnetwork},
the transfer matrix $A$ is unknown to the sink node
in the case of a noncoherent network coding system.
In this case,
the source node appends appropriate packet headers
 to the packets generated by the nested coset coding scheme.
The addition of packet headers
 is called the \textit{lifting construction} \cite{Silva2008}.
Since the information of global coding vectors are carried by the packet headers in the lifting construction,
this allows the scheme to be decoded when $A$ is unknown.

The lifting construction \cite{Silva2008}
of the nested coset coding scheme is described in detail as follows.
Let $\widetilde{m}$ be the degree of a field extension $\F_{q^{\widetilde{m}}}$, and let
$\phi_{\widetilde{m}}:\F_{q^{\widetilde{m}}} \rightarrow \F_q^{\widetilde{m} \times 1}$ be
an $\F_q$-linear isomorphism
that expands an element of $\F_{q^{\widetilde{m}}}$ to a column vector over $\F_q$
with respect to some fixed basis for $\F_{q^{\widetilde{m}}}$ over $\F_q$.
Suppose $m > n$.
Let $\widetilde{m} \triangleq m-n$,
and let $\c_1 \subseteq \F_{q^{\widetilde{m}}}^n$ and $\c_2 \subsetneqq \c_1$
be a linear code and its subcode, respectively.
By the nested coset coding scheme with $\c_1,\c_2$,
we generate $\widetilde{X} \in \F_{q^{\widetilde{m}}}^n$
 from a secret message $S \in \mathcal{S} = \F_{q^{\widetilde{m}}}^l$.
Then, expanding $\widetilde{X}=[X_1,\dots,X_n]\in\F_{q^{\widetilde{m}}}^n$ to an $\widetilde{m} \times n$ matrix
$\phi_{\widetilde{m}}(\widetilde{X}) \triangleq [\phi_{\widetilde{m}}(\widetilde{X}_1),\dots,\phi_{\widetilde{m}}(\widetilde{X}_n)] \in \F_{q}^{\widetilde{m} \times n}$
over the base field $\F_q$,
we construct $X \in \F_{q^m}^{n}$ of transmitted $n$ packets that is represented as
 $X^{\rm T} = \begin{bmatrix} I & \phi_{\widetilde{m}}(\widetilde{X})^{\rm T} \end{bmatrix}\in\F_q^{n \times m}$
 as a matrix over $\F_q$, where the identity matrix $I\in\F_q^{n \times n}$ is the packet header.
Hence, $\mathcal{X}_S$ and $\mathcal{P}_\mathcal{S}$ is given by
\begin{align}
\mathcal{X}_S
=
\mathcal{X}_{S,\textsf{lift}}
&\triangleq
\left\{
 X=\begin{bmatrix} I \\ \phi_{\widetilde{m}}(\widetilde{X}) \end{bmatrix} : \widetilde{X} \in \psi(S)
\right\},
\nonumber\\
\mathcal{P}_\mathcal{S}
 = \mathcal{P}_{\textsf{lift}}
 &\triangleq
\left\{
X \in \mathcal{X}_{S,\textsf{lift}}
:
S \in \F_{q^{\widetilde{m}}}^l
\right\},\label{eq:xslift}
\end{align}
where $X\in\F_{q^m}^n$ is regarded as an $m \times n$ matrix over $\F_q$.
Here, recall that we defined $\colq{X}\subseteq\F_{q^m}$ for $X=[X_1,\dots,X_n]\in\F_{q^m}^n$
 as an $\F_q$-linear subspace of $\F_{q^m}$ spanned by $X_1,\dots,X_n$,
and note that 
 $\dimq \colq{X} = n$ is always guaranteed for all $X\in\mathcal{X}_{S,\mathsf{lift}}$ and all $\mathcal{X}_{S,\mathsf{lift}}\in\mathcal{P}_{\mathsf{lift}}$ by the packet header $I$.

\begin{remark}
The packet headers of the lifting construction
do not convey the information generated from the secret message,
and convey only the information of the global coding vectors (and errors).
Thus, appending packet headers does not affect
 the security given in \Sect{sect:universalsecurity}.
\end{remark}
The following proposition shows that
in this setting of the lifting construction of the nested coset coding scheme in the noncoherent system,
the universal error correction capability is exactly expressed in terms of
the first RGRW $M_{R,1}(\c_1,\c_2)$ as in the coherent system.
\begin{proposition}\label{prop:errorcorrectioncap_noncoherent}
Assume $m > n$, and consider
the $t$-error $(n \times m)_q$ linear network in \Def{def:terror}.
Consider the lifting construction of the nested coset coding scheme with $\c_1\subseteq\F_{q^{\widetilde{m}}}^n$
 and $\c_2\subsetneqq\c_1$ for $\widetilde{m}=m-n$, as described in \Sect{sect:noncoherent}.
Then, the scheme is universally
 $t$-error-$\rho$-erasure-correcting
 if and only if $M_{R,1}(\c_1,\c_2) > 2t + \rho$.
\end{proposition}
\begin{IEEEproof}
See Appendix \ref{sect:appendix_errorcorrection},
where the detailed proof itself is given in Appendix \ref{sect:noncoherentcapability}.
\end{IEEEproof}\vspace{1.5ex}
This proposition also implies that
by applying the lifting construction,
the correction capability of the nested coset coding scheme
is maintained even over the noncoherent network coding system.


\section{A Construction of Secure Network Coding and Its Analysis}\label{sect:example}
This section proposes a construction of the nested coset coding scheme
with $\c_1$ and $\c_2$.
We also show its universal security performance and universal error correction capability
as an example of the analyses in \Sect{sect:universalsecurity} and \Sect{sect:errorcorrection} using
the RDLP and the RGRW.
By adding the error correction, the proposed scheme is an extension of
the universal strongly secure network coding scheme based on an MRD code with a systematic generator matrix,
presented in our earlier conference paper \cite{Kurihara2012}.
As well as the scheme of Silva and Kschischang \cite{Silva2011},
the proposed scheme guarantees
the universal equivocation $\uequivocation_{\dim \c_1 - l, P_{S,X}}=H(S)$
when the conditional distribution of $X$ given $S$ is uniform on $\psi(S)$,
and it is universally $t$-error-$\rho$-erasure-correcting when $n-\dim \c_1 + 1 > 2t+\rho$.
Moreover, unlike Silva et al.\@'s scheme, our scheme guarantees the universal maximum strength
 $\maxustrong = \dim \c_1 -1$, which means that
 no part of the secret message is deterministically revealed from the eavesdropped information
 observed from at most $\dim \c_1 -1$ links over any underlying network code.
An explicit construction of the nested coset coding scheme satisfying $\maxustrong = \dim \c_1 -1$
had remained an open question \cite{Silva2008},
and hence we solve this open question by the proposed scheme.

For the sake of simplicity, this section considers
the fundamental case of a coherent network coding system.
In the case of noncoherent network coding,
we can simply customize the proposed scheme by the lifting construction
 as we described in \Sect{sect:noncoherent}

\subsection{Theorems for Nested Coset Coding Scheme with MRD codes}\label{sect:app}
In this subsection, we first introduce some theorems for the nested coset coding scheme
using MRD codes $\c_1\subseteq\F_{q^m}^n$ and $\c_2\subsetneqq\c_1$.
These theorems will be used in the next subsection to clarify the security performance and error correction capability of our proposed scheme.
We note that they can be also used to reveal the performance
of the scheme proposed by Silva and Kschischang \cite{Silva2011}.
This will be briefly explained in \Sect{sect:comparison}.

First, we present the following two theorems that are established regardless of the choice of $\psi$
in the nested coset coding scheme.
For an arbitrary linear code $\c_1\subseteq \F_{q^m}^n$ and an MRD code $\c_2\subsetneqq\c_1$ with $m\geq n$,
since the dual of an MRD code is also MRD \cite{Gabidulin1985,Loidreau2008}, we have $K_{R,\mu}(\dc_2,\dc_1)=[\mu-\dim\c_2]^+$
($0\leq\mu\leq\dim\c_1$) by \Prop{prop:mrdrdip}.
Thus, for the universal equivocation $\uequivocation_{\mu,P_{S,X}}$,
we immediately have the following theorem
from \Prop{prop:equivocation2}.
\begin{theorem}\label{thm:apps1}
Assume $m \geq n$.
Let $\c_1\subseteq\F_{q^m}^n$ be an arbitrary linear code and let $\c_2 \subsetneqq \c_1$ be its subcode.
Suppose that $\c_2$ is an MRD code.
Write $l = \dim \c_1 - \dim \c_2$.
Then, for the nested coset coding scheme with $\c_1$ and $\c_2$ in \Def{def:nestedcoding},
the universal equivocation is in the range of
\begin{align*}
&H(S) - D(X\|U_{\psi(S)}|S) - \left[\mu-\dim \c_2\right]^+\\
&\qquad \leq \uequivocation_{\mu,P_{S,X}}\\
&\qquad \leq l - \left[\mu-\dim \c_2\right]^+,
\end{align*}
for $0 \leq \mu \leq \dim \c_1$.
\hfill\IEEEQED
\end{theorem}
This theorem shows that if $X$ is uniform,
the universal equivocation is $\uequivocation_{\dim \c_2,P_{S,X}} = H(S)$.
Also for the universal error correction capability,
we immediately have the following theorem from \Coro{coro:mrdrgrw2}
and \Thm{thm:errorcorrectioncap}.
\begin{theorem}\label{thm:apps2}
Assume $m \geq n$.
Let $\c_1\subseteq\F_{q^m}^n$ be a linear code and let $\c_2 \subsetneqq \c_1$ be its subcode.
Suppose that $\c_1$ is an MRD code.
Then, the nested coset coding scheme with $\c_1$ and $\c_2$ in \Def{def:nestedcoding}
is universally $t$-error-$\rho$-erasure-correcting
if and only if $2t+\rho < n-\dim\c_1+1$.
\hfill\IEEEQED
\end{theorem}

Next, we present a theorem for the universal maximum strength,
which is dependent on the setting of $\psi$ unlike \Thm{thm:apps1} and \Thm{thm:apps2}.
We have the following theorem immediately
from \Coro{coro:mrdrgrw2} and \Thm{thm:universalalpha}
since the dual of an MRD code is also MRD.
\begin{theorem}\label{thm:apps3}
Assume $m \geq n$.
Let $\c_1\subseteq\F_{q^m}^n$ be a linear code and let $\c_2 \subsetneqq \c_1$ be its subcode.
Write $l = \dim \c_1 - \dim \c_2$.
Let a bijective function $\psi:\F_{q^m}^l \rightarrow \c_1/\c_2$ be fixed in such a way that
for all $\mathcal{Z} \subseteq \{1,\dots,l\}$,
an $\F_{q^m}$-linear subspace $\c_{3,\mathcal{Z}}$ defined in \Eq{eq:defunion} is an MRD code with
$\dim \c_{3,\mathcal{Z}}=\dim\c_1-|\mathcal{Z}|$ and
$d_R(\c_{3,\mathcal{Z}})=n-\dim\c_1-|\mathcal{Z}|+1$.
Then, the nested coset coding scheme with $\c_1$, $\c_2$ and $\psi$ in \Def{def:nestedcoding}
guarantees the universal maximum strength $\maxustrong = \dim \c_1 -1$.
\hfill\IEEEQED
\end{theorem}

In the next subsection,
we present an explicit construction of the nested coset coding scheme
that satisfies 
all the assumptions in
Theorems \ref{thm:apps1}, \ref{thm:apps2} and \ref{thm:apps3} simultaneously.

\subsection{Description of the Proposed Scheme}\label{sect:description}
Recall that the punctured code and shortened code of a code $\c\in\F_{q^m}^N$
to $\mathcal{J} \subseteq \{1,\dots,N\}$
are respectively defined by $P_\mathcal{J}(\c) = \left\{P_\mathcal{J}(c) : c\in \c \right\}$
and $\c_\mathcal{J} = \left\{ P_\mathcal{J}(c) : c=[c_1,\dots,c_N] \in \c, c_i = 0 \text{ for } i \notin \mathcal{J} \right\}$, where $P_\mathcal{J}(c)$ for $c\in\F_{q^m}^N$ represents a vector of length $|\mathcal{J}|$ obtained by removing the $t$-th components of $c$ for $t\notin\mathcal{J}$.
Assume that the degree $m$ of the field extension $\F_{q^m}$ satisfies
$m \geq l+n$.
Then, the proposed scheme generates
the transmitted $n$ packets $X\in\F_{q^m}^n$
by the following setting of the nested coset coding scheme.

First, we set the linear codes $\c_1,\c_2 \subseteq \F_{q^m}^n$.
Let $\d$ be an $[l+n, k]$ MRD code over $\F_{q^m}$ with
$\dim \d = k (\geq l)$
and a systematic generator matrix
$G = \begin{bmatrix} I & P \end{bmatrix} \in \F_{q^m}^{k \times (l+n)}$.
Let $\mathcal{L} \triangleq \{l+1,\dots,l+n\}$ be an index set.
Define $\c_1 \triangleq P_\mathcal{L}(\d)$
 as a punctured code of $\d$ to the index set $\mathcal{L}$.
Also define $\c_2\triangleq \d_\mathcal{L}$
 as a shortened code of $\d$ to the index set $\mathcal{L}$.
Here we note the following facts for $\c_1,\c_2$.
Since an MRD code over $\F_{q^m}^N$ with $m \geq N$ is also an MDS code over $\F_{q^m}$ \cite{Gabidulin1985},
a $k \times k$ matrix over $\F_{q^m}$ consisting of
arbitrary $k$ columns of $G$
is always nonsingular,
and hence $\dim P_\mathcal{L}(\d) = \dim \c_1 = k$.
Also, since the MRD code $\d$ is also MDS \cite{Gabidulin1985},
the shortening of $\d$ to $\mathcal{L}$
simply reduces the dimension of $\d$ over $\F_{q^m}$ by $l$,
\ie $\dim \d_\mathcal{L} = \dim \c_2 = k-l$.
Also, we should note that from the definition of the punctured code and shortened code,
we have $\c_2 \subseteq \c_1$, and $\dim \c_1 - \dim \c_2 = \dim \c_1/\c_2 = l$.

Next, we set the bijective function $\psi:\F_{q^m}^l \rightarrow \c_1/\c_2$.
We define submatrices of the systematic generator matrix $G$ of $\d$ as follows.
\begin{align*}
G \triangleq 
\begin{array}{rcccl}
\ldelim[{3}{0pt}[]& I & \Delta G & \rdelim]{3}{0pt}[] &\rdelim\}{1}{0pt}[\ $l$ rows] \\[2ex]
& O &G_2 & & \rdelim\}{1}{0pt}[\ $k-l$ rows]\\[-1.5ex]
& \underbrace{\hspace{1em}}_{\text{$l$ columns}} & \underbrace{\hspace{1em}}_{\text{$n$ columns}}&
\end{array}
\end{align*}
Then, we set $\psi$ by $\Delta G\in\F_{q^m}^{l \times n}$ as follows.
\begin{align}
\psi(S) \triangleq S \Delta G + \c_2 \in \c_1/\c_2. \label{eq:proposedmapping}
\end{align}
We note that $G_1 \triangleq \left[ \begin{smallmatrix} \Delta G \\ G_2 \end{smallmatrix} \right] \in \F_{q^m}^{k \times n}$
is the generator matrix of $\c_1$,
and $G_2 \in \F_{q^m}^{(k-l)\times n}$ is the generator matrix of $\c_2$.
Also note that since $\rank \Delta G = l$ from $\dim \c_1 - \dim \c_2 = \rank G_1 - \rank G_2 = l$,
$\psi$ is bijective.

In our scheme, we execute the nested coset coding scheme with these settings of $\c_1$, $\c_2$ and $\psi$,
and generate the transmitted packets $X$.
Then, the source node transmits $X$ over the network as described in \Sect{sect:basicnetwork},
and the sink node receives $Y$ and attempts to obtain the secret message from $Y$.
The universal security performance and the universal error correction capability of our scheme
is clarified in the next subsection.

\begin{remark}
Consider the case where $\d$ and $G$ are not an MRD code and its systematic generator matrix but
a Reed-Solomon code and its systematic one respectively in the above settings.
Then, this nested coset coding scheme becomes
the strongly-secure secret sharing scheme of Nishiara and Takizawa \cite{Nishiara2009}.
Similarly to the relation between the wiretap channel II and
secure network coding, our scheme can be viewed as a
generalization of their scheme \cite{Nishiara2009} for network coding.
\end{remark}

\subsection{Analyses on the Proposed Scheme}
As an example of applications of the analyses of \Sect{sect:universalsecurity} and \Sect{sect:errorcorrection} using the RDIP and the RGRW,
this subsection presents
the analyses on the proposed scheme described in the previous subsection.
We first reveal the security performance and error correction capability of our scheme.
Next, we discuss the required packet length in our scheme.
Finally, we summarize the comparison of the proposed scheme with the scheme of Silva and Kschischang \cite{Silva2011}.

\subsubsection{Security Performance and Error Correction Capability of the Proposed Scheme}
Here, we analyze the security performance
and the error correction capability of the proposed scheme
using the theorems presented in \Sect{sect:app}.

First, in order to show that assumptions in Theorems \ref{thm:apps1}--\ref{thm:apps3}
are satisfied in our scheme,
we introduce the following lemmas about a shortened code and a punctured code
of an MRD code.
\begin{lemma}\label{lma:shorteningdimension}
Let $m \geq N$, and $\c \subseteq \F_{q^m}^{N}$ be an MRD code of length $N$ over $\F_{q^m}$.
For a subset $\mathcal{I} \subseteq \{1,\dots,N\}$
satisfying $\mathcal{I} \supseteq \{\dim \c + 1 ,\dots, N\}$,
let $\c_\mathcal{I} \subseteq \F_{q^m}^{|\mathcal{I}|}$
 be a shortened code of $\c\subseteq\F_{q^m}^{N}$ to $\mathcal{I}$.
Then, $\c_\mathcal{I}$ is an MRD code
with $\dim \c_\mathcal{I}=\dim \c -N+|\mathcal{I}|$ and $d_R(\c_\mathcal{I})= N -\dim\c + 1$.
\end{lemma}
\begin{IEEEproof}
Since the MRD code $\c$ is also MDS \cite{Gabidulin1985}
and $\mathcal{I} \supseteq \{\dim\c+1,\dots,N\}$,
the shortening of $\c$ to $\mathcal{I}$
simply reduces the dimension of $\c$ over $\F_{q^m}$ by $N-|\mathcal{I}|$,
\ie $\dim \c_\mathcal{I} = \dim \c -N + |\mathcal{I}|$.

Since $m \geq N$ and shortened codes can be viewed as subcodes,
we have
\begin{align*}
d_R(\c_{\mathcal{I}})
 &\geq d_R(\c)\\*
 &=N-\dim \c+1.
\end{align*}
On the other hand,
from $m \geq N$ and the Singleton-type bound for the rank distance
 given in \Prop{prop:singleton},
we have
\begin{align*}
d_R(\c_\mathcal{I})
&\leq |\mathcal{I}| - \underbrace{\dim \c_\mathcal{I}}_{=\dim \c-N+|\mathcal{I}|} + 1
\\*
&=N-\dim \c + 1.
\end{align*}
Therefore, we have $d_R(\c_\mathcal{I}) = N-\dim \c +1$.
\end{IEEEproof}\vspace{1.5ex}
\begin{lemma}\label{lma:puncturingdimension}
Let $ m \geq N$,
and $\c \subseteq \F_{q^m}^{N}$ be an MRD code of length $N$ over $\F_{q^m}$.
For a set $\mathcal{I} \subseteq\{1,\dots,N\}$ satisfying
 $|\mathcal{I}|\geq N-\dim \c$ and $|\mathcal{I}| \geq \dim \c$,
let $P_\mathcal{I}(\c) \subseteq \F_{q^m}^{|\mathcal{I}|}$
 be a punctured code of $\c$ to $\mathcal{I}$.
Then, $P_\mathcal{I}(\c)$ is an MRD code with
$\dim P_\mathcal{I}(\c) = \dim \c$
and $d_R(P_\mathcal{I}(\c))=|\mathcal{I}|-\dim \c + 1$.
\end{lemma}
\begin{IEEEproof}
Since an MRD code is also MDS,
a $\dim \c \times \dim \c$ matrix over $\F_{q^m}$ consisting of
arbitrary $\dim \c$ columns of the generator matrix of $\c$
is always nonsingular.
Thus, the dimension of a punctured code $P_\mathcal{I}(\c)$ of length $|\mathcal{I}| (\geq \dim \c)$
is $\dim P_\mathcal{I}(\c) = \dim \c$.

The puncturing of $\c$ to $\mathcal{I}$ reduces the minimum rank distance of $\c$
by at most $N-|\mathcal{I}|$ $(\leq \dim \c)$ from the definition of rank distance \cite{Gabidulin1985}.
This implies that $d_R(P_\mathcal{I}(\c)) \geq d_R(\c) - N + |\mathcal{I}|$.
From $m \geq N$, we thus have
\begin{align*}
d_R(P_\mathcal{I}(\c))
&\geq d_R(\c) - N + |\mathcal{I}|\\*
&=|\mathcal{I}| - \dim \c + 1.
\end{align*}
On the other hand,
from $m \geq N$ and the Singleton-type bound for the rank distance
 given in \Prop{prop:singleton},
we have
\begin{align*}
d_R(P_\mathcal{I}(\c))
&\leq |\mathcal{I}| - \dim P_{\mathcal{I}}(\c) + 1\\*
&=|\mathcal{I}| - \dim \c + 1.
\end{align*}
Therefore, we have $d_R(P_\mathcal{I}(\c)) =|\mathcal{I}| - \dim \c + 1$.
\end{IEEEproof}\vspace{1.5ex}

By the above lemmas, we finally derive the following propositions for the universal security
performance and the universal error correction capability in our scheme,
and show that our scheme satisfies Theorems \ref{thm:apps1}, \ref{thm:apps2} and \ref{thm:apps3}
simultaneously.
\begin{proposition}\label{prop:ourschemeequivocation}
Consider the nested coset coding scheme proposed in \Sect{sect:description}.
Then, the universal equivocation $\uequivocation_{\mu,P_{S,X}}$ of the scheme is
in the range of
\begin{align*}
& H(S)- D(X\|U_{\psi(S)}|S) - \left[\mu-\dim \c_2\right]^+\\
&\qquad \leq
\uequivocation_{\mu,P_{S,X}}\\
&\qquad \leq
l - \left[\mu-\dim \c_2\right]^+,
\end{align*}
for $0 \leq \mu \leq \dim \c_1=k$.
\end{proposition}
\begin{IEEEproof}
From \Lma{lma:shorteningdimension},
since $m \geq l+n$,
we have $\dim \c_2 = k-l$,
and $\c_2$ is an MRD code with $d_R(\c_2) = n + l - k + 1$.
Thus, from \Thm{thm:apps1}, we have the proposition.
\end{IEEEproof}\vspace{1.5ex}
\begin{proposition}\label{prop:ourschemeerror}
The nested coset coding scheme proposed in \Sect{sect:description}
is universally $t$-error-$\rho$-erasure-correcting
if and only if $2t+\rho < n-k+1$.
\end{proposition}
\begin{IEEEproof}
From \Lma{lma:puncturingdimension},
since $m \geq l+n$,
we have $\dim \c_1 = \dim \d = k$,
and $\c_1$ is an MRD code with $d_R(\c_1) = n - k + 1$.
Therefore, the proposition is proved from \Thm{thm:apps2}.
\end{IEEEproof}\vspace{1.5ex}
\begin{proposition}\label{prop:universalalphaproposed}
The nested coset coding scheme proposed in \Sect{sect:description}
has the universal maximum strength $\maxustrong = k-1$.
\end{proposition}
\begin{IEEEproof}
Recall $\dim \d = k$.
For a subset $\mathcal{Z} \subseteq \{1,\dots,l\}$,
let $\overline{\mathcal{Z}} \triangleq \{1,\dots,l,l+1,\dots,l+n\}\backslash\mathcal{Z}$.
Denote by $\d_{\overline{\mathcal{Z}}} \subseteq \F_{q^m}^{l+n-|\mathcal{Z}|}$
 a shortened code of $\d$ to ${\overline{\mathcal{Z}}}$.
Since $\d \subseteq \F_{q^m}^{l+n}$ is an MRD code with $m \geq l+n$
and $\overline{\mathcal{Z}} \supseteq \{l+1,\dots,l+n\} \supseteq \{ k + 1,\dots, l+n\}$
 from $l \leq k$,
$\d_{\overline{\mathcal{Z}}}$ is an MRD code
with $\dim \d_{\overline{\mathcal{Z}}} = k - |\mathcal{Z}|$
and $d_R(\d_{\overline{\mathcal{Z}}}) = l+n-k +1$
from \Lma{lma:shorteningdimension}.
Since $\psi$ in the proposed scheme is specified by the systematic generator matrix $G$ of $\d$ as \Eq{eq:proposedmapping},
we can see that for $\mathcal{Z}$,
$\c_{3,\mathcal{Z}}$ in \Eq{eq:defunion}
can be defined as a punctured code of $\d_{\overline{\mathcal{Z}}}$
to an index set $\{l+1-|\mathcal{Z}|,\dots,l+n-|\mathcal{Z}|\}$,
\ie it is obtained by eliminating first $l-|\mathcal{Z}|$ coordinates
of codewords in $\d_{\overline{\mathcal{Z}}}$.
Hence, from \Lma{lma:puncturingdimension},
$\c_{3,\mathcal{Z}} \subseteq \F_{q^m}^n$ is an MRD code
with $\dim \c_{3,\mathcal{Z}} = k - |\mathcal{Z}|$
and $d_R(\c_{3,\mathcal{Z}}) = n - k - |\mathcal{Z}| + 1$.
Therefore, we have the proposition from \Thm{thm:apps3}.
\end{IEEEproof}\vspace{1.5ex}

Here we note that in the proposed scheme,
the exact value of $\maxustrong$ derived in \Prop{prop:universalalphaproposed}
coincides with the upper and lower bounds of $\maxustrong$ that are respectively given by 
\Prop{prop:universalalphaupper} and \Prop{prop:universalalphalower}.
The reason is as follows.
For $i \in \{1,\dots,l\}$ and $\bari{i}=\{1,\dots,l+n\}\backslash\{i\}$,
define the punctured code $\d_{1,i} \triangleq P_{\bari{i}}(\d)$
and the shortened code $\d_{2,i} \triangleq \d_{\bari{i}}$.
Then, $\d_{2,i}$ is MRD with $\dim \d_{2,i} = k-1$ from \Lma{lma:shorteningdimension}.
Since the dual of an MRD code is also MRD,
we have $M_{R,1}(\dd_{2,i},\dd_{1,i})=n-\dim\dd_{2,i}+1=k$ from \Coro{coro:mrdrgrw2}.
Thus, we obtain $\maxustrong \geq k-1$ from \Prop{prop:universalalphalower}.
On the other hand,
the subcode $\c_{3,\{i\}}$ is MRD with $\dim \c_{3,\{i\}} = k-1$
as shown in the proof of \Prop{prop:universalalphaproposed}.
We thus have $M_{R,1}(\dc_{3,\{i\}},\dc_1)=n-\dim\dc_{3,\{i\}}+1=k$.
Therefore $\maxustrong \leq k-1$ holds from \Prop{prop:universalalphaupper}.

\subsubsection{Required Packet Length $m$}\label{sect:packetlength}
Assume that $m < N$ in \Lma{lma:shorteningdimension} and \Lma{lma:puncturingdimension}.
Then, \Lma{lma:shorteningdimension} and \Lma{lma:puncturingdimension} do not always hold.
We give here a specific case in which \Lma{lma:shorteningdimension} and \Lma{lma:puncturingdimension} do not hold.
Considering the case where $m < N$ and additionally $m < |\mathcal{I}|$ in \Lma{lma:shorteningdimension}, we have
\begin{align*}
d_R(\c_\mathcal{I})\leq \frac{m}{|\mathcal{I}|}(|\mathcal{I}|-\dim \c_\mathcal{I})+1=\frac{m}{|\mathcal{I}|}(N-\dim\c)+1,
\end{align*}
by the Singleton-type bound for rank distance.
Since $m < |\mathcal{I}|$,
this clearly shows that \Lma{lma:shorteningdimension} does not hold in the case.
Also, when $m < N$ and $m < |\mathcal{I}|$ in \Lma{lma:puncturingdimension},
we have
\begin{align*}
d_R(P_\mathcal{I}(\c)) \leq \frac{m}{|\mathcal{I}|}(|\mathcal{I}| - \dim P_{\mathcal{I}}(\c)) + 1 = \frac{m}{|\mathcal{I}|}(|\mathcal{I}| - \dim \c) +1,
\end{align*}
and hence \Lma{lma:puncturingdimension} does not hold in the case.
Hence, we can see that $m \geq N$ is a necessary condition
for \Lma{lma:shorteningdimension} and \Lma{lma:puncturingdimension} to hold.
This also implies that Propositions
\ref{prop:ourschemeequivocation}--\ref{prop:universalalphaproposed}
 do not always hold
if the packet length is $m < l+n$ in our scheme.
Thus, the assumption $m \geq l+n$ is a necessary condition for our scheme
to always satisfy Propositions
\ref{prop:ourschemeequivocation}--\ref{prop:universalalphaproposed}
simultaneously.

\subsubsection{A Comparison of the Security and the Error-Correction Capability}\label{sect:comparison}

Here we summarize the comparison of our scheme with the scheme of Silva and Kschischang.
First we present a comparison about the security and error correction capability.
The scheme of Silva and Kschischang \cite{Silva2011} is the nested coset coding scheme using
a linear code $\c_1\subseteq\F_{q^m}^n$ and its subcode $\c_2 \subsetneqq \c_1$
where both $\c_1$ and $\c_2$ are MRD with $m \geq n$.
This immediately yields that \Thm{thm:apps1} and \Thm{thm:apps2}
are simultaneously established in their scheme as in our scheme.
However, their scheme does not specify the bijective function $\psi$ in such a way that
the condition in \Thm{thm:apps3} is always satisfied,
and hence their scheme does not always guarantee the universal maximum strength
$\maxustrong = \dim \c_1 -1$.
On the other hand, our scheme simultaneously satisfies Theorems \ref{thm:apps1}--\ref{thm:apps3}
as shown in Propositions \ref{prop:ourschemeequivocation}--\ref{prop:universalalphaproposed}.
Especially, one specific reason why our scheme satisfies \Prop{prop:universalalphaproposed}
is that the bijective function $\psi$ in our scheme is specified
by the systematic generator matrix $G$ of $\d$ as \Eq{eq:proposedmapping}.
Therefore, we can see that our scheme clearly has the advantage over their scheme in terms of the strong security.

Next we give a comparison about the required packet length.
In \cite[Theorem 8]{Silva2009},
Silva et al.\@ showed that there exist cases where their scheme in \cite{Silva2011}
satisfies the universal maximum strength $\maxustrong=\dim \c_1 -1$,
and that the sufficient condition on the existence of such a case is $m \geq (l+n)^2/8 + \log_q 16 l$ for packet length.
In contrast, we have demonstrated an explicit construction of the nested coset coding scheme
satisfying $\maxustrong=\dim \c_1 -1$ whenever $m \geq l+n$ is satisfied.
Furthermore, we always have $l+n < (l+n)^2/8 + \log_q 16 l$
 for $l \geq 1$ and $n \geq 2$.
Therefore, our condition for the packet length is less demanding than that
of Silva et al.\@'s sufficient condition.


\section{Conclusion}\label{sect:conc}
In this paper,
we have introduced two relative code parameters,
the relative dimension/intersection profile (RDIP) of a linear code $\c_1 \subseteq \F_{q^m}^n$
 and its subcode $\c_2\subsetneqq\c_1$ and
the relative generalized rank weight (RGRW) of $\c_1$ and $\c_2$.
We have also elucidated some basic properties of the RDIP and the RGRW.
We have clarified
the relation between the RGRW and the Gabidulin's rank distance \cite{Gabidulin1985},
that between the RGRW and the relative generalized Hamming weight \cite{Luo2005},
and that between the RGRW and the relative network generalized Hamming weight \cite{Zhang2009}.
As applications of the RDIP and the RGRW,
the security performance and the error correction
capability of secure network coding based on the nested coset coding scheme with $\c_1$ and $\c_2$
have been analyzed and clarified.
We have revealed that the security performance and the error correction capability,
guaranteed independently of the underlying network code,
are expressed in terms of the RDIP and the RGRW.
Further, we have proposed an explicit construction of the nested coset coding scheme,
and have analyzed its universal security performance and universal error correction capability
by using the RDIP and the RGRW.
As well as the scheme of Silva and Kschischang \cite{Silva2011},
the proposed scheme guarantees,
independently of the underlying network code,
that no information of the secret message is obtained from any $\mu \leq \dim \c_2$ tapped links
when the transmitted packets are uniformly distributed over $\c_1$,
and that the secret message is correctly decodable against any $t$ error packets injected somewhere
in the network and $\rho$ rank deficiency of the transfer matrix of the sink node
whenever $n-\dim\c_1+1<2t+\rho$ holds.
Moreover, our scheme also always guarantees that
no part of the secret message is revealed to the adversary with $\mu \leq \dim \c_1 -1$ tapped links
when the secret message and transmitted packets are uniformly distributed,
unlike Silva et al.\@'s scheme \cite{Silva2009,Silva2011}.

\Sect{sect:example} of this paper presented only one instance of the nested coset coding scheme
 that has specific universal security performance and universal error correction capability,
 \ie specific values of RDIP and RGRW.
We believe that the security scheme should be designed according to the system requirements and environments.
Hence, how to design a pair of a linear code $\c_1$ and its subcode $\c_2$ from arbitrarily given RDIP and RGRW
is left as an important open problem for future work.
Another possible avenue is to derive other types of bounds of the RGRW,
\eg generalizing the Gilbert-Varshamov bound of the rank distance \cite{Gadouleau2007} for the RGRW, etc.

Recall that the theory of the RDIP and RGRW established in this paper
is similar to the theory of the GHW \cite{Wei1991} that was proposed to investigate the security performance
of coding schemes on the Wiretap Channel II \cite{Ozarow1984}.
The specific coding schemes based on maximum distance separable codes
have been already known as optimal ones in the Wiretap Channel II.
However, the theory of the GHW is not regarded as unnecessary,
because it is important and required to reveal the security performance
of \textit{any} coding schemes on the Wiretap Channel II.
As a conclusion of this paper,
we allege that
this importance of the GHW is exactly same as
what we have established in this paper about the RDIP and RGRW for network coding.

\appendices

\section{Proof of \Lma{lma:lmamutual}}\label{sect:lmamutual_proof}

We first give the following lemma
that will be used to prove \Lma{lma:lmamutual}.
\begin{lemma}\label{lma:generalizedforney}
Let $\c_1 \subseteq \F_{q^m}^n$ be a linear code and $\c_2 \subsetneqq \c_1$ be its subcode.
For an arbitrary subspace $V \subseteq \F_{q^m}^n$,
we have
\begin{align*}
&\dim (\c_1 \cap V)
-
\dim \left( \c_2 \cap V \right)\\
&\quad=
\dim \c_1/\c_2
-
\dim(\dc_2\cap V^\perp) + \dim(\dc_1\cap V^\perp).
\end{align*}
\end{lemma}
\begin{IEEEproof}
For a linear subspace $\c \subseteq \F_{q^m}^n$, we have
$\dim \c + \dim (\dc \cap V^\perp) =  \dim V^\perp + \dim (\c \cap V)$.
Thus, by letting $\c = \c_1$ and $\c=\c_2$ in this equation,
we obtain
\begin{align*}
0 &= \dim \c_1 + \dim (\dc_1 \cap V^\perp) - \dim V^\perp - \dim (\c_1 \cap V),
\end{align*}
and
\begin{align*}
0 &= \dim \c_2 + \dim (\dc_2 \cap V^\perp) - \dim V^\perp - \dim (\c_2 \cap V),
\end{align*}
respectively.
Therefore, the lemma is established by these equalities since $\dim \c_1 - \dim \c_2 = \dim \c_1/\c_2$.
\end{IEEEproof}\vspace{1.5ex}

Next we recall that for random variables $A\in\mathcal{A}$ and $B \in \mathcal{B}$,
we have the following relations among the conditional entropy
 and the (conditional) relative entropy \cite[Ch.~2, p.~27]{Cover2006}:
\begin{align}
H(A) &= \log |\mathcal{A}| - D(A\|U_\mathcal{A}), \label{eq:kl1}\\
H(A|B) &= \E_{B} \left[\log |\mathcal{A}(B)|\right]
- D(A\|U_{\mathcal{A}(B)}|B), \label{eq:kl2}
\end{align}
where $A \in \mathcal{A}(b)$ with probability one given $B=b$ and
$\E_B$ denotes the expectation over the probability distribution $P_B$.
In the following, we will use these relationships to prove the lemma.

Recall that for each $S=s$, a coset
 $\psi(s) \in \c_1/\c_2$ is uniquely determined.
Also observe that for given $W=w$ as a realization of $W$,
there exists a unique coset
$\mathcal{X}(w) = \{ x \in \c_1 : Bx^{\rm T} = w^{\rm T} \} \in \c_1/(\row{B}^\perp\cap\c_1)$.
Observe that
$X$ belongs to $\psi(s) \cap \mathcal{X}(w)$ when $S=s$ and $W=w$,
and that 
\begin{align*}
|\psi(s) \cap \mathcal{X}(w)|
=|\psi(0) \cap \mathcal{X}(0)|
&=|\c_2 \cap (\row{B}^\perp \cap \c_1)|\\*
&= |\c_2 \cap \row{B}^\perp|.
\end{align*}
Hence we have
\begin{align*}
 \log_{q^m} \left| \psi(s) \cap \mathcal{X}(w) \right|
&= \log_{q^m} \left| \c_2 \cap \row{B}^\perp \right|\\
&= \dim \left( \c_2 \cap \row{B}^\perp \right),
\end{align*}
for any $s$ and $w$.
Thus, by \Eq{eq:kl2}, we have
\begin{align}
H(X|S,W)
&= \dim \left( \c_2 \cap \row{B}^\perp \right) - D(X \| U_{\psi(S) \cap \mathcal{X}(W)}| S, W).
 \label{eq:lmamutual1}
\end{align}

Also observe that
$X$ is distributed over $\mathcal{X}(w)$ when $W=w$,
and that
\begin{align*}
\log_{q^m}|\mathcal{X}(w)|
=\log_{q^m} \left| \c_1 \cap \row{B}^\perp \right|
= \dim \left(\c_1 \cap \row{B}^\perp\right), 
\end{align*}
for any $w$.
Thus, by \Eq{eq:kl2}, we obtain
\begin{align}
H(X|W)
&=
\dim \left( \c_1 \cap \row{B}^\perp \right)
-D(X \| U_{\mathcal{X}(W)} |W).
 \label{eq:lmamutualx}
\end{align}

Recall that $X$ is distributed over
 a coset $\psi(s)\in\c_1/\c_2$ for fixed $S=s$,
and $\log_{q^m} |\psi(s)| = \dim \c_2$ for any $s$.
Thus, by \Eq{eq:kl2}, we have
\begin{align}
H(X|S)
= \dim \c_2
-D(X \| U_{\psi(S)} | S).
\label{eq:lmamutual2}
\end{align}

Let a subspace $\mathcal{W} = \{x B^{\rm T} : x \in\c_1\}$.
For the cardinality of $\mathcal{W}$,
we have 
\begin{align*}
 \log_{q^m}|\mathcal{W}| = \dim \mathcal{W} = \dim \c_1 - \dim (\c_1 \cap \row{B}^\perp).
\end{align*}
Thus, by \Eq{eq:kl1}, we have
\begin{align}
H(W)
 &= \log_{q^m} |\mathcal{W}| - D(W \|U_\mathcal{W})\nonumber\\*
 &= \dim \c_1 - \dim (\c_1 \cap \row{B}^\perp) - D(W \| U_\mathcal{W}). \label{eq:lmamutual3}
\end{align}

Recall that for given $B$ and fixed $X=x$, $W=xB^{\rm T}$ is uniquely determined. This implies $H(W|X) = 0$.
Thus, by $H(W|S,X) \leq H(W|X)=0$ and the nonnegativity of the entropy function \cite[Ch.~2, p.~14]{Cover2006},
we have
\begin{align}
H(W|S,X) = 0. \label{eq:lmamutualadd}
\end{align}

By expanding $I(S;W)$ and substituting
\Eq{eq:lmamutual1}, \Eq{eq:lmamutual2}, \Eq{eq:lmamutual3} and \Eq{eq:lmamutualadd}
into the expanded equation \Eq{eq:expanded},
we obtain
\begin{align}
&I(S;W)\nonumber\\*
&\!=\underbrace{I(S,X;W)}_{=H(W)-H(W|S,X)} - \underbrace{I(X;W|S)}_{=H(X|S)-H(X|S,W)}
\nonumber\\*
&\!= \underbrace{H(W)}_{\substack{= \dim \c_1 - \dim (\c_1 \cap \row{B}^\perp) - D(W\| U_\mathcal{W}) \\ \text{(by \Eq{eq:lmamutual3})}}}
- \underbrace{H(W|S,X)}_{\substack{=0 \\ \text{(by \Eq{eq:lmamutualadd})}}}
\nonumber\\*
&\quad\ 
- \underbrace{H(X|S)}_{\substack{= \dim \c_2 -D(X \| U_{\psi(S)} | S) \\ \text{(by \Eq{eq:lmamutual2})}}}
+ \underbrace{H(X|S,W)}_{\substack{= \dim \left( \c_2 \cap \row{B}^\perp \right) - D(X \| U_{\psi(S) \cap \mathcal{X}(W)}|S,W) \\ \text{(by \Eq{eq:lmamutual1})}}}
\label{eq:expanded}\\*
&\!=
\underbrace{\dim \c_1 \!-\! \dim \c_2}_{=l}
\!-\! \dim \left(\c_1 \!\cap\! \row{B}^\perp\right)
\!+\! \dim \left( \c_2 \!\cap\! \row{B}^\perp \right)\nonumber\\*
&\qquad
+D(X \| U_{\psi(S)} | S)
- D(W\| U_\mathcal{W})
- D(X \| U_{\psi(S) \cap \mathcal{X}(W)}|S,W)
\nonumber\\*
&\!\leq
\underbrace{l \!-\! \dim \left(\c_1 \!\cap\! \row{B}^\perp\right) \!+\! \dim \left( \c_2 \!\cap\! \row{B}^\perp \right)}_{\substack{= \dim \left(\dc_2 \cap \row{B} \right) - \dim \left(\dc_1 \cap \row{B} \right) \text{ (by \Lma{lma:generalizedforney})}}}
+ D(X \| U_{\psi(S)} | S),
\nonumber
\end{align}
which proves \Eq{eq:lmamutual}.

On the other hand, observe that
the number of possible $S$
for any given $W=w$ is exactly equal to $q^{m \cdot  \dim (\c_1 \cap \row{B}^\perp)}/q^{m  \cdot \dim (\c_2 \cap \row{B}^\perp)}$.
Also recall that the relative entropy is nonnegative \cite[Ch.~2, p.~26]{Cover2006}.
Thus, by applying \Eq{eq:kl2} to the set $\{s \in \F_{q^m}^l : w = xB^{\rm T}, x \in \psi(s) \}$ that depends on the realization $W = w$,
we have the following inequality.
\begin{align*}
H(S|W) \leq \dim (\c_1 \cap \row{B}^\perp)
-\dim (\c_2 \cap \row{B}^\perp).
\end{align*}
Thus,
\begin{align*}
I(S;W) &= H(S) - H(S|W) \\
&\geq H(S) - \dim (\c_1 \cap \row{B}^\perp)
-\dim (\c_2 \cap \row{B}^\perp)\\
&= 
H(S) - l + \dim(\dc_2 \!\cap\! \row{B}) - \dim(\dc_1 \!\cap\! \row{B})\\ &\hspace{34ex}\text{(by \Lma{lma:generalizedforney})}
\\
&= \dim(\dc_2 \!\cap\! \row{B}) \!-\! \dim(\dc_1 \!\cap\! \row{B}) 
\!-\!D(S\| U_{\F_{q^m}^l}),\\ &\hspace{40ex}\text{(by \Eq{eq:kl1})}
\end{align*}
which proves \Eq{eq:lmamutualxx}.
Thus, we have the statement 1) in the lemma.

Here, we show the equalities in \Eq{eq:lmamutual} and \Eq{eq:lmamutualxx} for the uniformly distributed $S$ and $X$.
Assume that $S$ is uniform over $\F_{q^m}^l$.
Then, from \Eq{eq:kl1}, we have $D(S \| U_{\F_{q^m}^l}) = l-H(S) = 0$.
Also, when $X$ is uniform over $\c_1$, \ie uniform over $\psi(S)$,
we have $H(X|S)=\dim\c_2$
and hence $D(X \| U_{\psi(S)} |S) = \dim \c_2 - H(X|S) = 0$ from \Eq{eq:kl2}.
Therefore, the equalities in \Eq{eq:lmamutual} and \Eq{eq:lmamutualxx} hold,
and we have the statement 2) in the lemma.

Finally, we show the statement 3)
for the distribution of $S$ that assigns a positive probability to every element in $\F_{q^m}^l$.
Let $P_S$ be a distribution of $S$ such that all elements in $\F_{q^m}^l$ have positive probabilities,
and assume that $I(S;W)=0$ holds for $P_S$.
Recall that the mutual information is expressed in terms of the relative entropy \cite[Ch.~2, pp.~18--19]{Cover2006} as
\begin{align*}
0=I(S;W) = \sum_{s}P_S(s) D( W_{S=s} || W ).
\end{align*}
where $W_{S=s}$ is the random variable with the distribution $P_{W|S=s}$.
Thus, $D(W_{S=s} || W)=0$, \ie $P_{W|S=s} = P_W$, simultaneously holds for all $s\in\F_{q^m}^l$
from the nonnegativity of relative entropy \cite[Ch.~2, p.~26]{Cover2006}.

Here, consider another random variable $S'$ with an arbitrary distribution $P_{S'}$,
and the corresponding random variable $W'$.
Here we assume that
the conditional probability of $W'$ given $S'$ is the same as
that of $W$ given $S$, which means that
$P_{W'|S'=s}= P_{W|S=s} = P_W = P_{W'}$ for all $s$.
By $I(S'|W') = \sum_{s} P_{S'}(s) D(W'_{S'=s} || W')$, we see
$I(S'|W')=0$.
In particular, for the uniformly distributed $S'$
we have $I(S';W')=0$ and $D(S' \| U_{\F_{q^m}^l})=0$,
and hence we have
\begin{align*}
0=I(S';W') \geq \dim(\dc_2 \cap \row{B}) - \dim(\dc_1 \cap \row{B}),
\end{align*}
from \Eq{eq:lmamutualxx}.
Therefore, since 
\begin{align*}
 \dim(\dc_2 \cap \row{B}) - \dim(\dc_1 \cap \row{B}) \geq 0,
\end{align*} holds,
we have $\dim(\dc_2 \cap \row{B}) - \dim(\dc_1 \cap \row{B}) = 0$.
\hfill\IEEEQED

\section{Proof of \Prop{prop:universalalphalower}}\label{sect:universalalphalower_proof}

From the definition of $\c'_1$,
$\d_{2,i}$ is a subcode of $\d_{1,i}$ with dimension
 $\dim \d_{2,i}=\dim \d_{1,i} - 1 = \dim \c_1 - 1$ over $\F_{q^m}$
for each $i \in \{1,\dots,l\}$.
Let $\mathcal{L}\triangleq\{1,\dots,l\}$
and $S_{\mathcal{L}\backslash\{i\}} \triangleq [S_1,\dots,S_{i-1},S_{i+1},\dots,S_l]$ for
 $1\leq i \leq l$.
For $S_i \in\F_{q^m}$, define a coset
\begin{align*}
\tau(S_i)
&\triangleq \left\{[S_{\mathcal{L}\backslash\{i\}},X]:
 S_{\mathcal{L}\backslash\{i\}} \in\F_{q^m}^{l-1}
 \text{ and }
 X \in \psi([S_1,\dots,S_l])\right\}
\\
&\hspace{35ex}\in\d_{1,i}/\d_{2,i}.
\end{align*}
Here we define
$\vecxi \triangleq P_{\bari{i}}([S,X])= [S_{\mathcal{L}\backslash\{i\}},X] \in \d_{1,i}$.
Recall that $S_1,\dots,S_l$ are mutually independent and
 uniformly distributed over $\F_{q^m}$.
Thus, $\vecxi$ can be regarded as the one generated from a secret message $S_i \in \F_{q^m}$
by a nested coset coding scheme with $\d_{1,i}$ and $\d_{2,i}$
according to the uniform distribution over $\tau(S_i)$,
that is, $\vecxi \in \tau(S_i)$
 is chosen uniformly at random from $\tau(S_i) \in \d_{1,i}/\d_{2,i}$.
Therefore, 
we have $I(S_i;D\vecxi^{\rm T})=0$
for any $D\in\F_q^{\mu \times (n+l-1)}$ whenever $\mu < M_{R,1}(\dd_{2,i},\dd_{1,i})$
from \Coro{prop:perfectsecrecy}.

For an arbitrary subset $\mathcal{R} \subseteq \mathcal{L}\backslash\{i\}$,
define a matrix $F_\mathcal{R}$ that consists of $|\mathcal{R}|$ rows of
 an $(l-1)\times(l-1)$ identity matrix, satisfying
$[S_j : j  \in  \mathcal{R}]^{\rm T}  =  F_\mathcal{R} S_{\mathcal{L}\backslash\{i\}}^{\rm T}$.
Here we note that $F_\mathcal{R}\in\F_{q^m}^{|\mathcal{R}|\times (n+l-1)}$ is defined as a matrix over the base field $\F_q$.
For an arbitrary matrix $B \in \F_q^{k \times n}$ ($0  \leq  k  \leq  n$),
let $\mu=|\mathcal{R}|+k$ and $D = \left[\begin{smallmatrix}F_\mathcal{R} & O \\ O & B\end{smallmatrix}\right]\in\F_q^{(|\mathcal{R}|+k)\times(n+l-1)}$.
Then,
since $D\vecxi^{\rm T}=\left[\begin{smallmatrix} [S_j : j  \in  \mathcal{R}]^{\rm T}\\ BX^{\rm T} \end{smallmatrix}\right]$,
we have the following equality from the foregoing proof.
\begin{align}
0=I(S_i;D\vecxi^{\rm T}) = I(S_i; S_\mathcal{R},BX^{\rm T}), \label{eq:x201}
\end{align}
whenever $|\mathcal{R}|  +  k  <  M_1(\dd_{2,i},\dd_{1,i})$.
Let $\mathcal{R}' \triangleq \mathcal{R}\cup\{i\} = \{r_1,\dots,r_{|\mathcal{R}|+1}\}$.
Since $S_1,\dots,S_l$ are mutually independent,
the mutual information between $S_{\mathcal{R}'}$ and $BX^{\rm T}$ is given by
\begin{align*}
I(S_{\mathcal{R}'};BX^{\rm T})
 &=
H(S_{\mathcal{R}'}) - H(S_{\mathcal{R}'}|BX^{\rm T})\\*
&=
\sum_{j=1}^{|\mathcal{R}|+1} H(S_{r_j})
-
\sum_{j=1}^{|\mathcal{R}|+1}
H(S_{r_j}|BX^{\rm T},S_{\{r_1,\dots,r_{j-1}\}})
\\*
&=
\sum_{j=1}^{|\mathcal{R}|+1}
 I(S_{r_j} ; BX^{\rm T}, S_{\{r_1,\dots,r_{j-1}\}}),
\end{align*}
from the chain rule \cite[Ch.~2, p.~16]{Cover2006}.
Since the mutual information is nonnegative \cite[Ch.~2, p.~27]{Cover2006},
we have $I(S_{\mathcal{R}'};BX^{\rm T})=0$
if and only if $I(S_{r_j} ; BX^{\rm T}, S_{\{r_1,\dots,r_{j-1}\}})=0$
for all $r_j \in \mathcal{R}'$.
By substituting $i=r_j$ in \Eq{eq:x201},
we always have $I(S_{r_j} ; BX^{\rm T}, S_{\{r_1,\dots,r_{j-1}\}})=0$
only for $r_j$ if $|\mathcal{R}|  +  k  <  M_1(\dd_{2,r_j},\dd_{1,r_j})$.
Thus, we always have $I(S_{\mathcal{R}'};BX^{\rm T})=0$ for arbitrary $k$ and $\mathcal{R}'$
whenever $|\mathcal{R}|  +  k  <  \min\left\{ M_1(\dd_{2,i},\dd_{1,i}) : 1 \leq i \leq l \right\}$ holds.
Therefore, we prove that the universal $\ustrong$-security is attained
whenever $\ustrong < \min\left\{
M_{R,1}(\dd_{2,i},\dd_{1,i}) : 1 \leq i \leq l
\right\}$,
and we have \Eq{eq:universalalphamin}.
\hfill\IEEEQED

\section{Derivation of the Main Theorems of Universal Error-Correction Capability}\label{sect:appendix_errorcorrection}
In this appendix,
we first briefly review Silva et al.\@'s approach \cite[Section III]{Silva2009a}.
Next, by using their approach,
 we analyze the error correction capability of the nested coset coding scheme
 over the coherent network coding system
 and derive \Thm{thm:errorcorrectioncap}.
We finally extend the analysis to the noncoherent systems
and also derive \Prop{prop:errorcorrectioncap_noncoherent}.
Here we note that these derivations of theorems in \Sect{sect:errorcorrection}
 are natural generalizations of the work in \cite{Silva2009a} to the error correction of the nested coset coding scheme.

\subsection{Brief Review of Silva et al.\@'s Approach}\label{sect:silvaapproach}
First we give a brief review of the approach of \cite[Section III]{Silva2009a}.
Consider a transmission of data over a channel in which there exists an adversary.
Let the channel be specified by
 a finite input alphabet $\mathcal{P}$ (\eg a code),
 a finite output alphabet $\mathcal{Q}$ (\eg a vector space),
 and a collection of fan-out sets $\mathcal{Q}_P \subseteq \mathcal{Q}$
 for all $P \in \mathcal{P}$ (\eg a collection of cosets).
For each input $P \in \mathcal{P}$, the output $Q$ of the channel
is constrained to be in $\mathcal{Q}_P$ but is otherwise arbitrarily chosen
by an adversary.
A decoder for $\mathcal{P}$ is any function
$\hat{P}:\mathcal{Q} \rightarrow \mathcal{P} \cup \{f\}$,
where $f \notin \mathcal{P}$ denotes a decoding failure, \ie detected errors.
When $P\in\mathcal{P}$ is transmitted and $Q\in\mathcal{Q}_P$ is received,
a decoder is said to be \textit{successful} if $\hat{P}(Q)=P$.
We also say that a decoder is \textit{infallible}
if it is successful for all $Q \in \mathcal{Q}_P$
and all $P \in \mathcal{P}$.

Assume that the fan-out sets for a input $P$ is given as
\begin{align*}
\mathcal{Q}_P = \left\{
Q \in \mathcal{Q} : \Delta(P,Q) \leq t
\right\},
\end{align*}
for some  $\Delta: \mathcal{P} \times \mathcal{Q} \rightarrow \mathbb{N}$.
The value $\Delta(P,Q)$ is called the \textit{discrepancy} between $P$ and $Q$ for the given channel,
which represents the minimum effort required for an adversary in the channel to
transform $P$ to $Q$.
The value $t$ represents the maximum effort of the adversary allowed in the channel.
The problem is to decode $P$ from $Q$ by correcting at most $t$ discrepancy.
Then, the \textit{minimum-discrepancy decoder} is defined by
\begin{align*}
\hat{P} = \argmin_{P \in \mathcal{P}} \Delta(P,Q).
\end{align*}
The relation between the discrepancy function
 and the error correction capability of this decoder was given in \cite{Silva2009a}
as follows.
\begin{definition}[{\cite[Definition 1]{Silva2009a}}]\label{def:silvanormal}
For a discrepancy function $\Delta: \mathcal{P} \times \mathcal{Q} \rightarrow \mathbb{N}$,
let $\delta(P,P')=\min\left\{ \Delta(P,Q)+\Delta(P',Q): Q \in \mathcal{Q} \right\}$.
Then, $\Delta$ is said to be \textit{normal} if,
for all $P,P'\in\mathcal{P}$ and all
 $0 \leq i \leq \delta(P,P')$,
there exists some $Q \in \mathcal{Q}$ such that $\Delta(P,Q)=i$
 and $\Delta(P, Q)=\delta(P,P')-i$.
\end{definition}
\begin{theorem}[{\cite[Proposition 1, Theorem 3]{Silva2009a}}]\label{thm:silvacorrection}
Let $\delta(\mathcal{P}) = \min \left\{ \delta(P,P') : P,P'\in\mathcal{P}, P \neq P'\right\}$.
Suppose $\Delta(\cdot,\cdot)$ is normal.
Then, the minimum discrepancy decoder $\hat{P}$
is infallible
if and only if $t \leq \lfloor (\delta(\mathcal{P}) -1)/2 \rfloor$.
\end{theorem}

\subsection{How to Prove \Thm{thm:errorcorrectioncap}}\label{sect:coherentcapability}

By applying the above approach \cite[Section III]{Silva2009a}
to the secure network coding over the coherent network coding system
in \Sect{sect:coherent},
this subsection derives \Thm{thm:errorcorrectioncap},
\ie the universal error correction
capability of the nested coset coding scheme with $\c_1,\c_2$ for given $A$,
 expressed in terms of the first RGRW.

Recall that the received packets $Y$ are given by
 $Y^{\rm T}=AX^{\rm T}+DZ^{\rm T}$ in the setup of \Sect{sect:nestedcoding},
and that $X\in\F_{q^m}^n$ is chosen from a set
 $\mathcal{X}_S \in \mathcal{P}_{\mathcal{S}}$
 corresponding to $S \in \mathcal{S}$ by
a certain coding scheme defined in \Def{def:generalcoding}.
Note that we do not restrict the coding scheme to the nested coset coding scheme here.
From now on, we write $\mathcal{X}\triangleq\mathcal{X}_S$ for the sake of simplicity.
Suppose that the transfer matrix $A$ is known to the sink node as in \Sect{sect:coherent}.
Here, we define the discrepancy function between
$\mathcal{X}$ and $Y$ for given $A$ by
\begin{align}
&\Delta_A (\mathcal{X}, Y)\nonumber\\
& \triangleq 
\min \left\{r \!\in\! \mathbb{N} :
\exists D \!\in\! \F_{q}^{N \times r},
\exists Z \!\in\! \F_{q^m}^{r},
\exists X \!\in\! \mathcal{X},
Y^{\rm T} \!=\! AX^{\rm T} \!+\! DZ^{\rm T}\right\}.
\label{eq:defdiscrepancy}
\end{align}
This definition of $\Delta_A(\mathcal{X},Y)$ represents 
the minimum number $r$
of error packets $Z$ required to be injected in order to transform
at least one element of $\mathcal{X}$ into $Y$,
as \cite[(9)]{Silva2009a}.
For the discrepancy function $\Delta_A(\mathcal{X},Y)$,
the minimum discrepancy decoder is given as
\begin{align*}
\hat{\mathcal{X}} = \argmin_{\mathcal{X} \in \mathcal{P}_\mathcal{S}} \Delta_A(\mathcal{X},Y).
\end{align*}
Note that ``the minimum discrepancy decoder $\hat{P}$ is infallible''
 in \Thm{thm:silvacorrection}
means that for the discrepancy function $\Delta_A(\mathcal{X},Y)$,
``any $t$ error packets can be corrected by the coding scheme for given $A$ using
 the minimum discrepancy decoder $\hat{\mathcal{X}}$.''
In the following, we will show that $\Delta_A(\mathcal{X},Y)$ is normal.

We define the \textit{$\Delta$-distance} \cite{Silva2009a}
between
$\mathcal{X}$ and $\mathcal{X}'$, induced by
$\Delta_A(\mathcal{X},Y)$, as
\begin{align}
\delta_A(\mathcal{X},\mathcal{X}')
\triangleq
\min
\left\{\Delta_A(\mathcal{X},Y)+ \Delta_A(\mathcal{X}',Y) : Y \in \F_{q^m}^N\right\},
\label{eq:defdelta}
\end{align}
for $\mathcal{X},\mathcal{X}'\in\mathcal{P}_\mathcal{S}$.
Let $\delta_A(\mathcal{P}_\mathcal{S})$ be the minimum $\Delta$-distance given by
\begin{align*}
\delta_A (\mathcal{P}_\mathcal{S})
\triangleq
\min \left\{ \delta_A(\mathcal{X},\mathcal{X}') :
 \mathcal{X},\mathcal{X}' \in \mathcal{P}_\mathcal{S}, \mathcal{X}\neq\mathcal{X}'
\right\}.
\end{align*}
\begin{lemma}[{\cite[Lemma 4]{Silva2009a}}]\label{lma:silvadiscrepancy}
\begin{align*}
\min\left\{
r \!\in\! \mathbb{N} :
D \!\in\! \F_{q}^{N \times r},
Z \!\in\! \F_{q^m}^{r},
Y^{\rm T} \!=\! AX^{\rm T} \!+\! DZ^{\rm T}\right\}
=
d_R(XA^{\rm T},Y).
\end{align*}
\end{lemma}\vspace{1.5ex}
\begin{lemma}\label{lma:discrepancy}
$\Delta_A (\mathcal{X}, Y) = \min \left\{
d_R(XA^{\rm T},Y) : X \in \mathcal{X}
\right\}$.
\end{lemma}
\begin{IEEEproof}
From \Lma{lma:silvadiscrepancy}, we have
\begin{align*}
&\Delta_A (\mathcal{X}, Y)\\
& =
\min \left\{r \!\in\! \mathbb{N} :
D \!\in\! \F_{q}^{N \times r},
Z \!\in\! \F_{q^m}^{r},
X \!\in\! \mathcal{X},
Y^{\rm T} \!=\! AX^{\rm T} \!+\! DZ^{\rm T}\right\}
\\
&=
\min
\left\{
\min
\left\{
r \!\in\! \mathbb{N} :
D \!\in\! \F_{q}^{N \times r},
Z \!\in\! \F_{q^m}^{r},
Y^{\rm T} \!=\! AX^{\rm T} \!+\! DZ^{\rm T}\right\}
: X \!\in\! \mathcal{X}
\right\}
\\
&=
\min \left\{
d_R(XA^{\rm T},Y) : X \in \mathcal{X}
\right\}.
\end{align*}
\end{IEEEproof}\vspace{1.5ex}

\begin{lemma}\label{lma:deltadistance}
For $\mathcal{X}, \mathcal{X}' \in \mathcal{P}_\mathcal{S}$, we have
\begin{align}
\delta_A(\mathcal{X},\mathcal{X}')
&=
\min
\left\{
d_R(XA^{\rm T},X'A^{\rm T}):
X \in \mathcal{X},
X' \in \mathcal{X}'
\right\}. \label{eq:deltalma}
\end{align}
\end{lemma}
\begin{IEEEproof}
First we have
\begin{align}
&\delta_A(\mathcal{X},\mathcal{X}')\nonumber\\
&=
\min
\left\{\Delta_A(\mathcal{X},Y)+ \Delta_A(\mathcal{X}',Y) : Y \in \F_{q^m}^N\right\}
\nonumber\\
&= \min
\Big\{
\min \left\{
d_R(XA^{\rm T},Y) : X \in \mathcal{X}
\right\}+ \nonumber\\
&\hspace{15ex} 
\min \left\{
d_R(X'A^{\rm T},Y) : X' \in \mathcal{X}'
\right\}
: Y \in \F_{q^m}^N
\Big\}\nonumber\\
&=
\min \left\{
d_R(XA^{\rm T},Y) + d_R(X'A^{\rm T},Y):
X \in \mathcal{X},
X' \in \mathcal{X}',
Y \in \F_{q^m}^N
\right\}.\label{eq:triangle}
\end{align}
The rank distance satisfies 
the triangle inequality
$d_R(XA^{\rm T},X'A^{\rm T}) \leq d_R(XA^{\rm T},Y) + d_R(X'A^{\rm T},Y)$
for $\forall Y\in\F_{q^m}^N$ \cite{Gabidulin1985}.
This lower bound can be achieved by choosing,
\eg $Y =X A^{\rm T}$.
Therefore, from \Eq{eq:triangle}, we have \Eq{eq:deltalma}.
\end{IEEEproof}\vspace{1.5ex}

\begin{lemma}\label{lma:normal}
The discrepancy function $\Delta_A(\mathcal{X},Y)$ is normal.
\end{lemma}
\begin{IEEEproof}
Let $\mathcal{X},\mathcal{X}'\in\mathcal{P}_\mathcal{S}$ and let
 $0 \leq i \leq d=\delta_A(\mathcal{X},\mathcal{X}')$.
Then, $d=\min \left\{ d_R(XA^{\rm T},X'A^{\rm T}): X \in \mathcal{X}, X' \in \mathcal{X}'\right\}$
from \Lma{lma:deltadistance}.
Let $\bar{X}\in\mathcal{X}$ and $\bar{X'}\in\mathcal{X}'$ be vectors
 satisfying $d=d_R(\bar{X}A^{\rm T},\bar{X}'A^{\rm T})$.
Here, we can always find two vectors $W,W'\in\F_{q^m}^n$ such that
$W+W' = (\bar{X}'-\bar{X})A^{\rm T}$, $\dimq\colq{W}=i$ and
 $\dimq\colq{W'}=d-i$,
as shown in the proof of \cite[Theorem 6]{Silva2009a}.
Taking $\bar{Y}=\bar{X}A^{\rm T}+W=\bar{X}'A^{\rm T}-W'$,
we have
$d_R(\bar{X}A^{\rm T},\bar{Y}) = i$ and $d_R(\bar{X}'A^{\rm T},\bar{Y}) = d-i$.
We thus obtain
$\Delta_A(\mathcal{X},\bar{Y}) \leq i$ and
$\Delta_A(\mathcal{X}',\bar{Y}) \leq d-i$ from \Lma{lma:discrepancy}.
On the other hand, since $\delta_A(\mathcal{X},\mathcal{X}')=d$,
we have $\Delta_A(\mathcal{X},Y) + \Delta_A(\mathcal{X}',Y) \geq d$
for any $Y \in \F_{q^m}^n$ from \Eq{eq:defdelta}.
Therefore, 
$\Delta_A(\mathcal{X},\bar{Y}) = i$ and $\Delta_A(\mathcal{X}',\bar{Y})
 = d-i$
hold.
\end{IEEEproof}\vspace{1.5ex}
As \cite[Theorem 7]{Silva2009a} obtained by \cite[Theorems 3 and 6]{Silva2009a},
we have \Prop{prop:deltacorrection} from \Thm{thm:silvacorrection} and \Lma{lma:normal}
by the approach of Appendix \ref{sect:silvaapproach}.

\begin{proposition}\label{prop:deltacorrection}
Consider the $t$-error $(n \times m)_q$ linear network in \Def{def:terror}.
Suppose that for a secret message $S\in\F_{q^m}^l$,
the transmitted $n$ packets $X\in\mathcal{X}$ are generated by a coding scheme defined in \Def{def:generalcoding}.
Then, the minimum discrepancy decoder for $\Delta_A(\mathcal{X},Y)$
is infallible for any fixed $A$
if and only if $t \leq \lfloor (\delta_A(\mathcal{P}_\mathcal{S})-1)/2 \rfloor$.
\hfill\IEEEQED
\end{proposition}
This proposition implies that
the coding scheme given in \Def{def:generalcoding} is guaranteed to
determine the unique set $\mathcal{X}$ against
any $t$ packet errors for any fixed $A$
if and only if
$\delta_A(\mathcal{P}_\mathcal{S}) > 2t$.
Here we note that if $\mathcal{X}$ is uniquely determined, $S$ is also uniquely
determined from \Def{def:generalcoding}.

In the following, we restrict the coding scheme to the nested coset coding scheme with $\c_1$ and $\c_2$,
and present a special case of \Prop{prop:deltacorrection} expressed in terms of the RGRW.
That is, we set $\mathcal{S}=\F_{q^m}^l$
 and $\mathcal{P}_\mathcal{S} = \c_1/\c_2$ as defined in \Def{def:nestedcoding}.

\begin{lemma}\label{eq:cosetdelta}
\begin{align*}
 \delta_A(\c_1/\c_2) =
\min\{
d_R(XA^{\rm T}, X'A^{\rm T}) : X,X' \in \c_1, X' - X \notin \c_2
\}.
\end{align*}
\end{lemma}
\begin{IEEEproof}
\begin{align*}
&\delta_A(\c_1/\c_2)\\*
&=\min \left\{ \delta_A(\mathcal{X},\mathcal{X}') :
 \mathcal{X},\mathcal{X}' \in \c_1/\c_2, \mathcal{X}\neq\mathcal{X}'
\right\}\\*
&=
\min \Big\{
 \min \left\{
d_R(XA^{\rm T},X'A^{\rm T}) : X \!\in\! \mathcal{X},X' \!\in\! \mathcal{X}'
\right\} :\\
&\hspace{32ex} \mathcal{X},\mathcal{X}' \!\in\! \c_1/\c_2, \mathcal{X} \!\neq\! \mathcal{X}'
\Big\}\\*
&=
\min\Big\{
d_R(XA^{\rm T},X'A^{\rm T}) \!:\!
X \!\in\! \mathcal{X} \!\in\! \c_1/\c_2,
X' \!\in\! \mathcal{X}' \!\in\! \c_1/\c_2, \mathcal{X} \!\neq\! \mathcal{X}'
\Big\}\\*
&=
\min \left\{
d_R(XA^{\rm T},X'A^{\rm T}) : X,X' \in \c_1, X'-X \notin \c_2
\right\}.
\end{align*}
\end{IEEEproof}\vspace{1.5ex}
\begin{lemma}[{\cite[Ch.~4, p.~211]{Meyer2001}, \cite{Silva2009a}}]\label{lma:meyer}
For an arbitrary vector $x\in\F_{q^m}^n$ and an arbitrary matrix $A \in \F_q^{N \times n}$,
we have $\dimq \colq{xA^{\rm T}} \geq \left[\dimq \colq{x} + \rank A - n\right]^+$.
\end{lemma}
\begin{lemma}\label{lma:meyer2}
Fix $x\in\F_{q^m}^n$ and $\rho \in \{0\dots,n\}$ arbitrarily.
Then, there always exists $A\in\F_q^{N \times n}$ with $\rank A = n-\rho$
that satisfies the equality $\dimq \colq{xA^{\rm T}} = \left[\dimq \colq{x} - \rho \right]^+$
in \Lma{lma:meyer}.
\end{lemma}
\begin{IEEEproof}
First represent an $n$-dimensional vector $x\in\F_{q^m}^n$ over $\F_{q^m}$
as an $m \times n$ matrix over the base field $\F_q$, denoted by $M_x \in \F_q^{m\times n}$.
Here we note that $\dimq \colq{xA^{\rm T}} = \rank M_x A^{\rm T}$.
We define by $\langle M_x \rangle \subseteq \F_q^n$ and $\langle A \rangle \subseteq \F_q^n$
row spaces of $M_x$ and $A$ over $\F_q$, respectively.
The rank of $M_xA^{\rm T}$ is given by
$\rank M_x A^{\rm T} = \rank A - \dim \left( \langle M_x \rangle^\perp \cap \langle A \rangle\right)$
\cite[Ch.~4, p.~210]{Meyer2001},
where $\langle M_x \rangle^\perp\in\F_q^n$ is the dual of $\langle M_x \rangle$
over $\F_q^n$.
If $\dim \langle M_x \rangle^\perp \leq n- \rho$, \ie if $\rank M_x = \dimq\colq{x} \geq \rho$,
we can always choose $A$ satisfying $\rank A=n-\rho$ and
$\langle A \rangle \supseteq \langle M_x \rangle^\perp$.
Then, for such $A$,
we have $\langle M_x \rangle^\perp = \langle M_x \rangle^\perp \cap \langle A \rangle$
and hence
\begin{align*}
\rank M_x A^{\rm T}
&= \underbrace{\rank A}_{n-\rho}
 - \dim \underbrace{\left( \langle M_x \rangle^\perp \cap \langle A \rangle\right)}_{=\langle M_x \rangle^\perp}\\
&= n- \rho - \underbrace{\dim \langle M_x \rangle^\perp}_{=n-\rank M_x}\\
&= \rank M_x - \rho\\
&= \dimq\colq{x}-\rho.
\end{align*}
On the other hand,
if $\dim \langle M_x \rangle^\perp > n- \rho$, \ie if $\rank M_x = \dimq\colq{x} < \rho$,
we can always choose $A$ satisfying $\rank A=n-\rho$ and
$\langle A \rangle \subsetneqq \langle M_x \rangle^\perp$.
Then, for such $A$,
we have $\langle A \rangle = \langle M_x \rangle^\perp \cap \langle A \rangle$
and hence
\begin{align*}
\rank M_x A^{\rm T}
&= \rank A - \underbrace{\dim \left( \langle M_x \rangle^\perp \cap \langle A \rangle\right)}_{=\rank A}
= 0.
\end{align*}
Therefore, the lemma is established.
\end{IEEEproof}\vspace{1.5ex}
For the rank deficiency $\rho = n - \rank A$,
we have $\left[ d_R(X,X') - \rho\right]^+ \leq d_R(XA^{\rm T},X'A^{\rm T})$ from \Lma{lma:meyer},
and there always exists $A \in \F_q^{N \times n}$ depending on $(X,X')$ such that the equality holds
from \Lma{lma:meyer2}.
Thus, from \Lma{eq:cosetdelta}, we have
the following inequalities for an arbitrarily fixed $A$ with $\rank A = n - \rho$.
\begin{align*}
&\min_{A \in \F_q^{N \times n}: \rank A = n-\rho}
\delta_A(\c_1/\c_2)\\*
&\quad =\left[ \min \left\{
d_R(X,X') : X,X' \in \c_1, X'-X \notin \c_2
\right\} - \rho
\right]^+
\\*
&\quad =
\left[
\min \left\{
d_R(X,0) : X \in \c_1, X \notin \c_2
\right\}-\rho
\right]^+
\\*
&\quad =
\left[
M_{R,1}(\c_1,\c_2) -\rho
\right]^+. \hspace{10ex} \text{(by \Lma{lma:rankdistance})}
\end{align*}
Thus, for $1 \leq i \leq \rho$, we have
\begin{align*}
\min_{A: \rank A = n-\rho} \delta_A(\c_1/\c_2)
<
\min_{A: \rank A = n-(\rho-i)} \delta_A(\c_1/\c_2),
\end{align*}
and hence we obtain
\begin{align*}
\min_{A: \rank A \geq n-\rho}\delta_A(\c_1/\c_2)
&=
\min_{A: \rank A = n-\rho} \delta_A(\c_1/\c_2)
\\
&= \left[M_{R,1}(\c_1,\c_2) - \rho\right]^+.
\end{align*}
Therefore, from \Prop{prop:deltacorrection} for $\mathcal{P}_\mathcal{S}=\c_1/\c_2$,
\Thm{thm:errorcorrectioncap} is proved.
\hfill\IEEEQED


\subsection{How to Prove \Prop{prop:errorcorrectioncap_noncoherent}}\label{sect:noncoherentcapability}
In this subsection,
we extend the analysis for the coherent network coding system, given in Appendix \ref{sect:coherentcapability},
to one for the noncoherent system in the setup of \Sect{sect:noncoherent}.
We derive \Prop{prop:errorcorrectioncap_noncoherent}, \ie an expression for the error correction capability of the lifting construction \cite{Silva2008}
 of the nested coset coding scheme in terms of the RGRW.
Here we recall that only one sink node has been assumed without loss of generality.

As in Appendix \ref{sect:coherentcapability},
we first consider the correction capability
of the generalized coding scheme defined in \Def{def:generalcoding}.
Recall that in the noncoherent network coding system,
the transfer matrix $A$ at the sink node is unknown.
Define the discrepancy function between $\mathcal{X} = \mathcal{X}_S \in \mathcal{P}_\mathcal{S}$ and $Y$
for unknown $A$ with at most $\rho$ rank deficiency,
as follows:
\begin{align}
&\Delta_{\rho} (\mathcal{X},Y)\nonumber\\
&\triangleq
\min \Big\{r \in \mathbb{N} :
\exists D \in \F_{q}^{N \times r},
\exists Z \in \F_{q^m}^{r},
\exists A \in \F_q^{N \times n},
\exists X \in \mathcal{X},\nonumber\\
&\hspace{23ex} Y^{\rm T} = AX^{\rm T} + DZ^{\rm T},
\rank A \geq n-\rho \Big\}
\nonumber\\
&=
\min\left\{
 \Delta_A(\mathcal{X},Y) :
 A \in \F_q^{N \times n},
 \rank A \geq n-\rho
\right\},\label{eq:defdelta_noncoherent}
\end{align}
where the second equality is obtained by \Eq{eq:defdiscrepancy}.
The definition of $\Delta_\rho(\mathcal{X},Y)$ represents the minimum number $r$ of error packets $Z$ required
to be injected in order to transform at least one element of $\mathcal{X}$ into $Y$, for at least one transfer matrix $A$ satisfying $\rank A\geq n- \rho$.
For $\Delta_\rho(\mathcal{X},Y)$, the minimum discrepancy decoder is given as
\begin{align}
\hat{X} = \argmin_{\mathcal{X} \in \mathcal{P}_\mathcal{S}} \Delta_\rho(\mathcal{X},Y).
\label{eq:noncoherentdecoder}
\end{align}
We also define $\Delta$-distance between $\mathcal{X}$ and $\mathcal{X}'$,
 induced by $\Delta_\rho(\mathcal{X},Y)$, as
\begin{align*}
&\delta_{\rho} (\mathcal{X},\mathcal{X}')\\
&\triangleq
\min \left\{
\Delta_{\rho}(\mathcal{X},Y)
+ \Delta_{\rho}(\mathcal{X}',Y) : Y \in \F_{q^m}^N
\right\}\\
&=
\min \Big\{
\Delta_A(\mathcal{X},Y)
+ \Delta_{A'}(\mathcal{X}',Y) :
A,A' \in \F_q^{N \times n}, Y \in \F_{q^m}^N,\\
&\hspace{23ex}\rank A \geq n- \rho,
\rank A' \geq n- \rho
\Big\},
\end{align*}
where the second equality is obtained by \Eq{eq:defdelta_noncoherent}.
Let $\delta_\rho(\mathcal{P}_\mathcal{S})$ be the minimum $\Delta$-distance given by
\begin{align*}
\delta_\rho(\mathcal{P}_\mathcal{S})
\triangleq
\min \left\{ \delta_\rho(\mathcal{X},\mathcal{X}') :
 \mathcal{X},\mathcal{X}' \in \mathcal{P}_\mathcal{S}, \mathcal{X}\neq\mathcal{X}'
\right\}.
\end{align*}
Observe that from \Lma{lma:discrepancy}, we can rewrite $\Delta_{\rho}(\mathcal{X},Y)$ as
\begin{align}
&\Delta_{\rho}(\mathcal{X},Y)\nonumber\\*
&=
\min\left\{
 d_R(XA^{\rm T},Y):
 X \in \mathcal{X},
 A \in \F_q^{N \times n},
 \rank A \geq n-\rho
\right\}.\label{eq:discrepancy_noncoherent}
\end{align}
Also, from \Lma{lma:deltadistance}, we have
\begin{align}
&\delta_{\rho} (\mathcal{X},\mathcal{X}')\nonumber\\
&=
\min
\Big\{
 d_R(XA^{\rm T},X'A'^{\rm T}):
 X \in \mathcal{X},
 X' \in \mathcal{X}',
 A, A'\in \F_q^{N \times n},\nonumber\\
&\hspace{21ex} \rank A \geq n-\rho,
 \rank A' \geq n-\rho
\Big\}. \label{eq:deltadistance_noncoherent}
\end{align}
\begin{lemma}\label{lma:normal_noncoherent}
The discrepancy function $\Delta_{\rho}(\mathcal{X},Y)$ is normal.
\end{lemma}
\begin{IEEEproof}
Let $\mathcal{X},\mathcal{X}'\in\mathcal{P}_\mathcal{S}$
and let $0 \leq i \leq d=\delta_\rho(\mathcal{X},\mathcal{X}')$.
Let $A,A'\in\F_q^{N \times n}$ be fixed matrices that minimize \Eq{eq:deltadistance_noncoherent},
and then suppose that $\bar{X}\in\mathcal{X}$ and $\bar{X'}\in\mathcal{X}'$ are vectors
 satisfying $d=d_R(\bar{X}A^{\rm T},\bar{X}'A'^{\rm T})$.
Here, we can always find two vectors $W,W'\in\F_{q^m}^n$ such that
$W+W' = \bar{X}'A'^{\rm T}-\bar{X}A^{\rm T}$,
 $\dimq\colq{W}=i$ and
 $\dimq\colq{W'}=d-i$,
as shown in the proof of \cite[Theorem 13]{Silva2009a}.
Taking $\bar{Y}=\bar{X}A^{\rm T}+W=\bar{X}'A'^{\rm T}-W'$,
we have
$d_R(\bar{X}A^{\rm T},\bar{Y}) = i$ and $d_R(\bar{X}'A'^{\rm T},\bar{Y}) = d-i$.
We thus obtain
$\Delta_\rho(\mathcal{X},\bar{Y}) \leq i$ and
$\Delta_\rho(\mathcal{X}',\bar{Y}) \leq d-i$ from \Eq{eq:discrepancy_noncoherent}.
On the other hand, since $\delta_\rho(\mathcal{X},\mathcal{X}')=d$,
we have $\Delta_\rho(\mathcal{X},Y) + \Delta_\rho(\mathcal{X}',Y) \geq d$
for any $Y \in \F_{q^m}^n$ from \Eq{eq:defdelta_noncoherent}.
Therefore, 
$\Delta_\rho(\mathcal{X},\bar{Y}) = i$ and $\Delta_\rho(\mathcal{X}',\bar{Y})
 = d-i$
hold.
\end{IEEEproof}\vspace{1.5ex}
As \cite[Theorem 14]{Silva2009a}, we have \Prop{prop:deltacorrection_noncoherent} from \Thm{thm:silvacorrection}
and \Lma{lma:normal_noncoherent} by the approach of Appendix \ref{sect:silvaapproach}.

Then, we have the following proposition by the approach of Appendix \ref{sect:silvaapproach}.
\begin{proposition}\label{prop:deltacorrection_noncoherent}
Consider the $t$-error $(n \times m)_q$ linear network in \Def{def:terror}.
Suppose that for a secret message $S\in\mathcal{S}$,
the transmitted $n$ packets $X\in\mathcal{X}$ are generated by a coding scheme defined in \Def{def:generalcoding}.
Then, the minimum discrepancy decoder for $\Delta_\rho(\mathcal{X},Y)$
is infallible
if and only if $t \leq \lfloor (\delta_\rho(\mathcal{P}_\mathcal{S})-1)/2 \rfloor$.
\hfill\IEEEQED
\end{proposition}
This proposition implies that
the coding scheme given in \Def{def:generalcoding} is guaranteed to
determine the unique set $\mathcal{X}$ against
any $t$ packet errors if and only if
$\delta_\rho(\mathcal{P}_\mathcal{S}) > 2t$.

In the following,
 we restrict $X$ to that generated by the lifting construction \cite{Silva2008}
 of the nested coset coding scheme, as described in \Sect{sect:noncoherent},
and we shall express the error correction capability given in \Prop{prop:deltacorrection_noncoherent}
in terms of the RGRW.
Recall that in the lifting construction of the nested coset coding scheme,
$\c_1 \subseteq \F_{q^{\widetilde{m}}}^n$ and $\c_2 \subsetneqq \c_1$
are a linear code and its subcode for $\widetilde{m} = m-n$, respectively.
Also recall that
$\mathcal{S} = \F_{q^{\widetilde{m}}}^l$,
$\mathcal{X}_S=\mathcal{X}_{S,\textsf{lift}}$
and
$\mathcal{P}_\mathcal{S} = \mathcal{P}_{\textsf{lift}}$
defined in \Eq{eq:xslift}.
We will consider the error correction capability in this setup.
Here, we introduce the following proposition given in \cite{Silva2009a}.
\begin{proposition}[{\cite[Proposition 18]{Silva2009a}}]\label{prop:silvaprop18}
For $X,X'\in\F_{q^m}^n$, we have
\begin{align*}
&\min
\Big\{
d_R(XA^{\rm T},X'A'^{\rm T}) : A, A' \in \F_q^{N \times n},\\*
&\hspace{20ex} \rank A \geq n - \rho, \rank A' \geq n - \rho
\Big\}\\
&=
\left[
\dimq\left(\colq{X} \!+\! \colq{X'}\right)
 \!-\! \min \left\{ \dimq\colq{X}, \dimq\colq{X'} \right\}
 \!-\! \rho
\right]^{+}.
\end{align*}
\end{proposition}
From this proposition, we have the following lemma.
\begin{lemma}\label{lma:rankdistancenoncoherent}
\begin{align*}
\delta_\rho(\mathcal{P}_{\textsf{lift}})=
\min\left\{
\left[d_R(\widetilde{X},\widetilde{X}') - \rho \right]^{+}
: \widetilde{X}, \widetilde{X}' \in \c_1, \widetilde{X}'- \widetilde{X} \notin \c_2
\right\}.
\end{align*}
\end{lemma}
\begin{IEEEproof}
Since the transmitted packets are generated by the lifting construction of the nested coset coding scheme,
we have $\dimq \colq{X} = n$ for all $X\in\mathcal{X}$ and for all $\mathcal{X} \in \mathcal{P}_{\textsf{lift}}$.
For $X\in\mathcal{X}\in\mathcal{P}_{\textsf{lift}}$
and $X'\in\mathcal{X}'\in\mathcal{P}_{\textsf{lift}}$,
we thus have
\begin{align*}
&\dimq\left( \colq{X}+\colq{X'}\right)
 - \min \left\{ \underbrace{\dimq\colq{X}}_{=n}, \underbrace{\dimq\colq{X'}}_{=n} \right\}
 - \rho
\\
&\qquad=
\rank
\begin{bmatrix}
I & I\\
\phi_{\widetilde{m}}(\widetilde{X}) & \phi_{\widetilde{m}}(\widetilde{X}')
\end{bmatrix}
-
\min\{n, n\} - \rho
\\
&\qquad=
\underbrace{
\rank
\begin{bmatrix}
I & 0\\
\phi_{\widetilde{m}}(\widetilde{X}) & \phi_{\widetilde{m}}(\widetilde{X}') - \phi_{\widetilde{m}}(\widetilde{X})
\end{bmatrix}
}_{=n+ \rank(\phi_{\widetilde{m}}(\widetilde{X}') - \phi_{\widetilde{m}}(\widetilde{X}))}
-
n - \rho
\\
&\qquad=
\rank (
 \underbrace{\phi_{\widetilde{m}}(\widetilde{X}')-\phi_{\widetilde{m}}(\widetilde{X})}_{=\phi_{\widetilde{m}}(\widetilde{X'}-\widetilde{X})}
) -\rho
\\
&\qquad=
\dimq \colq{\widetilde{X}'-\widetilde{X}}
-\rho
\\
&\qquad=d_R(\widetilde{X},\widetilde{X}') - \rho,
\end{align*}
where in the first equality,
$X$ and $X'\in\F_{q^m}^n$ are regarded as $m \times n$ matrices over $\F_q$,
$X^{\rm T} = \begin{bmatrix} I & \phi_{\widetilde{m}}(\widetilde{X})^{\rm T} \end{bmatrix}$
and
$X'^{\rm T} = \begin{bmatrix} I & \phi_{\widetilde{m}}(\widetilde{X'})^{\rm T} \end{bmatrix}$,
respectively.
Thus, by combining \Prop{prop:silvaprop18} and \Eq{eq:deltadistance_noncoherent},
 we have the following equation
for $\mathcal{X}_{S,\textsf{lift}},\mathcal{X}_{S',\textsf{lift}}\in\mathcal{P}_{\textsf{lift}}$.
\begin{align*}
&\delta_\rho(\mathcal{X}_{S,\textsf{lift}},\mathcal{X}_{S',\textsf{lift}})\\*
&\quad =
\min\left\{
\left[d_R(\widetilde{X},\widetilde{X}') - \rho \right]^{+}:
X  \in \mathcal{X}_{S,\textsf{lift}},
X' \in \mathcal{X}_{S',\textsf{lift}}
\right\}
\\*
&\quad =
\min\left\{
\left[d_R(\widetilde{X},\widetilde{X}') - \rho \right]^{+}:
\widetilde{X} \in \psi(S), \widetilde{X}' \in \psi(S')
\right\}. \hspace{1ex} \text{(by \Eq{eq:xslift})}
\end{align*}
Therefore, we finally have
\begin{align*}
&\delta_\rho(\mathcal{P}_{\textsf{lift}})
\\*
&=
\min \left\{ \delta_\rho(\mathcal{X},\mathcal{X}') :
 \mathcal{X},\mathcal{X}' \in \mathcal{P}_{\textsf{lift}}, \mathcal{X}\neq\mathcal{X}'
\right\}\\*
&=
\min\Biggl\{
\min\left\{
\left[d_R(\widetilde{X},\widetilde{X}') - \rho \right]^{+}:
\widetilde{X} \in \psi(S), \widetilde{X}' \in \psi(S')
\right\}
:\\*
&\hspace{23ex}
\psi(S),\psi(S') \in \c_1/\c_2, \psi(S) \neq \psi(S')
\Biggr\}\\*
&=
\min\Big\{
\left[d_R(\widetilde{X},\widetilde{X}') - \rho \right]^{+}:
\widetilde{X} \in \psi(S) \in \c_1/\c_2,\\*
&\hspace{24ex}\widetilde{X}' \in \psi(S') \in \c_1/\c_2,
\psi(S) \neq \psi(S')
\Big\}\\*
&=
\min\left\{
\left[d_R(\widetilde{X},\widetilde{X}') - \rho \right]^{+}:
\widetilde{X}, \widetilde{X}' \in \c_1,
\widetilde{X}'-\widetilde{X} \not\in \c_2
\right\}.
\end{align*}
\end{IEEEproof}\vspace{1.5ex}
From \Lma{lma:rankdistancenoncoherent}, we have
\begin{align*}
\delta_\rho(\mathcal{P}_{\textsf{lift}}) + \rho
&=
\min\left\{
d_R(\widetilde{X},\widetilde{X}')
: \widetilde{X}, \widetilde{X}' \in \c_1, \widetilde{X}'- \widetilde{X} \notin \c_2
\right\}\\
&=
\min\left\{
d_R(\widetilde{X},0)
: \widetilde{X} \in \c_1, \widetilde{X} \notin \c_2
\right\}\\
&=
M_{R,1}(\c_1,\c_2). \hspace{10ex} \text{(by \Lma{lma:rankdistance})}
\end{align*}
Thus, from \Prop{prop:deltacorrection_noncoherent} for $\mathcal{P}_\mathcal{S}=\mathcal{P}_{\textsf{lift}}$,
\Prop{prop:errorcorrectioncap_noncoherent} is proved.
\hfill\IEEEQED

\section*{Acknowledgment}
The authors would like to thank Prof.\@ Terence H.\@ Chan for
pointing out the fact that the conditions for zero mutual
information are independent of the probability distribution
of secret messages.
The authors would also like to thank the Associate Editor and the anonymous reviewers for their many helpful comments which have greatly improved the presentation of this paper.
A part of this research was done during the
first author's stay at Palo Alto Research Center (PARC), CA, USA.
He greatly appreciates the support by Dr.\ Ersin Uzun.
Another part of this research was also done during the
second author's stay at Aalborg University, Denmark.
He greatly appreciates the support by Profs.\ Olav Geil and Diego Ruano.



\begin{IEEEbiographynophoto}{Jun Kurihara}
(M'13) received the B.E. degree in computer science, the M.E. degree in communication engineering and the Ph.D. degree in electrical and electronic engineering, all from Tokyo Institute of Technology, Tokyo, Japan, in 2004, 2006 and 2012 respectively.
He joined KDDI Corp., Tokyo, Japan in April 2006. Since July 2006, he has been with KDDI R\&D Laboratories, Inc., Saitama, Japan as a researcher.
From 2013 to 2014, he was a visiting researcher at Palo Alto Research Center (PARC), Palo Alto, CA, USA.
His research interests include coding theory, networking architecture and security.
He received the Best Paper Award from IEICE in 2014.
\end{IEEEbiographynophoto}
\begin{IEEEbiographynophoto}{Ryutaroh Matsumoto}
(M'00) was born in Nagoya, Japan, on November
29, 1973. He received the B.E. degree in computer science, the M.E. degree
in information processing, and the Ph.D. degree in electrical and electronic
engineering, all from Tokyo Institute of Technology, Japan, in 1996, 1998 and
2001, respectively. He was an Assistant Professor from 2001 to 2004, and has
been an Associate Professor since 2004 in the Department of Communications
and Computer Engineering, Tokyo Institute of Technology. His research interests
include error-correcting codes, quantum information theory, information
theoretic security, and communication theory. Dr. Matsumoto received the
Young Engineer Award from IEICE and the Ericsson Young Scientist Award
from Ericsson Japan in 2001. He received the Best Paper Awards from IEICE
in 2001, 2008, 2011 and 2014.
\end{IEEEbiographynophoto}
\begin{IEEEbiographynophoto}{Tomohiko Uyematsu}
(M'95--SM'05) received the B.E., M.E.~and Dr.Eng.~degrees
from Tokyo Institute of Technology in 1982, 1984 and 1988,
respectively. From 1984 to 1992, he was with the Department of
Electrical and Electronic Engineering of Tokyo Institute of Technology,
first as research associate, next as lecturer, and lastly as associate
professor. From 1992 to 1997, he was with School of Information Science
of Japan Advanced Institute of Science and Technology as associate
professor. Since 1997, he returned to Tokyo Institute of Technology as
associate professor, and currently he is with the Department of
Communications and Computer Engineering as professor.
In 1992 and 1996, he was a visiting researcher at Sup\'elec, France
and Delft University of Technology, Netherlands,
respectively. He was Technical Program Committee Co-Chair for the 2012 International Symposium on Information Theory and its Applications. He served as an Associate Editor for the IEEE Transactions on Information Theory in 2010--2013.
He received the Achievement Award in 2008, and the Best
Paper Award six times both from IEICE. His current
research interests are in the areas of information theory, especially
Shannon theory and multi-terminal information theory.
\end{IEEEbiographynophoto}

\end{document}